\documentclass[twocolumn, twocolappendix]{aastex63}

\usepackage{graphicx}
\usepackage{natbib}
\usepackage{gensymb}
\hypersetup{linkcolor=blue,citecolor=blue,filecolor=blue,urlcolor=blue}

\usepackage{longtable}
\usepackage{threeparttablex}
\usepackage{multirow}
\usepackage{footmisc}

\usepackage{amsmath}
\usepackage{verbatim}

\newcommand{\HI}{H{\sc\ i}}
\newcommand{\HII}{H{\sc\ ii}}
\newcommand{\OI}{O{\sc\ i}}
\newcommand{\AlII}{Al{\sc\ ii}}
\newcommand{\SiII}{Si{\sc\ ii}}
\newcommand{\SiIII}{Si{\sc\ iii}}
\newcommand{\SiIV}{Si{\sc\ iv}}
\newcommand{\SII}{S{\sc\ ii}}
\newcommand{\CII}{C{\sc\ ii}}
\newcommand{\CIIS}{C{\sc\ ii*}}
\newcommand{\CIV}{C{\sc\ iv}}
\newcommand{\FeII}{Fe{\sc\ ii}}
\newcommand{\PII}{P{{\sc\ ii}}}
\newcommand{\msun}{\rm M_\odot}
\newcommand{\msunyr}{\msun~{\rm yr^{-1}}}
\newcommand{\kms}{{\rm km \, s^{-1}}}
\newcommand{\mkms}{km s$^{-1}$}

\newcommand{\logNaod}{\log N_{\rm AOD}}

\newcommand{\mb}{B_{\rm MS}}
\newcommand{\ml}{L_{\rm MS}}

\newcommand{\vgal}{v_{\rm IC1613}}
\newcommand{\vesc}{v_{\rm esc}}
\newcommand{\ew}{W_{\rm r}}
\newcommand{\moutz}{\dot{M}_{\rm out, Z}}

\newcommand{\mstar}{M_*}

\newcommand{\blue}{\color{blue}}

\newcommand{\rvir}{\rm R_{200}}

\newcommand{\vlsr}{v_{\rm LSR}}
\newcommand{\mvlsr}{$\vlsr$}
\newcommand{\fluxunit}{erg~cm$^{-2}$~s$^{-1}$~\AA$^{-1}$}
\newcommand{\hsla}{\href{https://archive.stsci.edu/missions-and-data/hst-spectroscopic-legacy-archive-hsla}{HSLA}}

\shorttitle{The CGM and Outflows of IC1613}
\shortauthors{Zheng et~al.}

\begin{document}
\defcitealias{mcconnachie12}{M12}
\defcitealias{richter17}{Richter17}
\defcitealias{fox14}{Fox14}
\defcitealias{Zheng19b}{Zheng19}
\defcitealias{bordoloi14}{Bordoloi14}
\defcitealias{liang14}{Liang14}
\defcitealias{johnson17}{Johnson17}

\title{Characterizing the Circumgalactic Medium of the Lowest-Mass Galaxies: A Case Study of IC 1613}

.


\author[0000-0003-4158-5116]{Yong Zheng}
\affiliation{Department of Astronomy, University of California, Berkeley, CA 94720; {\blue yongzheng@berkeley.edu}}
\affiliation{Miller Institute for Basic Research in Science, University of California, Berkeley, CA 94720}


\author[0000-0003-2807-328X]{Andrew Emerick}
\altaffiliation{Carnegie Fellow in Theoretical Astrophysics}
\affiliation{Carnegie Observatories, Pasadena, CA, 91101, USA}
\affiliation{TAPIR, California Institute of Technology, Pasadena, CA, 91125, USA}

\author[0000-0002-1129-1873]{Mary E. Putman}
\affiliation{Department of Astronomy, Columbia University, New York, NY 10027, USA}

\author[0000-0002-0355-0134]{Jessica K. Werk}
\affiliation{Department of Astronomy, University of Washington, Seattle, WA 98195, USA}

\author[0000-0001-6196-5162]{Evan N. Kirby}
\affiliation{Astronomy Department, California Institute of Technology, Pasadena, CA 91125, USA}

\author[0000-0003-4797-7030]{Joshua Peek}
\affiliation{Space Telescope Science Institute, 3700 San Martin Drive, Baltimore, MD 21218, USA}

\begin{abstract}

Using 10 sightlines observed with the Hubble Space Telescope/Cosmic Origins Spectrograph, we study the circumgalactic medium (CGM) and outflows of IC1613, which is a low-mass ($M_*\sim10^8~M_\odot$), dwarf irregular galaxy on the outskirts of the Local Group. Among the sightlines, 4 are pointed towards UV-bright stars in IC1613, and the other 6 sightlines are background QSOs at impact parameters from 6 kpc ($<0.1R_{200}$) to 61 kpc ($0.6R_{200}$). We detect a number of Si{\sc\ ii}, Si{\sc\ iii}, Si{\sc\ iv}, C{\sc\ ii}, and C{\sc\ iv} absorbers, most of which have velocities less than the escape velocity of IC1613 and thus are gravitationally bound. The line strengths of these ion absorbers are consistent with the CGM absorbers detected in dwarf galaxies at low redshifts. Assuming that Si{\sc\ ii}, Si{\sc\ iii}, and Si{\sc\ iv} comprise nearly 100\% of the total silicon, we find 3\% ($\sim$8$\times$10$^3~{\rm M_\odot}$), 2\% ($\sim$7$\times$10$^3~{\rm M_\odot}$), and 32--42\% [$\sim$(1.0--1.3)$\times$10$^5~{\rm M_\odot}$] of the silicon mass in the stars, interstellar medium, and within $0.6R_{200}$ of the CGM of IC1613. We also estimate the metal outflow rate to be ${\rm \dot{M}_{out, Z}\geq1.1\times10^{-5}~M_\odot~yr^{-1}}$ and the instantaneous metal mass loading factor to be $\eta_{\rm Z}\geq0.004$, which are in broad agreement with available observation and simulation values. This work is the first time a dwarf galaxy of such low mass is probed by a number of both QSO and stellar sightlines, and it shows that the CGM of low-mass gas-rich galaxies can be a large reservoir enriched with metals from past and ongoing outflows.  


\end{abstract}


\section{Introduction}
\label{sec:intro}

Galaxies at redshift $\lesssim2.5$  have lost the majority of the metals produced over their star formation histories, giving rise to the so-called missing metals problem \citep[e.g., ][]{bouche07, peeples14}. For instance, \cite{peeples14} show that local star-forming galaxies with stellar masses $\mstar=10^{9.3-11.6}~\msun$ only contain 20--25\% of metals in their stars and interstellar medium (ISM). Detailed studies of single galaxies yield similar results. For example, \cite{Telford19} find that 62\% of the metal mass formed within $r<$19 kpc is missing from M31's disk based on resolved star-formation history analyses with data from the Panchromatic Hubble Andromeda Treasury \citep{dalcanton12}. The missing metals problem is found to be the most severe in low-mass dwarf galaxies, which contain fewer metals than their higher-mass counterparts according to the gas-phase \citep{tremonti04, lee06, andrews13} and stellar \citep{gallazzi05, kirby13} mass-metallicity relations. In the Local Group (LG), 
dwarf galaxies are found to have lost $\gtrsim96\%$ of the iron they have synthesized through star formation \citep{kirby11, kirby13}, with the missing iron located either in their ISM or CGM, and with some fraction possibly having escaped the galaxies altogether. While processes such as metal-poor gas infall or low star-formation efficiency could contribute to the low metal abundances \citep[e.g., ][]{brooks07, calura09}, low-mass dwarf galaxies are prone to lose more metals via outflows due to their shallow gravitational potential \citep[e.g.][]{maclow99,ma16,muratov17,christensen18,emerick18b,romano19}.

It remains to be seen if the rest of the metals, if not in the main bodies of the galaxies, are within their circumgalactic medium (CGM) or have been ejected into the intergalactic medium (IGM). Cosmological and idealized hydrodynamic simulations have widely explored the metal content in dwarf and higher-mass galaxies \citep[e.g.][]{brooks07, vogelsberger14, ma16}; however, only a few have focused on the distribution of metals in the CGM of low-mass dwarf galaxies with $\mstar\lesssim10^{8.5}~\msun$ \citep{muratov17, christensen18, Hafen19}.  
For example, \cite{muratov17} show that the metal content in a galaxy's CGM closely follows the star-formation and outflow activities; for a dwarf galaxy with $\mstar\sim10^{8.5}~\msun$,
the CGM has gained most of its current metal mass ($10^{6.7}~\msun$) since $z=1$, 
and at $z=0$ the metals in the CGM account for $\sim40\%$ of the total metal mass.
Similarly, \cite{christensen18} show that for galaxies with $\mstar\lesssim10^{8.5}~\msun$, less than $10\%$ of the metals are retained in stars, $\sim10-30\%$ of the metals are in the ISM, and the rest are either in the CGM of the galaxies or have escaped beyond the virial radii.

\begin{deluxetable}{ccc}
\tablenum{1}
\tablecaption{IC1613 Information \label{tb:ic1613}}
\tablewidth{0pt}
\tablehead{
\colhead{Variable} & \colhead{Value} & \colhead{References} 
}
\startdata
R.A. & 01h04m54.2s (16.2258$\degree$) & 1 \\
\hline
DEC  & +02d08m00s  (2.1333$\degree$)  & 1 \\
\hline
D$_\odot$ & 755$\pm42$ kpc  & 2, 3 \\
\hline
v$_{\rm helio}$ & -232 \mkms          & 2, 4 \\
\hline
\mvlsr &  -236 \mkms                  & -- \\
\hline
M$_{\rm HI}$  & $6.5\times10^7~\msun$ & 2, 4, 5 \\
\hline
$\sigma_{\rm HI}$ & $25.0\pm3.0~\kms$ & 2, 4 \\
\hline
12+log(O/H)                          & $7.73\pm0.04$ & 6 \\
\hline
M$_*$ & $10^8~\msun$  & 7 \\
\hline
M$_{h}$ & $4\times10^{10}~\msun$ & 8 \\
\hline
R$_{\rm 200}$ & 107 kpc & 9 \\
\hline
SFR &  0.0025 $\rm M_\odot yr^{-1}$ & 10 \\ 
\hline
$\theta$ & 37.9$\degree$ & 11 \\ 
\hline
\hline
\enddata
\tablecomments{
(1) Simbad. 
(2) \cite{mcconnachie12}. 
(3) \cite{bernard10}. 
(4) \cite{lake89}. 
(5) \cite{silich06}. 
(6) \cite{bresolin07}, from \HII\ regions. 
(7) \cite{mcconnachie12}, stellar mass, assuming a stellar mass to light ratio of 1. 
(8) Dark-matter halo mass, converted from $M_*$ based on the M$_*$-M$_h$ relation from \cite{moster10}. 
(9) Virial radius, defined with respect to 200 times the matter density $\rho_m\equiv \rho_c \Omega_m$.
(10) \cite{hunter04}. 
(11) Inclination angle determined from \HI\ observation using VLA \citep{hunter12}. 
}
\end{deluxetable}

%

Observationally, the quest to find baryons in dwarf galaxies' CGM has been limited to a few low-mass members in the LG \citep[e.g.][]{bowen97, richter17, Zheng19b} and at low redshift \citep{bordoloi14, liang14, johnson17}. For instance, \cite{bordoloi14} find a large reservoir of carbon with mass of $\geq1.2\times10^6~\msun$ in the CGM of 43 low-mass galaxies ($\mstar\sim10^{8.2-10.2}~\msun$) at redshift $\leq0.1$. Most of their \CIV\ detection occur within $0.5$ virial radius, beyond which no \CIV\ is detected at a sensitivity limit of 50--100 m\AA. In the LG,  \cite{Zheng19b} find a total mass of (0.2--1.0)$\times10^5~\msun$ detected in \SiII, \SiIII, and \SiIV\ in the CGM of the dwarf galaxy WLM. Their detection is deemed tentative given the uncertain contamination from the Magellanic Stream in the foreground. In this work, we will address the Magellanic contamination in the context of the CGM absorbers of gas-rich galaxies, including IC1613, in the LG. 

\begin{deluxetable*}{cccccccc}
\tablenum{2}
\tablecaption{Target Information \label{tb:starqso}}
\tablewidth{0pt}
\tablehead{
\colhead{Star ID} & \colhead{Target name} & \colhead{R.A. (J2000)} & \colhead{Dec (J2000)} & \colhead{$v_{\rm HI}^{[a]}$} & SNR$^{[b]}$ & \colhead{Spec. Type$^{[c]}$} & \colhead{PI, Program$^{[d]}$} \\
 & & (degree) & (degree) & (\mkms) & & &
}
\startdata
S1 & \href{http://simbad.u-strasbg.fr/simbad/sim-id?Ident=\%404088384\&Name=\%5bBUG2007\%5d\%20C\%2010\&submit=submit}{IC1613-C10}
& 16.1806 & +2.1732 & -239.4 & 12.5 & B1.5Ib & \href{http://archive.stsci.edu/proposal_search.php?mission=hst&id=15156}{Zheng, 15156} \\
S2 & \href{http://simbad.u-strasbg.fr/simbad/sim-id?Ident=\%404088419\&Name=\%5bBUG2007\%5d\%20B\%20\%207\&submit=submit}{IC1613-B7$^{[e]}$}
& 16.2581 & +2.1351 & -234.3 & 15.2 & O9I & \href{http://archive.stsci.edu/proposal_search.php?mission=hst&id=15156}{Zheng, 15156} \\
S3 & \href{http://simbad.u-strasbg.fr/simbad/sim-id?Ident=\%5BBUG2007\%5D+A13\&NbIdent=1\&Radius=2\&Radius.unit=arcmin\&submit=submit+id}{IC1613-A13}
& 16.2759 & +2.1791 & -231.7 & 11.8 & O3?O4v((f)) & \href{http://archive.stsci.edu/proposal_search.php?mission=hst&id=12867}{Lanz, 12867}\\
S4 & \href{http://simbad.u-strasbg.fr/simbad/sim-id?Ident=\%5BBUG2007\%5D+B11\&NbIdent=1\&Radius=2\&Radius.unit=arcmin\&submit=submit+id}{IC1613-B11}
& 16.1826 & +2.1128 & -242.0 & 10.5 & O9.5I & \href{http://archive.stsci.edu/proposal_search.php?mission=hst&id=12867}{Lanz, 12867}\\
\hline
\hline 
QSO ID & Target name & R.A. (J2000) & Dec (J2000) & $z^{[f]}$ & SNR$^{[b]}$ 
& $d_{\rm IC1613}^{[g]}$ & Program$^{[d]}$\\ 
 & & (degree) & (degree) & &  & (kpc)  & \\ 
\hline 
Q1 & \href{http://simbad.u-strasbg.fr/simbad/sim-id?Ident=LBQS+0100\%2B0205\&NbIdent=1&Radius=2\&Radius.unit=arcmin\&submit=submit+id}{LBQS-0100+0205} 
& 15.8041 & 2.3528 & 0.393 & 8.9 & 6.3 & \href{http://archive.stsci.edu/proposal_search.php?mission=hst&id=15156}{Zheng, 15156}\\
Q2 & \href{http://simbad.u-strasbg.fr/simbad/sim-id?Ident=LBQS+0101\%2B0009\&submit=submit+id}{LBQS-0101+0009}
& 15.9281 & 0.4270 & 0.394 & 7.8 & 22.8 & \href{http://archive.stsci.edu/proposal_search.php?mission=hst&id=15156}{Zheng, 15156}\\
Q3 & \href{http://simbad.u-strasbg.fr/simbad/sim-id?Ident=2MASX+J01022632-0039045\&submit=submit+id}{2MASX J01022632-0039045} 
& 15.6097 & -0.6513 & 0.296 & 8.6 & 37.6 & \href{http://archive.stsci.edu/proposal_search.php?mission=hst&id=15156}{Zheng, 15156}\\
Q4 & \href{http://simbad.u-strasbg.fr/simbad/sim-id?Ident=PG0044\%2B030\&submit=submit+id}{PG 0044+030} & 11.7746  & 3.3319 & 0.624 & 6.3  & 60.7 & \href{http://archive.stsci.edu/proposal_search.php?mission=hst&id=12275}{Wakker, 12275}\\
Q5 & \href{http://simbad.u-strasbg.fr/simbad/sim-id?Ident=\%401372357\&Name=QSO\%20J0110-0218\&submit=submit}{HB89-0107-025-NED05} & 17.5677  & -2.3142 & 0.956 & 11.7  & 61.2 & \href{https://www.stsci.edu/cgi-bin/get-proposal-info?id=11585&observatory=HST}{Crighton, 11585}\\
Q6 & \href{http://simbad.u-strasbg.fr/simbad/sim-id?Ident=LBQS+0107-0235\&submit=submit+id}{LBQS-0107-0235} & 17.5547  & -2.3314 & 0.957 & 12.2 & 61.4 & \href{https://www.stsci.edu/cgi-bin/get-proposal-info?id=11585&observatory=HST}{Crighton, 11585} \\
\enddata
\tablecomments{[$a$]: the systemic velocity of the ISM gas along the line of sight, measured from \HI\ 21cm emission from the VLA datacube. [$b$]: Signal-to-noise ratio (SNR) per resolution element for coadded spectra. Six pixels are assumed per resolution element for both G130M and G160M gratings. For each target, the SNR value is averaged over 8 absorption-line free locations at 1120, 1170, 1320, 1370, 1420, 1470, 1520, 1620 \AA. At each location, the SNR over a 10\AA\ spectral window is calculated. [$c$]: spectral types from Simbad for stellar sightlines. [$d$]: PI and Program ID for each sightline. [$e$]: This target was named as IC1613-010502-020805 in proposal GO15156. [$f$]: redshifts of the QSOs. [$g$]: Impact parameter, or transverse distance, between the target and IC1613. }
\end{deluxetable*}

\begin{figure*}[t]
\includegraphics[width=\textwidth]{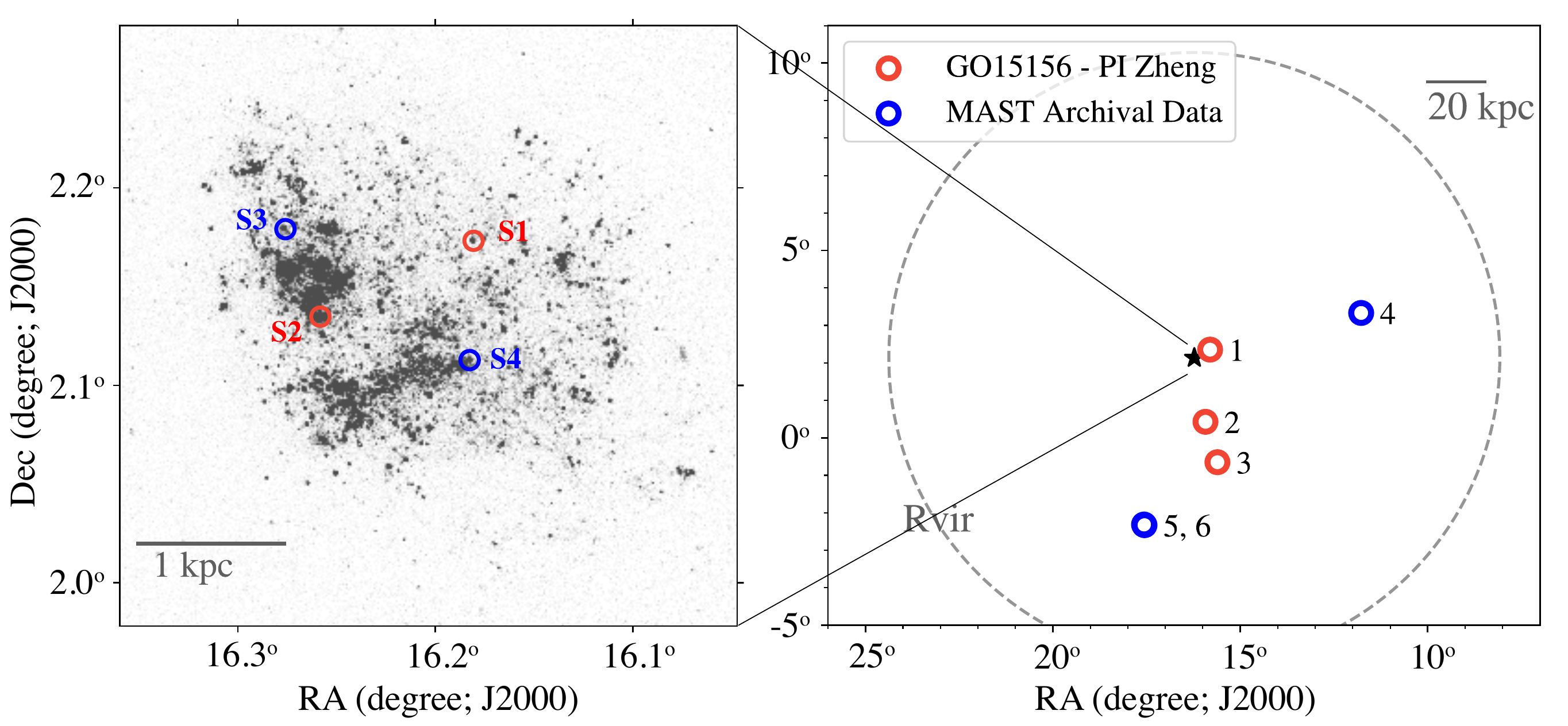}
\caption{Left: we show the locations of the 4 UV-bright stars in IC1613 against a FUV background image from the GALEX Ultraviolet Atlas of Nearby Galaxies \citep{galex}. Right: distribution of the 6 QSO sightlines within $\rvir$ of IC1613. Circles in red are new data observed with GO15156 (PI Zheng), and those in blue were retrieved from the STScI/MAST archive observed by previous programs (see Table \ref{tb:starqso}).
}
\label{fig:starqso}
\end{figure*}

IC1613 is a dwarf irregular galaxy on the outskirts of the LG. With 4 stellar sightlines in the galaxy and 6 QSO sightlines in its halo observed with HST/COS, we seek to understand (1) how the metals are distributed in the stars, ISM, and CGM of IC1613, and (2) how the metals travel to the CGM and what the instantaneous metal mass loading factor is. At $\mstar=10^8~\msun$ (see Table \ref{tb:ic1613}), IC1613 is among the lowest mass galaxies to have been studied in the context of the CGM metal content and outflows. And with 10 sightlines at $<0.6\rvir$ (see Figure \ref{fig:starqso}), it is one of the rare cases where the CGM is probed by numerous QSO sightlines, with the exception of the Milky Way \citep[e.g., ][]{putman12, richter17, zheng19a} and M31 \citep{howk17, lehner20}.

IC1613 is an excellent candidate to study the CGM and metal flows for a number of reasons. First, it is isolated from other galaxies in the LG, with the nearest neighbor (M33) 400 kpc away \citep{hunter04}. Therefore, the galaxy's halo does not overlap with other halos. Second, IC1613 is on the outskirts of the LG, in which case the gas in the galaxy has not been stripped off due to ram pressure and the galaxy's CGM is most likely to remain intact. Other galaxies that are close to the Milky Way or M31 have been found with their gas content largely stripped (\citealt{grcevich09}; Putman et al. 2020). Lastly, the galaxy has had a continuous and nearly constant star formation rate over the past $>$10 Gyrs
\citep{cole99, skillman03, skillman14, weisz14}, which is conducive to a metal-enriched CGM and the presence of current outflows. 




This paper is structured as follows: In \S\ref{sec:method}, we elaborate on the data reduction, including spectral co-addition, continuum normalization, Voigt-profile fitting, line measurements, and auxiliary \HI\ datasets. In \S\ref{sec:ic1613}, we study the connection between the detected absorbers and IC1613, and in \S\ref{sec:pv_ms} we discuss the presence of the Magellanic Stream in the foreground. In \S\ref{sec:metal_budget}, we estimate the metal budget of IC1613 and the outflow's instantaneous metal mass loading factor and compare the results to predicted values from simulations. In \S\ref{sec:discussion}, we compare our results with those of other dwarf galaxies at low redshifts and in the LG. We conclude in \S\ref{sec:summary}.

\section{Data and Measurements}
\label{sec:method}

In Table \ref{tb:ic1613}, we summarize IC1613's key properties that are used throughout this paper. The halo mass of IC1613, $M_h=4\times10^{10}~\msun$, is estimated from $\mstar$ using the $\mstar$--$M_h$ relation from \cite{moster10}. Note that at $\mstar=10^8~\msun$ the $\mstar$--$M_h$ relation is highly uncertain. Our derived mass is consistent with the allowed $M_h$ range derived for low-mass galaxies  ($M_*<10^8~\msun$; \citealt{gk14, gk17}). We arbitrarily define the boundary between the CGM and IGM as the galaxy's virial radius.  Following the definition used by COS-Halos \citep{werk13} and COS-Dwarfs \citep{bordoloi14}, we calculate the virial radius of IC1613 as ${\rm \rvir=(3/4\pi M_{halo}}/200\rho_m)^{1/3}=107$ kpc, where $\rho_{\rm m}=\rho_{\rm c}\Omega_{\rm m}$ is the cosmic critical matter density at $z=0$. Moreover, when examining the gas kinematics, we only consider CGM gas to be those absorbers with velocity less than the escape velocity of IC1613 at the corresponding impact parameter (see \S\ref{sec:ic1613}).


Our dataset includes 6 QSO sightlines (Q1--Q6) within the virial radius ($\rvir=107$ kpc or $\sim8\degree$ at $d_\odot=755$ kpc) of IC1613 and 4 UV-bright OB star sightlines (S1--S4) in the galaxy itself, as shown in Figure \ref{fig:starqso} and Table \ref{tb:starqso}. Among them, Q1, Q2, Q3, S1, and S2 (red dots in Figure \ref{fig:starqso}) were observed with HST/COS program GO \href{https://archive.stsci.edu/proposal_search.php?mission=hst&id=15156}{15156}. Because of a delayed guide-star acquisition failure, one of the HST visits for S2 that occurred on 11/25/2018 did not yield usable data. We filed a Hubble Observation Problem Report (HOPR 91429) and re-observed S2 for one more visit on 12/24/2018. Our following analysis of S2 includes data from the new visit and the usable spectra from the original observation.  We process all of the QSO and star spectra consistently as outlined below. 

The rest of the targets were retrieved from the STScI/MAST archive observed by previous programs (see Table \ref{tb:starqso}). The archival target list was decided on July 2019 when our last search for publicly available sightlines occurred. In addition to the adopted sightlines, we found a few other QSO spectra near IC1613 but decided not to use them because of the low SNR of the spectra. 

\subsection{Spectral Co-addition}
\label{sec:coadd}

We focus on data co-addition products from two community co-adding routines: \href{https://archive.stsci.edu/missions-and-data/hst-spectroscopic-legacy-archive-hsla}{
HST Spectroscopic Legacy Archive (HSLA) V2 Release} and \textsl{coadd\_x1d.pro} \citep{danforth10}. We decided to use HSLA when available, and otherwise use the coadded spectra processed by \textsl{coadd\_x1d.pro} that combines spectra from multiple exposures weighted by exposure times. We show in Appendix \ref{fig:appA_flux_offset} that these two methods yield consistent coadded flux levels and line profiles. The typical wavelength accuracy for the COS spectra is 15--20 $\kms$ (\href{https://hst-docs.stsci.edu/cosihb}{COS Instrument Handbook}). Because the COS spectra have been over-sampled with a native pixel size of $2.5~\kms$, after the co-addition, we bin the spectra by 3 pixels to improve the S/N by a factor of $\sqrt{3}$. 



\subsection{Continuum Normalization, Voigt-Profile Fitting, and Apparent Optical Depth Method}
\label{sec:cont_voigt}

We measure transitions of ionized metal species  that are commonly observed in a galaxy's CGM, including \SiII\ 1190/1193/1260/1526 \AA, \SiIII\ 1206 \AA, \SiIV\ 1393/1402 \AA, \CII\ 1334 \AA, and \CIV\ 1548/1550 \AA. We also detect \PII\ 1152 \AA, \SII\ 1250/1253/1259 \AA, \FeII\ 1144/1608 \AA, and \AlII\ 1670 \AA\ lines from sightlines S1-S4 but do not use these lines because they are typically related to a galaxy's ISM. Furthermore, we do not use \CIIS\ 1335 \AA\ because the IC1613's component of this line is always blended with the \CII\ 1334 \AA\ line from the Milky Way. 
The \OI\ 1302 \AA\ and \SiII\ 1304 \AA\ lines are not studied in this work due to the influence of the air-glow emission near 1302 \AA\ from \OI\ in the Earth's exosphere.

\begin{figure*}
    \centering
    \includegraphics[width=0.95\textwidth]{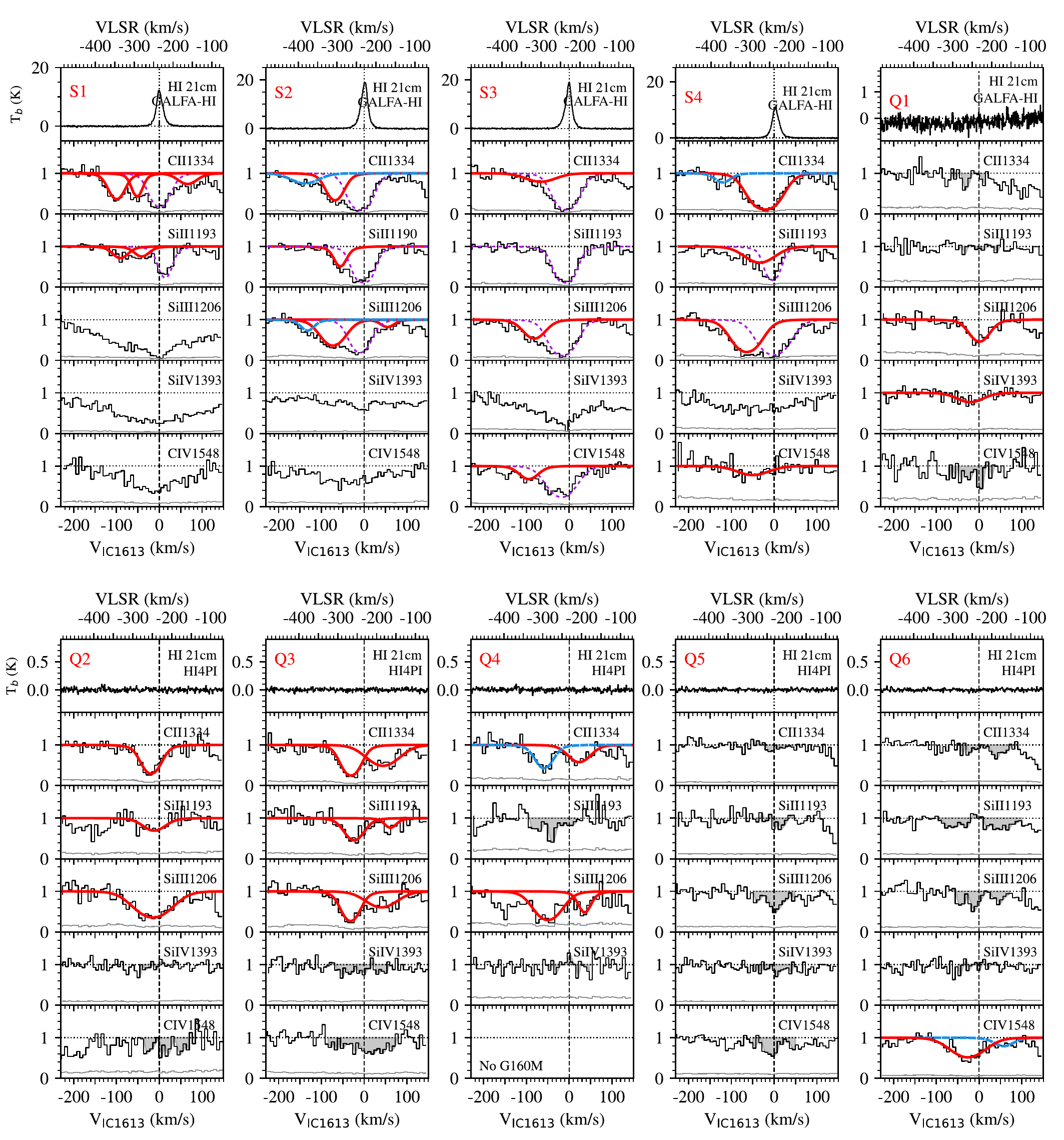}
    \caption{\HI\ 21cm emission and metal ion absorption lines measured towards S1--S4 and Q1--Q6. All the spectra are plotted in the LSR ($\vlsr$, top X axes) and in the rest frame of IC1613 ($v_{\rm IC1613}$, bottom X axes). We only show a subset of the ion lines in this figure, and include the full set of ion multiplets in Figures \ref{fig:s1_spec}--\ref{fig:q6_spec} in the Appendix. The red solid curves show the Voigt-profile components that are considered to be associated with IC1613 (i.e., ``CGM", ``CGM/Inflow", ``CGM/Outflow" in Table \ref{tb:lineresult}). The purple dotted curves indicate IC1613's ISM components, and the blue curves show absorbers that are unlikely to be related to IC1613 (i.e., ``Non-Association"). Toward Q1--Q6, when the Voigt-profile fitting does not yield robust results because low spectra SNR, we estimate the AOD column density with velocity integrated over the gray-shaded regions. The vertical line in each panel shows the systemic velocity of IC1613. For S1--S4, the galaxy's systemic velocity is estimated based on the peak \HI\ emission from VLA (see \S\ref{sec:hi} and Figures \ref{fig:s1_spec}-\ref{fig:s4_spec}) toward the corresponding sightline; for Q1--Q6, it is $\vlsr=-236~\kms$ as listed in Table \ref{tb:ic1613}. For S2, we show the \SiII\ 1190 line instead of the 1193 one because the latter is blended with an unknown feature and it does not yield reasonable Voigt-profile fits if included. } 
    \label{fig:all_line}
\end{figure*}

For continuum and Voigt-profile fitting, we used an IDL package developed for the COS-Halos survey as detailed in \cite{tumlinson13}. We briefly summarize the major procedures as follows. First, for each line, the continuum normalization is done over a spectral window of $\pm1000~\kms$ from its rest wavelength. Over this window, we manually mask any visible absorption features, and fit the rest with Legendre polynomials at low orders and determine the best continuum fit by minimizing the reduced $\chi^2$. We then proceed to conduct Voigt-profile fitting using the MPFIT package \citep{markwardt09}. For ions with multiple transition lines, the Voigt-profile fitting is run simultaneously among all the lines to ensure consistent fits. We also use reduced $\chi^2$ minimization to evaluate the best fit parameters, including column density ($\log N$), centroid velocity ($v$) in the rest frame of IC1613, and Doppler width ($b$). The best fit parameters are recorded in Table \ref{tb:lineresult} and the relevant measurements are noted as ``VP" in column (2). For each absorber, we also calculate its equivalent width ($\ew$) over a similar velocity range to the Voigt-profile fit result. We show the line profiles and fitting results in Figure \ref{fig:all_line} and Figures \ref{fig:s1_spec}--\ref{fig:q6_spec} in the Appendix.

For absorbers that do not have robust Voigt-profile fits, we calculate their column densities ($\logNaod$)
using the apparent optical depth method (AOD; see equation 6 in \citealt{savage96}). 
The AOD method is valid with a requirement that the absorption line has to be resolved and unsaturated \citep{savage91}, which is not a problem here since the detection is weak among most sightlines. For the \SiIV\ 1393/1402 and \CIV\ 1548/1550 doublets, we adopt results from the stronger line (1393 and 1548). For \SiII, because in many cases \SiII\ 1260 is blended with \SII\ 1259 from the Milky Way, we use the line measurements from \SiII\ 1193 instead. 
For each absorber, we decide the AOD velocity integration range based on visual inspection of the absorption line profile. For lines with no detection at the systemic velocity of IC1613, the velocity range is set to be [-50, 50] $\kms$ from the systemic velocity and we report $3\sigma$ upper limit values of $\logNaod$. 

For S1--S4, as shown in Figure \ref{fig:all_line}, the \SiIV\ line shows extended profiles. Because of line saturation, we are unable to find robust Voigt-profile fits and it is unpractical to calculate $\logNaod$ over the total velocity range which includes the ISM absorption. Thus, we do not use \SiIV\ detected in these stellar sightlines. A similar decision was applied to \SiIII\ in S1 and \CIV\ in S1 and S2. 


\subsection{Auxiliary \HI\ 21cm Data Sets}
\label{sec:hi}
Three \HI\ 21cm data sets are included to study the neutral gas in and around IC1613. We use the VLA observation from the Little THINGS survey \citep{hunter12} to probe the dense, cold \HI\ in the ISM of the galaxy. Furthermore, we use data cubes from the GALFA-\HI\ survey \citep{peek11, peek18} and \cite{hi4pi16} to probe more diffuse gas in the galaxy as well as along QSO sightlines in the halo.

The Little THINGS survey provides two data sets, one with ``natural weighting" and the other with ``robust weight". We adopt the natural weighting datacube because it has larger beam and is better at bringing out the diffuse \HI\ emission from the disk. The data cube is in Jy/beam, which we convert to brightness temperature in Kelvin as S(mJy/beam)=1.65$\times10^{-3}\delta_\alpha \delta_\beta T_B(K)$, where $\delta_\alpha=13.2$ arcsec and $\delta_\beta=11.0$ arcsec are the FWHM of the major and minor axes of the beam given in table 3 in \cite{hunter12}. 
The GALFA-\HI\ survey provides data with angular resolution of $\delta \theta=4\arcmin$, spectral resolution of $\delta v=0.184~\kms$, and brightness temperature sensitivity of 140 mK per \mkms\ velocity channel (1$\sigma$). The HI4PI survey provide lower angular and spectral resolutions ($\delta \theta =16.2\arcmin$, $\delta v=1.49~\kms$), but higher sensitivity ($\sim53$ mK per \mkms\ at 1$\sigma$). The \HI\ spectra from these data are shown in Figure \ref{fig:all_line} and Figures \ref{fig:s1_spec}--\ref{fig:q6_spec} in the Appendix. Generally, we do not find significant \HI\ detection except for those from the ISM of IC1613 as probed by stellar sightline S1--S4.

\section{Absorbers in the Rest Frame of IC1613}
\label{sec:ic1613}

In this section we investigate the ion absorbers' physical connection with IC1613, as all the sightlines (S1--S4, Q1--Q6) are within $0.6\rvir$ of the galaxy (see Figure \ref{fig:starqso}). We defer the discussion of potential foreground contamination to \S \ref{sec:pv_ms}. The distances to these absorbers are unknown except for their impact parameters with respect to IC1613, therefore our diagnosis is based on other measurements such as velocities and line widths.

\begin{center}
\begin{ThreePartTable}
\begin{TableNotes}
\label{tb:lineresult}
\item Note -- 
Col. (2): If Method = VP, the measurements are from Voigt-profile fits. If Method = AOD where the lines either do not yield robust Voigt-profile fits or there is no detection, we use \SiII\ 1193, \SiIII\ 1206, \SiIV\ 1393, \CII\ 1334, and \CIV\ 1548 to integrate for the AOD values. We do not use the stronger \SiII\ 1260 line because it is contaminated by the \SII\ 1259 line from the Milky Way. 
Col. (3): if Method=VP, $\vgal$ indicates the fitted centroid velocity in the rest frame of IC1613. If Method=AOD, $\vgal$ shows a velocity range as used in the AOD integration. 
Col. (4): if Method=VP, $b$ indicates the fitted Doppler width. No value is available if Method=AOD. 
Col. (5): if Method=VP, $\log N$ indicates the fitted column density. If Method=AOD, $\log N$ is estimated based the AOD method. We report a $3\sigma$ upper limit if there is no detection as often is the case in Q1--Q6.
Col. (6): Equivalent width integrated over the same velocity range as the $\log N$. We report $3\sigma$ upper limits for non-detection.
Col. (7): Origins of absorbers in the context of IC1613 as identified in \S\ref{sec:ic1613}. \\
Other notes: $[\ddagger]$: $v_{\rm sys}$ is the systemic velocity of the galaxy at the position of the sightline. For S1--S4, $v_{\rm sys}$ is estimated based on the peak emission of \HI\ 21cm emission (see \S\ref{sec:hi}). For Q1--Q6, $v_{\rm sys}=\vlsr=-236~\kms$ (see Table \ref{tb:ic1613}). $[*]$: \SiII\ 1193 in S2 is contaminated and does not yield good fit if included in the fitting, so we estimate $W_{\rm r}$ from \SiII\ 1190, and convert the value to \SiII\ 1193's with $W_{\rm r}(1193)=W_{\rm r}(1190)\frac{f_{1193}{\lambda_{\rm 1193}^2}}{f_{1190}{\lambda_{\rm 1190}^2}}$, where $f$ and $\lambda$ are the oscillator strength and wavelength, respectively. $[\mathparagraph]$: We consider this absorber to be a Non-Association because its $v_{\rm IC1613}$ value is $\sim20~\kms$ (19.6$~\kms$ for Q4/\CII, and 20.3 $\kms$ for Q6/\CIV) from the escape velocity at the corresponding impact parameter. Given the velocity uncertainty of COS (adopted as $20~\kms$ in this work) and the uncertainty of the absorber's centroid velocity ($\sim5~\kms$), we conservatively tag it as Non-Association but do not rule out its possibility to be related to IC1613.
\end{TableNotes}
\tablenum{3}
\begin{longtable*}{ccccccc}
\caption{Absorber Measurements}\\
\hline
Ion & Method & $\vgal$     & $b$     & $\log N$       & $W_r$    & \multirow{2}{*}{Origin tag}\\
    & & ($\kms$) & ($\kms$) & log(cm$^{-2}$) & (m\AA) &   \\
(1) & (2) & (3)     & (4)     & (5)            & (6)      & (7)   \\
\hline
\endfirsthead
\hline
Ion & Method & $~~~~~~~~~\vgal~~~~~~~~~$     & $~~~~~~~~~b~~~~~~~~~$     & $~~~~~~~~~\log N~~~~~~~~~$       & $~~~~~~~~~W_r~~~~~~~~~$    &  \multirow{2}{*}{\ \ \ Origin tag\ \ \ } \\
    & & (\mkms) & (\mkms) & log(cm$^{-2}$) & (m\AA)   \\
(1) & (2) & (3)     & (4)     & (5)            & (6)      & (7)        \\
\hline
\endhead
\endfoot
\hline
\insertTableNotes
\endlastfoot
\multicolumn{7}{c}{\textbf{S1: IC1613-C10 ($v_{\rm sys}=-239.4~\kms\ddagger$)}} \\
\hline 
\multirow{4}{*}{\CII}  & VP & -99.7$\pm$1.8 & 19.1$\pm$3.0  & 14.07$\pm$0.04 & 126.2$\pm$7.9 &  CGM/Outflow\\ 
                       & VP & -51.8$\pm$1.9 & 11.5$\pm$3.9  & 13.90$\pm$0.07 & 100.4$\pm$5.0 & CGM/Outflow\\ 
                       & VP & 1.0$\pm$1.7   & 22.5$\pm$3.4  & 14.42$\pm$0.04 & 258.7$\pm$7.2 & ISM\\ 
                       & VP & 65.9$\pm$7.8  & 25.8$\pm$14.7 & 13.59$\pm$0.19 & 73.0$\pm$7.1 & CGM/Inflow\\ 
\hline 
\multirow{3}{*}{\SiII} & VP & -90.7$\pm$2.3 & 18.9$\pm$3.6 & 13.02$\pm$0.04 & 40.8$\pm$7.6 & CGM/Outflow \\
                       & VP & -40.8$\pm$2.5 & 17.8$\pm$4.6 & 12.94$\pm$0.07 & 61.6$\pm$7.1 & CGM/Outflow \\
                       & VP & 13.6$\pm$0.6  & 11.0$\pm$0.8 & 14.00$\pm$0.08 & 109.3$\pm$6.8 & ISM \\
\hline
\hline 
\multicolumn{7}{c}{\textbf{S2: IC1613-B7 ($v_{\rm sys}=-234.3~\kms\ddagger$)}} \\
\hline 
\multirow{3}{*}{\CII}   &  VP & -138.9$\pm$8.5 & 39.2$\pm$12.4 & 13.70$\pm$0.10 & 77.7$\pm$8.1 & Non-Association\\ 
                        &  VP & -68.4$\pm$2.7  & 21.4$\pm$4.6  & 14.13$\pm$0.07 & 172.1$\pm$5.2 & CGM/Outflow\\ 
                        &  VP & -9.3$\pm$1.8   & 25.3$\pm$2.4  & 14.63$\pm$0.04 & 282.5$\pm$6.0 & ISM\\ 
\hline
\multirow{2}{*}{\SiII} & VP & -55.7$\pm$1.5 & 13.1$\pm$2.7 & 13.60$\pm$0.05 & 160.0$\pm$11.4$^{*}$ &  CGM/Outflow\\
                       & VP & -2.0$\pm$0.8  & 17.9$\pm$2.0 & 14.64$\pm$0.15 & 445.8$\pm$11.6$^{*}$ & ISM \\
\hline 
\multirow{4}{*}{\SiIII}& VP & -132.4$\pm$8.6 & 15.4$\pm$12.6 & 12.43$\pm$0.26  & 42.9$\pm$9.2 &  Non-Association \\ 
                       & VP & -73.1$\pm$5.5  & 31.4$\pm$12.8 & 13.15$\pm$0.12  &  169.0$\pm$7.7 & CGM/Outflow\\ 
                       & VP & -9.5$\pm$3.2   & 23.3$\pm$4.6 & 13.35$\pm$0.05  &  213.6$\pm$5.0 & ISM\\ 
                       & VP & 54.4$\pm$8.2   & 16.1$\pm$15.7 & 12.23$\pm$0.26  &  50.9$\pm$7.2 & CGM/Inflow\\ 
\hline 
\hline 
\multicolumn{7}{c}{\textbf{S3: IC1613-A13 ($v_{\rm sys}=-231.7~\kms\ddagger$)}} \\
\hline 
\multirow{2}{*}{\CII}  & VP & -66.7$\pm$43.9 & 43.2$\pm$37.6 & 13.65$\pm$0.52  & 49.5$\pm$9.4 &  CGM/Outflow \\
                       & VP & -7.2$\pm$3.2 & 27.9$\pm$3.4  & 14.61$\pm$0.07  & 324.2$\pm$8.3 & ISM \\
\hline
\SiII                  & VP & -9.5$\pm$0.6   & 19.0$\pm$0.9  & 14.30$\pm$0.04  & 237.2$\pm$8.6 & ISM \\
\hline 
\multirow{2}{*}{\SiIII}& VP & -81.6$\pm$7.7  &  28.1$\pm$9.2 & 12.89$\pm$0.12  & 169.1$\pm$7.8 &  CGM/Outflow \\
                       & VP & -15.7$\pm$2.9  &  28.5$\pm$3.4 & 13.55$\pm$0.05  & 229.3$\pm$6.4 & ISM \\
\hline
\multirow{2}{*}{\CIV}  & VP & -94.5$\pm$4.5 & 25.1$\pm$7.0 & 13.45$\pm$0.10  & 79.4$\pm$7.9 &  CGM/Outflow \\
                       & VP & -18.0$\pm$1.7 & 36.0$\pm$2.4 & 14.17$\pm$0.02  & 303.6$\pm$10.0 & ISM \\
\hline 
\hline 
\multicolumn{7}{c}{\textbf{S4: IC1613-B11 ($v_{\rm sys}=-242.0~\kms\ddagger$)}} \\
\hline 
\multirow{2}{*}{\CII} & VP & -120.0$\pm$6.3 & 20.6$\pm$10.0 & 13.44$\pm$0.12 & 62.6$\pm$10.7 &  Non-Association \\
                      & VP & -21.6$\pm$1.6  & 37.8$\pm$2.5  & 14.64$\pm$0.03 & 374.4$\pm$11.5 & CGM/Outflow\\
\hline 
\multirow{2}{*}{\SiII} & VP & -33.6$\pm$10.1 & 41.4$\pm$6.6 & 13.44$\pm$0.13 & 95.9$\pm$9.3 & CGM/Outflow \\
                       & VP & -3.6$\pm$0.9   & 12.2$\pm$1.9 & 14.33$\pm$0.17 & 164.9$\pm$8.1 & ISM \\
\hline 
\multirow{2}{*}{\SiIII} & VP & -64.2$\pm$14.6 & 35.2$\pm$10.6 & 13.39$\pm$0.24 & 242.7$\pm$8.9 &  CGM/Outflow \\
                        & VP & -1.1$\pm$12.1  & 36.4$\pm$8.7  & 13.53$\pm$0.18 & 300.8$\pm$9.7 & ISM \\
\hline 
\CIV                  & VP & -49.9$\pm$11.1 & 45.5$\pm$16.5 & 13.46$\pm$0.11 & 51.1$\pm$20.1 &  CGM/Outflow\\
\hline 
\hline 
\multicolumn{7}{c}{\textbf{Q1: LBQS-0100+0205 ($v_{\rm sys}=-236~\kms\ddagger$})}\\
\hline 
\CII   & AOD & [-70, 20]    & -  & $<$13.74 & $<$48.9 & Non-Detection \\
\hline 
\SiII  & AOD & [-50, 50]    & -            & $<$13.03       & $<$49.8 &  Non-Detection\\
\hline 
\SiIII & VP & 0.7$\pm$3.4 & 25.1$\pm$5.2 & 12.96$\pm$0.06 & 126.2$\pm$16.6 &  CGM \\
\hline 
\SiIV  & VP & -21.2$\pm$5.8 & 37.1$\pm$8.8 & 13.00$\pm$0.07 & 65.6$\pm$13.2  & CGM \\ 
\hline 
\CIV   & AOD & [-70, 50]    & -  & 13.57$\pm$0.09 & 117.3$\pm$27.5 & CGM \\
\hline 
\hline 
\multicolumn{7}{c}{\textbf{Q2: LBQS-0101+0009 ($v_{\rm sys}=-236~\kms\ddagger$)}} \\
\hline 
\CII   & VP & -19.3$\pm$2.1 & 23.6$\pm$3.1 & 14.21$\pm$0.05 & 189.9$\pm$14.8 & CGM \\ 
\hline 
\SiII  & VP & -12.0$\pm$4.3 & 30.0$\pm$5.4 & 13.19$\pm$0.06 & 77.8$\pm$16.1 & CGM \\
\hline 
\SiIII & VP & -14.8$\pm$4.5 & 50.1$\pm$6.2 & 13.30$\pm$0.05 & 271.3$\pm$23.1 & CGM \\ 
\hline 
\SiIV  & AOD & [-50, 50]     & -            & $<$12.79       & $<$35.1 & Non-Detection \\
\hline 
\CIV   & AOD & [-40, 75]    & -         & 13.64$\pm$0.07 & 144.4$\pm$23.2&  CGM \\
\hline 
\hline 
\multicolumn{7}{c}{\textbf{Q3: 2MASX-J0102-0039 ($v_{\rm sys}=-236~\kms\ddagger$)}} \\
\hline 
\multirow{2}{*}{\CII} & VP & -31.1$\pm$2.2  & 23.3$\pm$3.0 & 14.28$\pm$0.05 & 205.5$\pm$9.5 &  CGM \\ 
                      & VP & 43.5$\pm$5.7 & 44.4$\pm$8.0 & 14.15$\pm$0.06 & 185.2$\pm$9.9 & CGM \\ 
\hline 
\multirow{2}{*}{\SiII} &  VP & -23.5$\pm$1.4 & 24.0$\pm$2.1 & 13.48$\pm$0.03 & 124.1$\pm$12.0 & CGM \\
                       &  VP & 61.4$\pm$2.5  & 13.0$\pm$4.4 & 12.85$\pm$0.07 & 59.2$\pm$11.9 & CGM \\
\hline 
\multirow{2}{*}{\SiIII} & VP & -31.9$\pm$2.9 & 20.7$\pm$4.4 & 13.19$\pm$0.07 & 176.3$\pm$13.8 & CGM \\
                        & VP & 40.1$\pm$9.2  & 42.8$\pm$13.9& 12.92$\pm$0.09 & 123.1$\pm$10.8 & CGM \\
\hline 
\SiIV                   & AOD & [-90, 60] & - & 13.02$\pm$0.07 & 84.0$\pm$15.1 & CGM \\
\hline
\CIV                  & AOD & [-90, 80] & - &  13.78$\pm$0.05 & 205.7$\pm$22.4 & CGM \\ 
\hline 
\hline 
\multicolumn{7}{c}{{\textbf{Q4: PG0044+030 ($v_{\rm sys}=-236~\kms\ddagger$)}, G130M-only}} \\
\hline
\multirow{2}{*}{\CII} & VP  & -58.6$\pm$3.4 &  22.4$\pm$5.7 & 14.01$\pm$0.07 & 135.0$\pm$18.0 & Non-Association$^{\mathparagraph}$ \\ 
                      & VP  & 23.4$\pm$5.7  &  29.5$\pm$9.0 & 13.89$\pm$0.09 & 116.7$\pm$15.8 & CGM \\
\hline 
\SiII   & AOD & [-100, 20]  &  - & 13.31$\pm$0.10 & 102.5$\pm$25.2 & CGM \\ 
\hline 
\multirow{2}{*}{\SiIII} & VP & -49.9$\pm$4.7  & 34.5$\pm$7.5 & 13.26$\pm$0.07 & 229.8$\pm$21.4 & CGM \\ 
                        & VP & 36.3$\pm$4.6   & 13.4$\pm$8.0 & 12.79$\pm$0.16 & 89.1$\pm$17.4 & CGM \\ 
\hline 
\SiIV & AOD & [-50, 50] & - & $<$13.00 & $<$68.7& Non-Detection \\

\hline 
\hline 
\multicolumn{7}{c}{\textbf{Q5: HB89-0107-025-NED05 ($v_{\rm sys}=-236~\kms\ddagger$)}} \\
\hline
\CII  & AOD & [-50, 50]  & - & $<$13.38 & $<$29.7&  Non-Detection \\
\hline 
\SiII & AOD & [-50, 50]  & -  & $<$13.04 & $<$38.7&  Non-Detection \\
\hline 
\SiIII & AOD & [-50, 50] & - & 12.76$\pm$0.06 & 100.7$\pm$14.0 & CGM \\ 
\hline 
\SiIV & AOD & [-50, 50]  & - &  $<$13.02 & $<$34.2 &  Non-Detection \\
\hline 
\CIV & AOD & [-50, 50]   & - &  13.56$\pm$0.06 & 121.5$\pm$15.9 &  CGM \\ 
\hline 
\hline 
\multicolumn{7}{c}{\textbf{Q6: LBQS-0107-0235 ($v_{\rm sys}=-236~\kms\ddagger$)}} \\
\hline
\CII & AOD & [-50, 80] & - & 13.60$\pm$0.08 & 71.3$\pm$14.0 &  CGM \\
\hline 
\SiII & AOD & [-100, 100] & - & 13.30$\pm$0.07 & 111.0$\pm$17.7 &  CGM \\ 
\hline 
\SiIII & AOD & [-70, 75] & - & 12.77$\pm$0.06 & 105.1$\pm$15.4 &  CGM \\ 
\hline 
\SiIV & AOD & [-50, 50] & - & $<$12.80 & $<$42.6 & Non-Detection \\ 
\hline 
\multirow{2}{*}{\CIV} & VP & -26.8$\pm$3.7 &  47.9$\pm$5.3 & 13.91$\pm$0.04 & 221.5$\pm$10.7 &  CGM \\ 
                      & VP & 57.6$\pm$6.7  &  25.9$\pm$9.7 & 13.24$\pm$0.13 & 64.8$\pm$8.4 & Non-Association$^{\mathparagraph}$ \\ 
\hline
\hline
\end{longtable*}
\end{ThreePartTable}
\end{center}

\begin{figure*}[t]
    \centering
    \includegraphics[width=\textwidth]{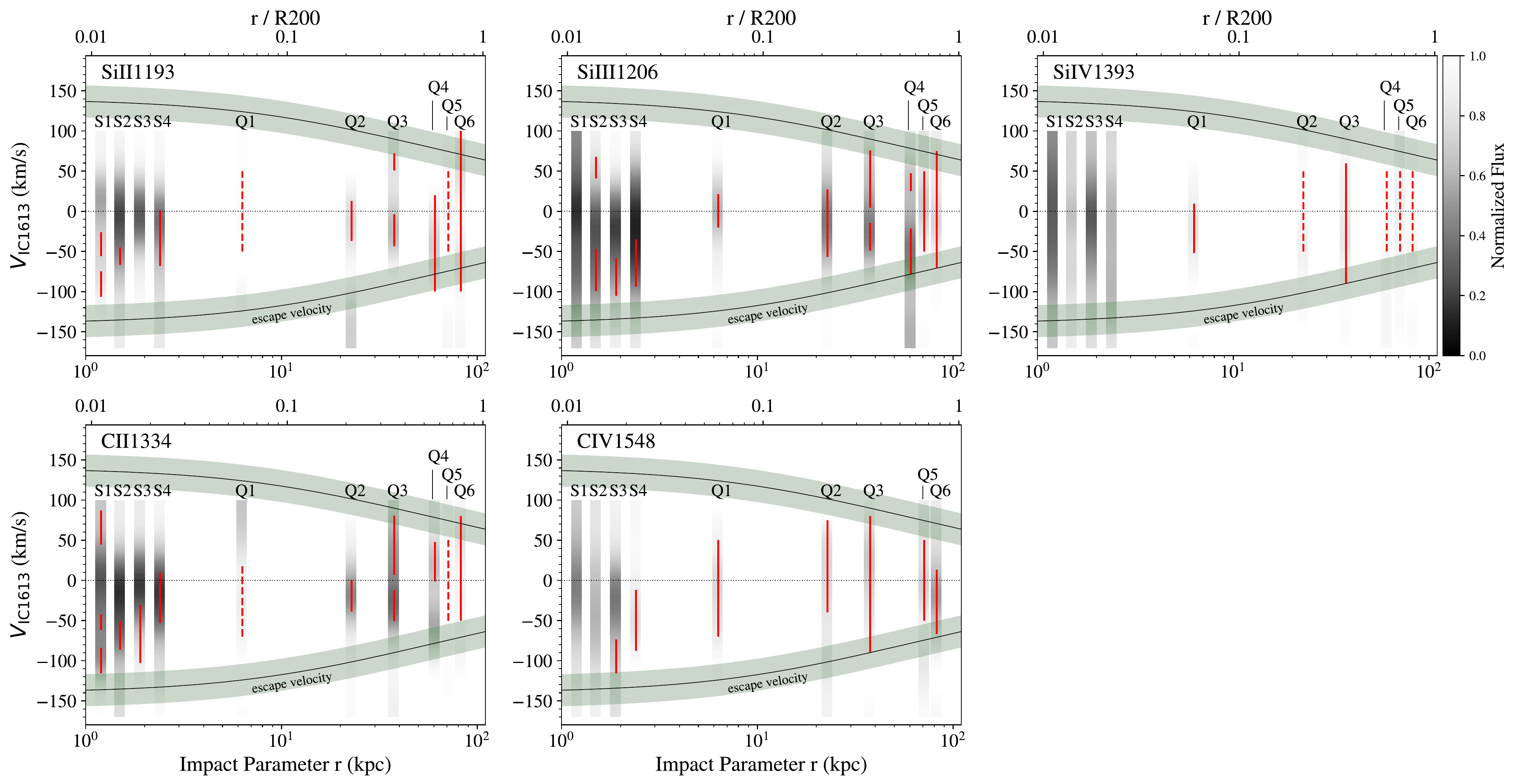}
    \caption{Ion absorbers that are most likely to be associated with IC1613. They are measurements tagged as ``CGM", ``CGM/Outflow", ``CGM/Inflow", or ``Non-Detection" in Table \ref{tb:lineresult} and \S\ref{sec:ic1613}. The red solid lines show the FWHM ($=1.667b$) of the Voigt-profile fits or the AOD velocity integration ranges of the detected absorbers, and the red dashed lines are for the non-detection $3\sigma$ upper limit ranges. For each target, the gray band shows the continuum-normalized line profile, with dark gray indicating strong absorption and vice versa. The green shades are escape velocities with $20~\kms$ uncertainty due to the COS spectral resolution. Stars S1--S4 are inside the galaxy; we place them at arbitrary but small $r$ to separate one from another. Q4--Q6 have similar impact parameters ($r\sim61$ kpc; Table \ref{tb:starqso}); we manually separate the gray bands slightly for better illustration. }
    \label{fig:v_dproj}
\end{figure*}

From Figure \ref{fig:all_line} and Table \ref{tb:lineresult}, we find that \SiII, \SiIII, \SiIV, \CII, and \CIV\ absorption are strong and commonly detected among the stars S1--S4; origins for the absorption include the ISM of IC1613 ($\vgal\sim0~\kms$), potential inflows ($\vgal>0~\kms$) and outflows ($\vgal<0~\kms$), and the galaxy's CGM ($|\vgal|<\vesc$), where $\vesc$ means the escape velocity of the galaxy and $\vgal$ means the velocity is relative to IC1613's systemic velocity. Toward the QSO sightlines Q1--Q6, absorbers appear to be weaker and the line strengths vary from sightline to sightline; they are likely to originate in the CGM of IC1613 if $|\vgal|<\vesc$. Based on the ion absorbers' positions and velocities relative to IC1613 and the line quality, we assign different tags to the absorbers tabulated in Table \ref{tb:lineresult} as follows.

\textsl{Origin tag = ``CGM"}: absorbers detected along Q1--Q6 that are most likely to originate in the CGM of IC1613. The velocities of these absorbers are $|\vgal|<|\vesc|-20~\kms$, where the $20~\kms$ value is to account for the COS spectral uncertainty.

\textsl{Origin tag = ``CGM/Outflow" or ``CGM/Inflow"}: absorbers detected along S1--S4 that are most likely to be either in the CGM of IC1613 or outflows or inflows near the galaxy. The ambiguity of the absorbers' locations is because these stellar sightlines are observed in a down-the-barrel manner. Specifically, absorbers with $\vesc+20$ $<\vgal<$ $-20~\kms$ are tagged as ``CGM/Outflow", and those with 20 $<\vgal<$ $\vesc-20~\kms$ are ``CGM/Inflow".

\textsl{Origin tag = ``Non-Detection"}: there is no detection of absorption within the designated velocity ranges. This tag is only for Q1--Q6, and we provide $3\sigma$ upper limits on the column densities and the equivalent widths. 

\textsl{Origin tag = ``ISM"}: absorbers detected along S1--S4 that are likely to be in the ISM of IC1613, with $|\vgal|<$ 20$~\kms$. Their Voigt profiles are shown in purple dotted curves in Figure \ref{fig:all_line}, which tend to be broader and stronger than the non-ISM components. We do not use these absorbers in our analyses.    

\textsl{Origin tag = ``Non-Association"}: absorbers that are unlikely to be associated with IC1613 because they are not gravitationally bound, $|\vgal|>|\vesc|-20~\kms$. The Voigt profiles of these absorbers are shown in blue curves in Figure \ref{fig:all_line}, which tend to be much weaker than other IC1613-associated counterparts. 
We do not use these absorbers in our analyses regarding the CGM metal content and outflows of IC1613.

Based on this tagging system, we show in Figure \ref{fig:v_dproj} the impact parameters and velocities of the absorbers tagged with ``CGM", ``CGM/Outflow", ``CGM/Inflow", or "Non-Detection" in the rest frame of IC1613. We also show the original line spectra as vertical gray bands to highlight the spread of the ion absorption. By design, the ion absorbers likely to be associated with IC1613 have velocities clustered within $\sim\pm100~\kms$, as limited by the range of the escape velocity. While it is necessary to use escape velocity to constrain whether an absorber is related to IC1613 given the complex gaseous environment in the LG (see \S\ref{sec:pv_ms}), we note that IC1613 may have high-velocity outflows escaping the disk (i.e., $\vgal>\vesc$) that are not gravitationally bound. Such outflows would not be recognized as ``CGM/Outflow" based on our criterion. Therefore, our estimates of the mean outflow velocities and other relevant properties (see \S\ref{sec:outflow_rate}) should be considered as conservative lower limits. 

The mean velocities of the ``CGM/Outflow" absorbers are $-45\pm20~\kms$ for \SiII, $-71\pm8~\kms$ for \SiIII, $-66\pm10~\kms$ for \CII, and $-63\pm20~\kms$ for \CIV\ respectively. The mean values are weighed by the measurement errors, and the uncertainties are the standard deviations of the velocities also weighted by the measurement errors. If corrected for the inclination of the galaxy ($\theta=37.9\degree$; see Table \ref{tb:ic1613}) and assuming that outflows are perpendicular to the galaxy's disk, the mean outflow velocity for each ion would increase by $1/{\rm cos}\theta=1.3$. Despite that there is detection of broad \SiIV\ absorption lines in all the stellar sightlines, we do not have an outflow velocity value for \SiIV\ because there is no robust Voigt-profile fit that can separate the ISM from the non-ISM components.

The detection rate (or covering fraction) $C_f$ of the ``CGM", ``CGM/Outflow", ``CGM/Inflow", and "Non-Detection" absorbers within $0.6\rvir$ is 82\% (9/11) for \SiII, 100\% (12/12) for \SiIII, 33\% (2/6) for \SiIV, 85\% (11/13) for \CII, and 100\% (7/7) for \CIV, respectively. And the detection limit for these ions is generally $W_{\rm r}\gtrsim50$ m\AA, although note that the detection limit depends on the spectral SNR. The mean column densities 
(as weighted by measurement errors) are $13.24\pm0.04$ dex for \SiII, $12.91\pm0.07$ dex for \SiIII, $13.01\pm0.05$ dex for \SiIV, $13.80\pm0.22$ dex for \CII, and $13.51\pm0.04$ dex for \CIV, respectively.

Lastly, the error-weighted mean Doppler width ($b$) and its standard deviation is $32\pm11~\kms$ for all the Voigt-profile fitted components tagged as ``CGM", ``CGM/Outflow", and ``CGM/Inflow". The $b$ value changes by $<10~\kms$ if we focus on a specific ion or outflow-only absorbers. Our derived $b$ values are consistent with those of \SiIII\ and \CIV\ measured toward two field dwarf galaxies (D1 and D2) with QSO sightlines at $<0.2\rvir$\footnote{The virial radii of \cite{johnson17}'s galaxies have been recalculated to be consistent with our definition of $\rvir$ using the galaxies' stellar masses (see \S\ref{sec:comp_lowz_dwarfs}).} by \cite{johnson17}. And they are on average larger than the $b$ values measured for the ionized gas near the Magellanic Stream \citep[$b<25~\kms$][]{fox20}, suggesting that our absorbers are unlikely to be associated with the Stream. We discuss in more details how the foreground Magellanic Stream impacts our diagnosis of the ion absorbers' origins in \S\ref{sec:pv_ms}.   


\section{The Magellanic System in the Foreground}
\label{sec:pv_ms}

\begin{figure*}[th!]
\centering
\includegraphics[width=0.95\textwidth]{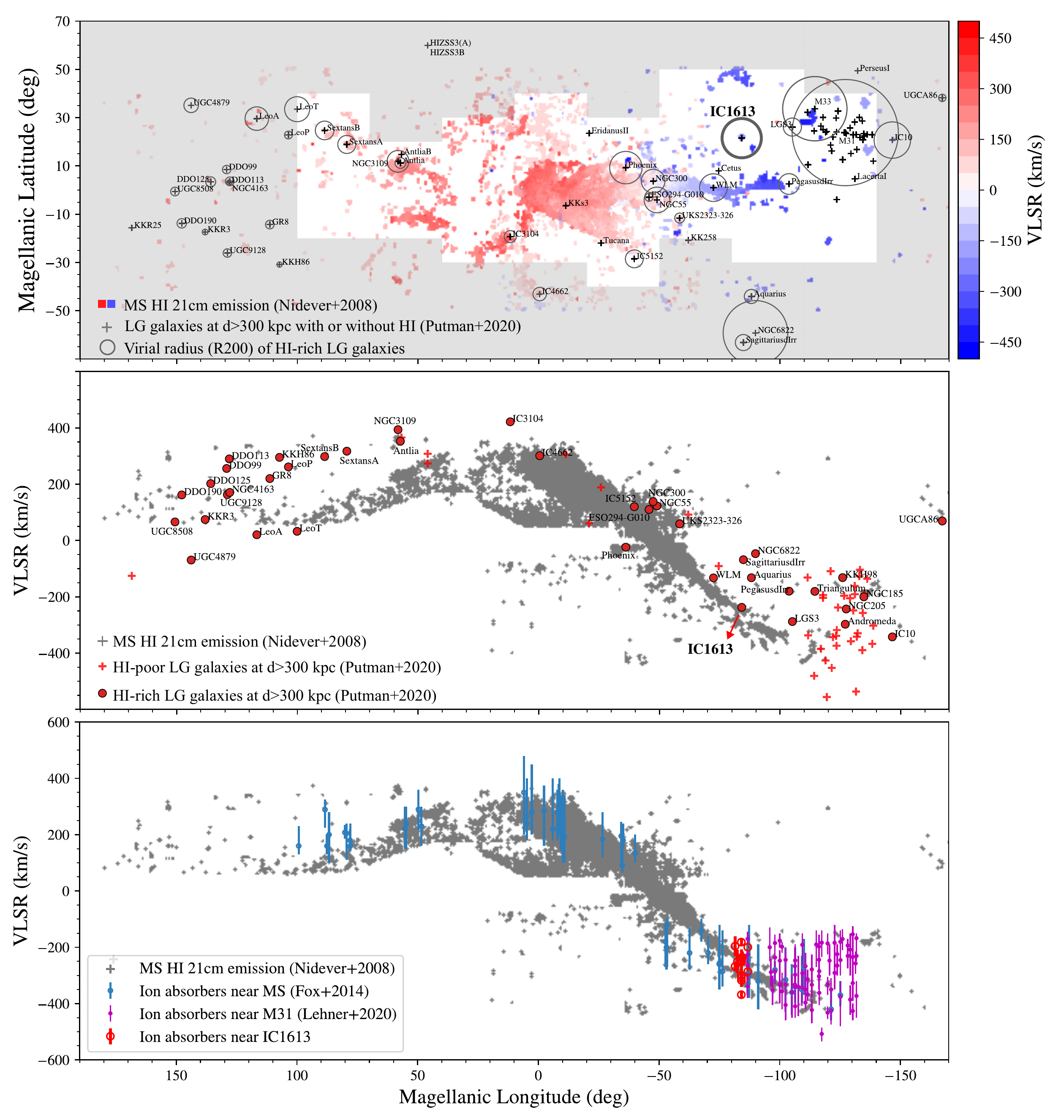}
\caption{\textbf{Top}: 81 LG galaxies with $d_\odot>300$ kpc (crosses; Putman et al. 2020, submitted) that are not physically connected to the Magellanic System (red and blue colors; \citealt{nidever08}). The LG galaxies include 77 dwarf galaxies and four more massive ones (M31, M33, NGC55, and NGC300). Data are shown in the Magellanic Coordinate system (\citealt{nidever08}; \textsl{gala} package, \citealt{gala-code}). For clarity, the names of the dwarfs clustering near M31 are not shown. The white region represents the ionized cross section of the Magellanic System as identified by \citetalias{fox14}. The virial radii ($\rvir$) of \HI-rich galaxies are indicated as black circles, which are used to determine the angular extent of their CGM (see Appendix \ref{sec:app_ms_lg}). 
\textbf{Middle}: coincident alignment between LG galaxies (red symbols) and the Magellanic \HI\ emission (grey dots; \citealt{nidever08}) on the position-velocity diagram. \textbf{Bottom}: similar coincident alignment between ion absorbers near IC1613 (red open circle; this work), M31 (magenta; \citealt{lehner20}), and the Magellanic \HI-emitting region (blue; \citetalias{fox14}). For data points from \citetalias{fox14}'s and \cite{lehner20}, the vertical bars show the minimum and maximum velocities used in their AOD measurements. For IC1613's, we show \SiIII's centroid velocities and the FWHM ($\equiv1.667b$). The middle and bottom panels show that an absorber's alignment with the Magellanic System does not necessarily lead to a physical connection between the two, and the angular extent of the Magellanic System's ionized gas should be revisited with more robust methods other than this position-velocity diagram.}  
\label{fig:ms_lg_dwarfs}
\end{figure*}

Hereafter we refer to the LMC/SMC, the Magellanic Stream, the Magellanic Bridge, and the Leading Arm as the Magellanic System. In Figure \ref{fig:ms_lg_dwarfs}, we show the Magellanic System in the so-called Magellanic Stream Coordinate System ($\ml$, $\mb$; \citealt{nidever08}), where the equator ($\mb=0\degree$) bisects the spine of the Stream and the LMC is at $\ml=0\degree$. IC1613 is located near the tail of the Stream at $\ml=-84.1\degree, \mb=21.5\degree$. It is isolated from other galaxies in the LG, and $\sim20\degree$ from the Magellanic Stream in projection.

The Magellanic System has been widely detected in \HI\ 21cm \citep{mathewson74, putman98, putman03, nidever08} and occupies $\sim$2700 square at N(\HI)$\geq10^{18}$ cm$^{-2}$ \citep{nidever10, donghia16}. H$\alpha$ emission from the Magellanic Stream is observed by WHAM \citep{haffner03} to extend $\sim2$ degrees from the Magellanic \HI\ bright regions \citep{barger17}. Ionized gas detected via UV absorption lines is thought to be distributed out to 30 degrees from the \HI, with a cross section of $\approx11,000$ deg$^2$, with the assumption that the ionized gas associated with the Magellanic System should have a line-of-sight velocity ($\vlsr$) aligned with the \HI\ at a given $\ml$ (\citealt{fox14}, hereafter \citetalias{fox14}; \citealt{richter17}). This is to say the ionized and neutral gas of the Magellanic System are assumed to occupy the same parameter space in the position ($\ml$) -- velocity ($\vlsr$) diagram. Here we examine this position-velocity criterion in the context of the CGM of IC1613.

In the top panel of Figure \ref{fig:ms_lg_dwarfs}, in red and blue colors we show the velocity ($\vlsr$) of the Magellanic System's \HI\ emission Gaussian-fitted components \citep{nidever08} as well as the positions of some LG galaxies (see below for selection criteria of these galaxies). The white area in this top panel shows the ionized cross section of the Magellanic System defined by \citetalias{fox14}; within this cross section, 81\% (56/69) of their QSO sightlines (not shown here) are detected with ion absorbers that are identified as Magellanic. We show in the bottom panel these ion absorbers (blue) on the $\ml$--$\vlsr$ diagram, which are indeed aligned with the Magellanic \HI\ emission (gray). We also overlay ion absorbers detected near IC1613 (red) in this bottom panel which appear to be largely consistent with the location of the Magellanic \HI. Furthermore, \cite{lehner20} find that 38\% (28/74) of their detected \SiIII\ absorbers towards M31 are aligned with the Magellanic \HI\ emission (magenta).

As we investigate further we find that an absorber's alignment with the Magellanic \HI\ on the $\ml$--$\vlsr$ diagram does not necessarily lead to a physical connection between the two. To demonstrate this, in the middle panel we show a number of galaxies in the LG that are near the Magellanic System in projection but are not physically connected to it. These galaxies are selected from the dwarf galaxy catalog compiled by Putman et al. (2020, submitted) and we also include four more massive LG members (M31, M33, NGC55, and NGC300). We only consider LG galaxies that are (1) with distance $d_\odot>300$ kpc from the Sun, and (2) have line of sight velocities. Criterion (1) is to exclude Milky Way satellites that could be considered physically associated with the Magellanic System based on proper motions and orbital history studies \citep[e.g., ][]{patel20}. Criterion (2) is a necessity for the $\ml$--$\vlsr$ diagram.

With criteria (1) \& (2), we find 81 LG galaxies near the Magellanic System in position-velocity space despite that they are physically not connected. In the middle panel, we calculate the separation between the LG galaxies and their closest \HI\ emission Gaussian components of the Magellanic System, and find that 73\% (59/81) of these galaxies are coincidentally aligned with the Magellanic System within $10~\kms$ in $\vlsr$ and $1\degree$ in $\ml$ and $\mb$. 
Without the prior knowledge of the distances to these LG galaxies (all at $d_\odot>300$ kpc), one may wrongly conclude that they are physically associated with the Magellanic System. Therefore, we argue that the alignment of an object with the Magellanic \HI\ on the $\ml$--$\vlsr$ diagram does not provide solid evidence that the object is originated from the System.

Because of the coincident alignment between the LG galaxies and the Magellanic System, we further show that potential CGM absorbers originated from \HI-rich galaxies in the LG will appear on a similar $\ml$--$\vlsr$ parameter space, further complicating the diagnosis of an absorber's origin. Because such an investigation is beyond the context of IC1613's CGM, we defer the relevant analysis to Appendix \ref{sec:app_ms_lg} to keep the main text focusing on IC1613. Briefly, in Appendix \ref{sec:app_ms_lg} we calculate the angular extent of the CGM of \HI-rich dwarf galaxies selected based on Criteria (1) \& (2) and show that the total cross section of these galaxies' CGM is non-negligible.

To conclude, we argue that the angular extent of the ionized cross section of the Magellanic System should be revisited using more robust methods other than the $\ml$--$\vlsr$ diagram. For example, a recent hydrodynamic simulation of the Magellanic System by \cite{lucchini20} predicts a broad ionized component encompassing both the Leading Arm and Magellanic Stream due to the interaction between a massive LMC corona with the Milky Way's CGM. They suggest that the column densities of the LMC-associated, highly-ionized gas should decrease with increasing impact parameters. It remains to be determined whether such a decreasing trend in column density can aid in better defining the angular extent of the Magellanic System. On the other hand, the ionized gas of the Magellanic System is likely to be confused with the CGM of \HI-rich LG galaxies if the QSO sightlines are within the galaxies' virial radii (see Appendix \ref{sec:app_ms_lg}). In the case of IC1613, as we discussed in \S\ref{sec:intro} and \S\ref{sec:ic1613}, the detected absorbers are most likely to be associated with the CGM of IC1613 given the star formation history of the galaxy, the proximity of the absorbers to the galaxy, and the larger $b$ values of the ion absorbers than other ionized gas near the Stream.

\section{The Metal Mass Budget and Mass Loading Factor of IC1613}
\label{sec:metal_budget}

In \S\ref{sec:ic1613}, we have identified ion absorbers that are most likely to be associated with IC1613. 
Here we will use the measurements of these absorbers to empirically estimate the silicon (Si) mass budget in the star, ISM, and CGM of IC1613 (see \S\ref{sec:empirical_mass_budget}), and then compare our estimates to predicted values from simulations (see \S\ref{sec:sim_mass_budget}). We will further estimate the metal outflow rate and the instantaneous metal mass loading factor in \S\ref{sec:outflow_rate}. 

\subsection{Metal Mass Budget Estimate}
\label{sec:empirical_mass_budget}

Given that there is no detection of \HI\ among the QSO sightlines (see Figure \ref{fig:all_line}), the CGM of IC1613 is likely to be fully ionized. We first estimate the total Si mass in the CGM assuming that \SiII, \SiIII, and \SiIV\ comprise nearly 100\% of the total Si and leveraging the fact that these ions are simultaneously detected in the COS spectra. We only use absorbers tagged as ``CGM" from Q1--Q6 in Table \ref{tb:lineresult}. We decide to exclude potential CGM absorbers detected in S1--S4 (i.e., those tagged with ``CGM/Outflow" or ``CGM/Inflow") because their impact parameters from the galaxy are ambiguous as the stellar sightlines were observed in a down-the-barrel manner. We note that including these absorbers would not change our mass estimate significantly\footnote{Assuming that these ``CGM/Outflow" and ``CGM/Inflow" absorbers have similar properties as those ``CGM" absorbers, we estimated their potential impact parameters by matching their $\log N$ values to the nearest ``CGM" $\log N$ values with known impact parameters. We then included these absorbers in Equation \ref{eq:msi_cgm} and found that they contributed a few thousand $\msun$, which is much less than the significant figure we adopted for the total estimated mass.}. 

We follow the same methodology as outlined in Section 4 of \citeauthor{Zheng19b} (2019; hereafter \citetalias{Zheng19b}) which estimated Si mass budget for the dwarf irregular galaxy WLM. The main difference from \citetalias{Zheng19b} is that here we are able to integrate the Si mass radially based on data from Q1--Q6, without assuming a radial profile or covering fraction.  By taking each absorber to represent the azimuthal average of concentric annuli around IC1613, the total Si mass can be derived as: 
\begin{equation}
\begin{array}{ll}
M_{\rm Si}^{\rm CGM}(\leq 0.6 \rvir) &= \sum \pi(r_{k}^2-r_{k-1}^2)m_{\rm Si}N_{\rm Si, k} \\
     & \approx(1.0-1.3)\times10^5~\msun 
\end{array}
\label{eq:msi_cgm}
\end{equation}
, where $m_{\rm Si}$ is the mass of a Si atom, $r_k$ is the impact parameter of each QSO with $k$ corresponding to the QSO's ID number in Table \ref{tb:starqso}, and $r_0$ is set as 0. Along each sightline, we have $N_{\rm Si, k}=N_{\rm SiII, k}+N_{\rm SiIII, k}+N_{\rm SiIV, k}$. In Table \ref{tb:fmetal} we record the Si mass estimated for each ($r_{k-1}, r_{k}$) annulus, as well as the Si mass locked in the stars and ISM as estimated below. 

\begin{deluxetable}{cccc}
\tablenum{4}
\tablecaption{Silicon Mass Fraction Radial Profile \label{tb:fmetal}}
\tablewidth{0pt}
\tablehead{
\colhead{$r$ (kpc)} & \colhead{Component} & \colhead{$M_{\rm si}/10^3~\msun$} & \colhead{$M_{\rm si}/M_{\rm Si}^{\rm tot}$}\\
\colhead{(1)} & \colhead{(2)} & \colhead{(3)} & \colhead{(4)}
}
\startdata
0 & stars & $\sim8$ & 3\% \\
\hline
0 & ISM   & $\sim7$ & 2\% \\ 
\hline 
(0, 6] & CGM ($<$Q1) & $\sim(0.5-0.8)$ & 0.2--0.3\% \\ 
\hline 
(6, 23] & CGM (Q1--Q2) & $\sim(12-14)$ & 4--5\% \\ 
\hline 
(23, 38] & CGM (Q2--Q3) & $\sim(40-50)$ & 13--16\% \\ 
\hline
(38, 61] & CGM (Q3--Q456) & $\sim(48-62)$ & 15--20\% \\ 
\hline
(0, 61] & CGM ($<0.6\rvir$) & $\sim(100-130)$ & 32--42\% \\
\hline
\hline
\enddata
\tablecomments{Col. (1): The impact parameter at which the Si mass is calculated. Col. (2): for the Si mass in the stars and ISM, we follow the same procedures outlined in \citetalias{Zheng19b}; the Si mass in the CGM probed by each QSO is computed with Eq \ref{eq:msi_cgm} without doing the total sum. Col. (3): for the CGM Si mass measured toward Q1--Q6, a mass range is given with the left bound estimated with ``CGM" absorbers, and the right bound with both ``CGM" and ``Non-Detection" absorbers for $3\sigma$ upper limit. Because the impact parameters of Q4--Q6 are very similar, we use the average of their impact parameters for $r_4$ and the corresponding mean column densities for $N_{\rm Si, 4}$ in Eq \ref{eq:msi_cgm}. Col. (4): Si mass fraction in the stars, ISM, and CGM related to the total amount of Si ever produced (See \S\ref{sec:empirical_mass_budget}).}
\end{deluxetable}

%

Same as \citetalias{Zheng19b}, we adopt $R=0.34$ for the fraction of mass returned to the ISM per stellar generation, and $R_*=1-R=M_*/M_{\rm tot, SF}=0.66$ for the fraction locked in stars since star formation, where $M_{\rm tot, SF}$ is the total mass formed with star formation. The stellar yield is $y_{\rm Si}\equiv M_{\rm Si}^{\rm gas}/M_*=0.003$, which is the ratio of the Si mass in gas to the total stellar mass. The $R$ and $y_{\rm Si}$ values were initially derived for WLM with the NuGrid collaboration yield set and the SYGMA simple stellar population model \citep{ritter18a, ritter18b}, which are applicable to IC1613 given that the two galaxies have similar gas-phase metallicity.
Below we follow \citetalias{Zheng19b}'s equations 3--7 to derive relevant Si masses, but refrain from explaining the details that go into each calculation.

The total Si mass in the gas, including those in the ISM, CGM, or beyond, is $M_{\rm Si}^{\rm gas}=y_{\rm Si}M_*=0.003\times10^8~\msun=3\times10^5~\msun$. The relative abundance of Si to H in IC1613's ISM is ${\rm 12+\log(Si/H)_{\rm IC1613}=6.55\pm0.07}$\footnote{12+${\rm \log(Si/H)_{\rm IC1613}=}$12+${\rm \log(O/H)_{\rm IC1613}}$+${\rm \log(Si/O)_\odot}$, where ${\rm \log(Si/O)_\odot=\log(Si/H)_\odot-\log(O/H)_\odot}$; we assume the ISM of IC1613 has the same element composition as the Sun, and adopt 12+${\rm \log(Si/H)_\odot=}$ 7.51$\pm$0.03 and 12+${\rm \log(O/H)_\odot}=$ 8.69$\pm$0.05 from \cite{asplund09}, and 12+${\rm \log(O/H)_{IC1613} =}$ 7.73$\pm$0.04 from \cite{bresolin07}.}, with which we can infer the Si mass in IC1613's ISM as $M_{\rm Si}^{\rm ISM}=M_{\rm HI}(m_{\rm Si}/m_{\rm H})({\rm Si/H})_{\rm IC1613}\sim7\times10^3~\msun$. Similarly, we can estimate the total amount of Si locked in the stars as $M_{\rm Si}^*=0.74M_*(m_{\rm Si}/m_{\rm H})({\rm Si/H})_{\rm IC1613}\sim8\times10^3~\msun$, where 0.74 is the hydrogen mass fraction.

\begin{figure}
    \centering
    \includegraphics[width=\columnwidth]{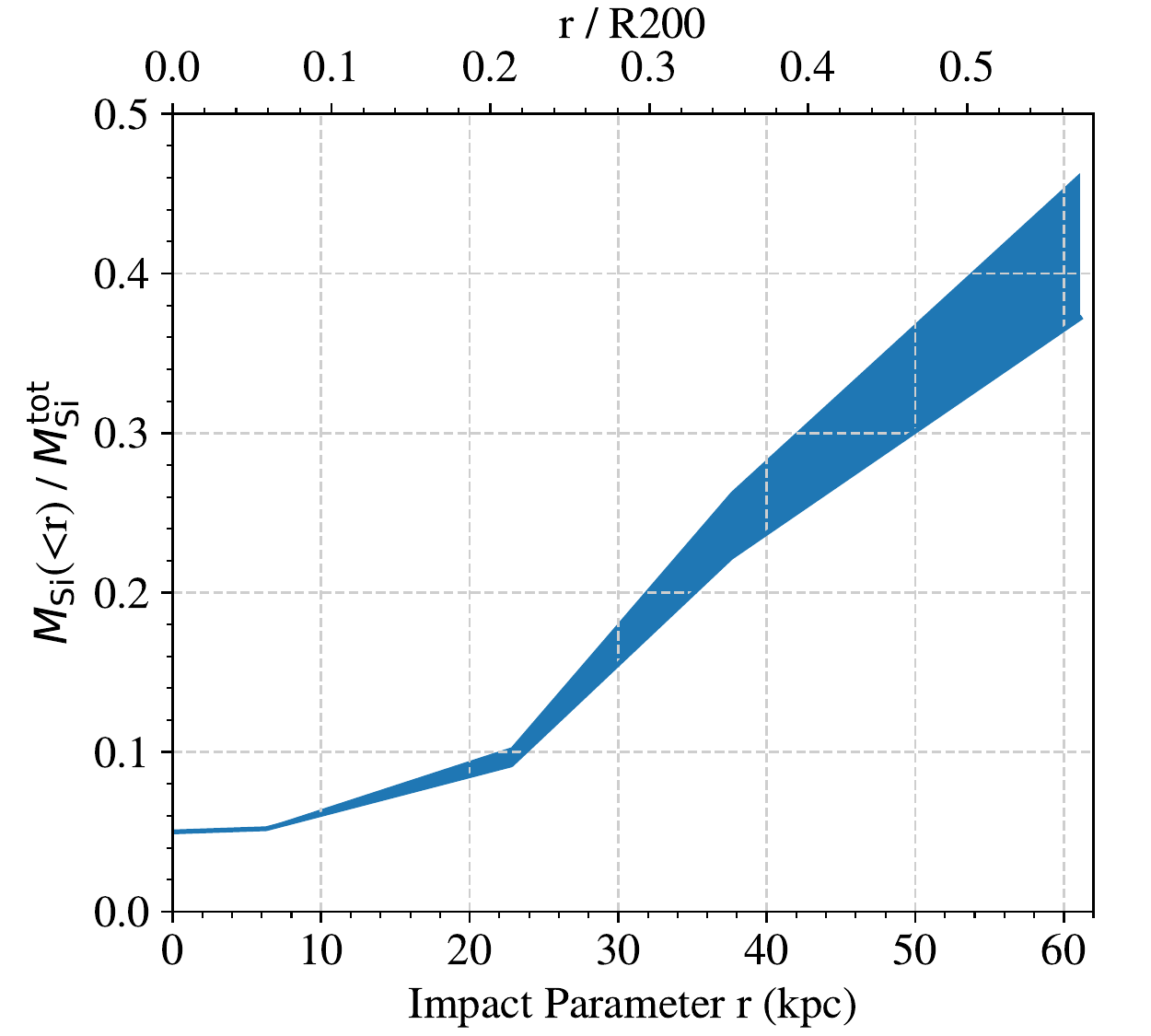}
    \caption{The cumulative mass fraction of Si in IC1613 as a function of impact parameter. At each $r$, the cumulative value is computed by summing up the $M_{\rm Si}/M_{\rm Si}^{\rm tot}$ values at $\leq r$ in Table \ref{tb:fmetal}. The boundaries of the shaded blue region represent the lower and upper bounds of the mass fractions based on the $M_{\rm Si}$ values in Table \ref{tb:fmetal}.}
    \label{fig:metal_fraction_cdf}
\end{figure}

\begin{figure*}[t!]
    \centering
    \includegraphics[width=\textwidth]{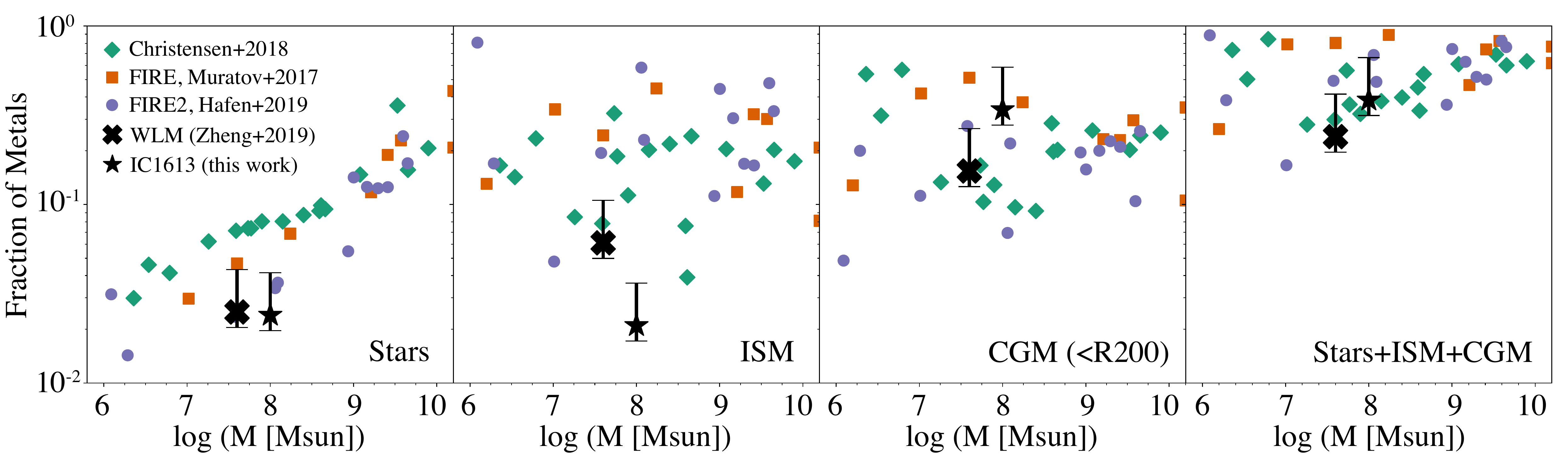}
    \caption{The mass fraction of metals in the stars, ISM, CGM, and their sum as a function of galaxy stellar mass for WLM \citep{Zheng19b}, IC1613 (this work), and a selection of recent suites of zoom-in simulations of individual, isolated galaxies from FIRE \citep{muratov17}, FIRE-2 \citep{Hafen19}, and \citet{christensen18}. The error bars on the observational data points reflect only potential variation in these quantities with choice of stellar yield in Si between models. We emphasize that the definition of ISM and CGM between simulations and these observations differs. See \S\ref{sec:sim_mass_budget} for more details. For IC1613, the CGM value is computed only for gas within $0.6\rvir$ as probed by our COS data. }
    \label{fig:metal_fraction}
\end{figure*}

The total amount of Si ever produced in IC1613 is $M_{\rm Si}^{\rm tot}=M_{\rm Si}^{*}+M_{\rm Si}^{\rm gas}\approx3.1\times10^5~\msun$. When considering the Si mass fraction, we find that $\sim$3\%, $\sim$2\%, and $\sim$32--42\% of the mass is in the stars, ISM, and within $0.6\rvir$ of the CGM, respectively. In Figure \ref{fig:metal_fraction_cdf}, we show the cumulative Si mass fraction in IC1613 and its CGM. At $d\sim0$ kpc, the galaxy itself contains $\sim5$\% of the Si in the stars and ISM. In the CGM, the Si mass fraction increases quickly with $r$ mainly because the mass is proportional to the surface area $\propto r^2$ (see Eq. \ref{eq:msi_cgm}).


\subsection{Mass Budget Comparison with Simulations}
\label{sec:sim_mass_budget} 

In Figure \ref{fig:metal_fraction}, we compare the Si mass budgets for IC1613 and WLM \citepalias{Zheng19b} to the predicted values for dwarf galaxies with $\mstar\sim10^{6-10}~\msun$ from the FIRE and FIRE2 simulations as analyzed in \citet{muratov17} and \citet{Hafen19} respectively, and the simulations of \citet{christensen18}. 
The left panel shows that the stellar metal mass fractions increase with $\mstar$ as a result of the stellar mass-metallicity relation. However, the simulated ISM metal mass fractions do not strongly correlate with $\mstar$ despite spanning four orders of magnitudes in $\mstar$. We suspect that even though these simulated galaxies follow a similar gas-phase mass-metallicity relation as their observational counterparts, the gas fractions in the galaxies decrease with $\mstar$ \citep{el-badry18}, resulting in the non-correction in the ISM panel.

When compared to observations, the fraction of metals locked in the stars and ISM in all simulations are a factor of $\sim$ 2 or more higher than observed in IC1613 and WLM. For instance, only 2--6\% of the Si are in the ISM of IC1613 and WLM, as compared to $\sim$2--60\% of the metals contained in the simulated ISM. The discrepancy is likely to be due to (1) different definitions of the ISM, (2) different assumptions on stellar yields and stellar evolution modeling, and (3) the specific simulation setup and feedback treatment that expels metals from galaxies to various degrees. For (1), both \citet{muratov17} and \citet{Hafen19} define the ISM as all gas within 0.1 virial radii. For a galaxy such as IC1613, defining the ISM as within 0.1$\rvir$ would include gas within 10 kpc. However, the half light radius of IC1613 is only 1.5 kpc \citep{mcconnachie12} and the \HI\ in its ISM extends to a radius of $\sim2.5$ kpc at a column density level\footnote{To derive the \HI\ extent of IC1613's ISM, we analyze the VLA's natural-weighted map cube of IC1613 from the LITTLE THINGS survey \citep{hunter12}. We generate an \HI\ column density map of the galaxy by integrating the data cube from $\vlsr=-360~\kms$ to $-120~\kms$ to include \HI\ emission within $\pm120~\kms$ of the systemic velocity of IC1613. We then smooth the column density map with Gaussian kernels and determine the extent of the \HI\ by estimating the size of the column density contour at $5\times10^{19}~{\rm cm^{-2}}$ over a velocity window of 240 $\kms$, which corresponds the rms value as listed in their table 3. \label{fnt_HI}} of $5\times10^{19}$ cm$^{-2}$. Therefore, \cite{muratov17} and \citet{Hafen19}'s ISM definition extends the ISM size by a factor of $\sim$4 and includes gas at higher temperatures that are typically not probed by \HI\ 21cm emission. Indeed, re-defining the ISM as gas within 2.5~kpc for all FIRE galaxies in this stellar mass range does lower the average ISM metal mass fraction from $\sim$0.24 to $\sim$0.13 (priv. comm. with FIRE). \cite{christensen18} defines the ISM as all gas with number density $> 0.1$~cm$^{-3}$, temperature $< 1.2 \times 10^{4}$~K, and within a cylindrical height of 3~kpc from the plane of the disk of their galaxies, which is more comparable for the particular properties of IC1613. Even so, our measurements are still low compared to the typical simulated values. Note that, the ISM definition would not change the values for the stellar metal fraction, which is similarly low for our observational estimates compared to what is expected from these simulations.

For (2), there are significant variations in the expected yield of Si depending on the choices of both nucleosynthetic yields and initial mass function (IMF). While this does not affect the results from the simulations as plotted since they are properly normalized by the total metals present in the computational domain, it does affect our observational estimates of the total amount of Si present. To understand the impact of this uncertainty, we bracket our observational estimates of WLM's and IC1613's metal fractions in Figure~\ref{fig:metal_fraction} with the lower bounds estimated with $y_{\rm Si}=1.64\times 10^{-3}$ as adopted in the FIRE simulations and the upper bounds with $y_{\rm Si}\sim 3.7 \times 10^{-3}$ from \citet{christensen18} for their choice of stellar yields and IMF. Note that, in our estimates, we use $y_{\rm Si}=3\times10^{-3}$ as discussed in \S\ref{sec:empirical_mass_budget}. Figure \ref{fig:metal_fraction} shows that varying $y_{\rm Si}$ values does result in a large range in the metal mass fraction in stars, ISM, and CGM, but the stellar and ISM values are still at the lower end of the prediction from simulated galaxies.

For (3), it is interesting that all simulations give broadly similar results in spite of their varying simulation setups and feedback recipes. It is beyond the scope of this work to explore deeply on what sets the scatters in the simulations, but we note that among all the simulated galaxies there are some with similarly low metal fractions as IC1613 and WLM. Therefore, it would be valuable to develop a larger observational sample of these types of measurements for a more statistically meaningful comparison across simulations. 

Lastly, in the CGM panel, we find that IC1613 and WLM contain as many metals as the simulations have predicted. No strong correlation is seen between the CGM metal mass fraction and $\mstar$. Unlike the ISM, neither the gas-phase mass-metallicity relationship nor the gas mass fraction of the CGM is well studied observationally. Relevant CGM properties in the simulated galaxies also await further investigation in order to fully understand the scatters and the non-correlation of the CGM metal fraction with $\mstar$.

\subsection{Metal Outflow Rate \& Instantaneous Metal Mass Loading Factor}
\label{sec:outflow_rate}

A number of ``CGM/Outflow" absorbers are detected toward stellar sightlines S1--S4 (see \S\ref{sec:ic1613}). Because these sightlines were observed in a down-the-barrel manner, the impact parameters of these absorbers from the galaxy are unknown, which means they could be absorbers in the CGM or outflows in the immediate region of the galaxy. Similar distance ambiguity in identifying absorbers' distances relative to host galaxies has also troubled other down-the-barrel studies of gas flows in extragalactic systems \citep[e.g., ][]{rubin12, rubin14, chisholm16, zheng17}. Hereafter, we assume that these absorbers probe outflowing material from IC1613 and estimate the metal outflow rate $\moutz$ and instantaneous metal mass loading factor $\eta_{\rm Z}$. Following the definition in \cite{christensen18}, $\eta_{\rm Z}\equiv \moutz/\dot{M}$ is the ratio of metal mass carried by outflows per unit time to the star-formation rate at the present day. Note that, $\eta_{\rm Z}$ is different from the effective metal mass loading factor or the instantaneous/effective gas mass loading factor that have been used in the literature\footnote{The effective metal mass loading factor is a cumulative quantity of $\eta_{\rm Z}$ integrated over time; it is the ratio of the total metal mass a galaxy has lost throughout its star-formation history to the total stellar mass ever formed. The instantaneous/effective gas mass loading factors are defined similarly, but with the nominator values from outflowing gas mass instead of metal mass \citep[e.g., ][]{christensen16, muratov17}. \label{fnt_4}}.

Given that S1--S4 are located at different corners of IC1613 (see Figure \ref{fig:starqso}), we assume a cylindrical geometry to represent the outflowing material with a radius of $R_{\rm out}=2.5$ kpc 
based on the \HI\ extent of the galaxy as calculated in \S\ref{sec:sim_mass_budget} and footnote \footref{fnt_HI}.
The metal outflow rate $\moutz$ for an ion X can be derived as the following:
\begin{equation}
\begin{array}{ll}
\moutz &= d{\rm M_{out}}/d{\rm t}=d({\rm C_{f} \rho_{x} \pi R_{out}^2 v_{out}t })/d{\rm t} \\
      &= {\rm C_{f} \pi R_{out}^2 v_{out} m_{x} N_{x}/D_{out}}
\end{array}
\end{equation}
. In the equation, ${\rm m_X}$ and $\rm N_{\rm X}$ are the atom mass and column density of ion X. $\rm C_{f}$ is the covering fraction, and we assume $\rm C_{f}=1$ as the outflow absorbers are commonly detected among the stellar sightlines. $\rm v_{out}$ is the outflow velocity corrected for the galaxy's inclination, with typical values summarized in \S\ref{sec:ic1613}. And $\rm D_{out}$ is the distance the outflows have reached. We adopt $D_{\rm out}=1$ kpc (or $\sim0.01\rvir$) for two considerations. First, in order to derive the instantaneous $\moutz$ and $\eta_{\rm Z}$ values, we assume the outflows to have been enriching the vicinity of the galaxy within the past $\sim$10--20 Myrs at current outflow velocities. This is reasonable given that IC1613 has a nearly continuous and constant star formation rate over the past $>$10 Gyrs \citep{cole99, skillman03, skillman14, weisz14}. Second, because ${\rm \rho_{x}\equiv m_x N_x/D_{out}}$, not only does ${\rm D_{out}}$ represent the distance the outflows have reached, it also indicates the physical size of an outflowing ion absorber. Though we do not have information on the typical absorber size in IC1613's CGM, a diameter of $\sim1$ kpc is typically seen from observations of CGM absorbers of $L\geq0.1L^*$ galaxies \citep{stocke13, werk14}. We find that the instantaneous outflow rate is $\moutz=1.1\times10^{-5}~\msunyr$ combining the measurements from \SiII\ and \SiIII\ outflow-like absorbers. The total star formation rate of IC1613 is $\dot{M}=2.5\times10^{-3}~\msun~{\rm yr^{-1}}$ as measured from the H$\alpha$ luminosity \citep{hunter04}. Therefore, the instantaneous metal mass loading factor is $\eta_{\rm Z}=\moutz/\dot{M}=0.004$ for \SiII\ and \SiIII.

We do not use the \SiIV\ lines because there are no robust Voigt-profile fits for these lines to effectively separate the ISM absorption from that of the outflows (\S\ref{sec:cont_voigt}). Instead, we run a grid of Cloudy models \citep{Ferland17} to estimate the \SiIV\ column density based on the measurements of \SiII\ and \SiIII, with the assumption that \SiII, \SiIII, and \SiIV\ are in the same phase. We implement an extragalactic UV background \citep{hm01} and add ionizing flux from the star formation in the galaxy as a function of impact parameter and escape fraction as in \cite{werk14} as radiation sources. We find our results are not sensitive to the details of the ionizing background, but only its overall shape. We examine the results at a metallicty of 0.1 solar \citep{bresolin07}, a star formation rate of $2.5\times10^{-3}~\msunyr$ \citep{hunter04}, and an escape fraction of 10\%. At N(\HI)$\leq1.5\times10^{19}$ cm$^{-2}$ for a line 30$~\kms$ wide as measured from the VLA data (see \S\ref{sec:hi}), the constraint from \SiIII/\SiII\ ion ratio yields a nearly constant ionization parameter $\log U\sim$(-3.3, -3.8) and a \SiIV\ column density $N_{\rm SiIV}\sim10^{10.9-11.7}$ cm$^{-2}$ that is well below the detection limit of our COS spectra. Therefore, there is only a negligible amount of \SiIV\ in the same phase as \SiII\ and \SiIII\ in IC1613's outflows. However, we cannot rule out the case that outflow-like \SiIV\ absorbers are present in a warmer phase given that \CIV\ is detected at $\vgal<-20~\kms$ along some of the stellar sightlines. The lower ionization states of carbon and silicon offer no constraints on the warmer-phase material, and the predicted \CIV/\SiIV\ ion ratio depends strongly on the warm-phase N(\HI), metallicity, and ionization state, none of which are known.

Without accurate N(\SiIV) values, the metal outflow rate ($\moutz=1.1\times10^{-5}~\msunyr$) and the instantaneous metal mass loading factor ($\eta_{\rm Z}=0.004$) are deemed lower limits. When compared with simulations of dwarf galaxies, \cite{christensen18} find $\eta_{\rm z}\sim0.004-0.01$ at a circular velocity of v$_c =40~\kms$, appropriate for a galaxy at the mass of IC1613 
(see also \citealt{muratov17}). Though there have been constraints for dwarf galaxies' outflow gas mass loading factors \citep{mcquinn19}, metal mass loading factors are rarely observationally determined. \cite{mcquinn19} show that the gas mass loading factors range from 0.2 to 7 for a sample of nearby low-mass galaxies ($\mstar\sim10^{7-9.3}~\msun$) based on H$\alpha$ emission line observations; however, as they noted, the gas mass loading factors and the metal mass loading factors are not directly comparable without the knowledge of the phases of the outflowing metals.

Lastly, we highlight that $\eta_{\rm Z}\geq0.004$ is consistent with the stellar yield $y_{\rm Si}$ ($=0.003$) adopted in \S\ref{sec:metal_budget} despite that they are derived under different sets of assumptions for IC1613. For every unit star formed, a fraction of $\geq0.004$ of the stellar mass is in the form of metal (Si) outflows. Furthermore, assuming a constant outflow rate over the lifetime of the galaxy ($T\sim14$ Gyrs) given its constant star-formation history \citep{skillman14}, the total amount of Si accumulated in the CGM would be $\moutz T\geq1.5\times10^5~\msun$. This is consistent with the Si mass in the CGM from \S\ref{sec:empirical_mass_budget} that we derive based on \SiII, \SiIII, and \SiIV\ column density measurements along Q1--Q6 sightlines.


\begin{figure*}[t]
    \centering
    \includegraphics[width=\textwidth]{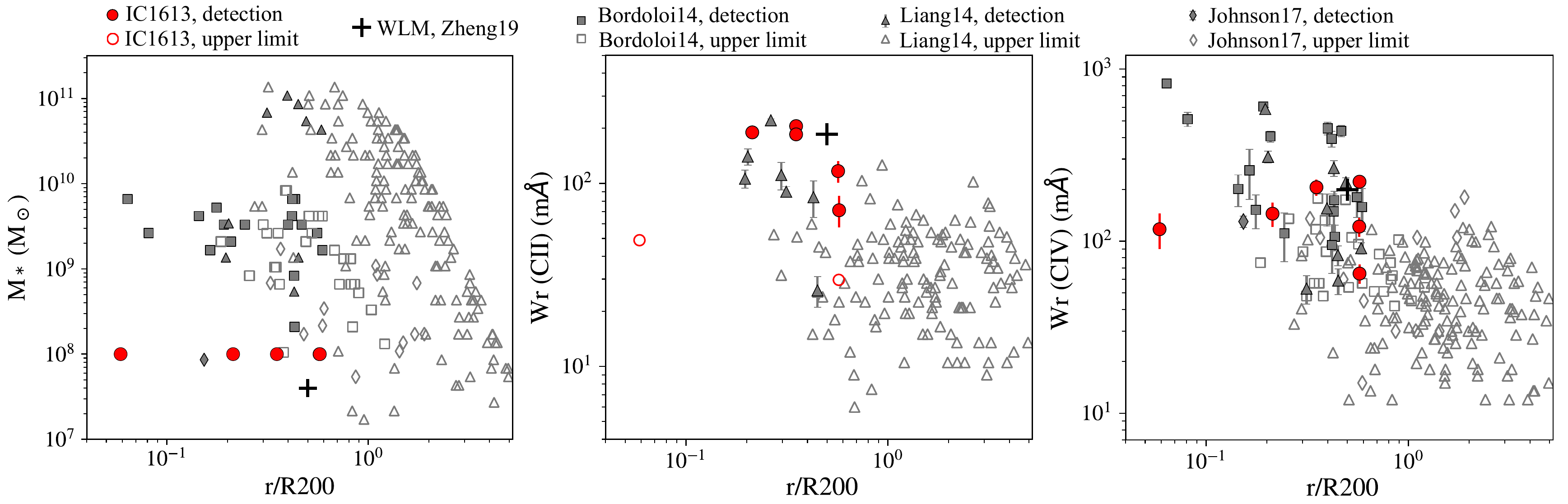}
    \caption{Left: Host galaxy stellar mass ($M_*$) v.s. impact parameter ($r/\rvir$) for \CIV\ absorbers from \citetalias{bordoloi14}, \citetalias{liang14}, \citetalias{johnson17}, \citetalias{Zheng19b}, and IC1613 (this work). We show that IC1613 and WLM probe a unique parameter space with low $M_*$ and small $r$ that has not been well studied before. Note that \citetalias{liang14}'s sample also includes 8 sightlines near host galaxies with $M_*<10^7~\msun$ or at $r/\rvir>5$ that we do not show in this figure, and all of them are non-detection. 
    Middle and right panels: equivalent width ($W_r$) as a function of impact parameter scaled with $\rvir$. We choose to use $\ew$ values instead of $\log N$ because it is the most common measurement among the three low-$z$ dwarf studies. For IC1613, we only use those measurements tagged as ``CGM" or ``Non-Detection" toward Q1--Q6. For absorbers from \citetalias{bordoloi14}, \citetalias{liang14}, \citetalias{johnson17}, and \citetalias{Zheng19b}, solid symbols show detection, and open ones indicate $3\sigma$ upper limits for non-detection.}
    \label{fig:Wr_dproj}
\end{figure*}

\section{Discussion}
\label{sec:discussion}

\subsection{The CGM of Other Low-Mass Galaxies}
\label{sec:comp_lowz_dwarfs}

We first compare our ion absorbers near IC1613 with those measured in and near the CGM of low-$z$ dwarf galaxies studied by \citeauthor{bordoloi14} (2014; hereafter, \citetalias{bordoloi14}), \citeauthor{liang14} (2014; hereafter, \citetalias{liang14}), and \citeauthor{johnson17} (2017; hereafter \citetalias{johnson17}), and a dwarf irregular galaxy WLM in the LG \citepalias{Zheng19b}. Since different initial mass functions (IMF) were used to derive the stellar mass ($M_*$) in different studies, we convert their $M_*$ values from the corresponding IMF (i.e., \citealt{salpeter55, chabrier03}) to that of \cite{kroupa01}. This is to be consistent with the IMF choice in our adopted $M_*$--$M_h$ relation from \cite{moster10} (see \S\ref{sec:method}). Specifically, using the rescaling factors recommended in \cite{madau14}, we multiply the $M_*$ values from \citetalias{bordoloi14} by 0.66 to convert from the Salpeter IMF to the Kroupa IMF. We multiply the $M_*$ values from \citetalias{liang14} and \citetalias{johnson17} by 1.08 to convert from the Chabrier IMF to the Kroupa IMF. 

In the left panel of Figure \ref{fig:Wr_dproj}, we show the range of the galaxy stellar mass $\mstar$ and impact parameter $r$ covered by these studies, and highlight that the sightlines near IC1613 and WLM probe a unique parameter space at $\mstar\lesssim10^8~\msun$ and $r\lesssim0.6\rvir$ that has not been well studied before.
\citetalias{bordoloi14} studied \CIV\ absorption in the CGM of 43 low-mass galaxies at $z\leq0.1$; their sample probes the inner CGM from $0.05\rvir$ to $0.5\rvir$, but focus on more massive galaxies with $\mstar\sim10^{8.2-10.2}~\msun$. \citetalias{liang14} studied Ly$\alpha$, \CII, \CIV, \SiII, \SiIII, and \SiIV\ absorbers within 500 kpc of 195 isolated galaxies at $z<0.176$. Their sample includes a wide range of galaxy stellar masses with $\mstar\sim10^{5.2-11.1}~\msun$, but $90\%$ of the sightlines are at $>0.6~\rvir$ and do not have detection. \citetalias{johnson17} studied 18 star-forming field dwarfs with $\mstar\sim10^{7.7-9.2}~\msun$ and $r/\rvir\sim0.2-2$; while most of their sightlines find non detection of metal lines, the one at $\mstar=10^{7.9}~\msun$ and $r=0.15\rvir$ show \SiIII\ and \CIV\ absorbers with similar equivalent widths as those near IC1613 and WLM. 


In the middle and right panels of Figure \ref{fig:Wr_dproj}, we show the $\ew$ values of \CII\ and \CIV\ as a function of impact parameter scaled with $\rvir$. For consistency, we recalculate $\rvir$ for all the galaxy halos from \citetalias{bordoloi14}, \citetalias{liang14}, \citetalias{johnson17}, and \citetalias{Zheng19b} using our $\rvir$ definition as detailed in \S\ref{sec:intro}. This definition is consistent with what is used by \citetalias{bordoloi14} and \citetalias{Zheng19b}, but systematically larger than those adopted by \citetalias{liang14} and \citetalias{johnson17}. The latter defines $\rvir$ based on the critical density with an over-density factor $\Delta_c$ from \cite{bryan98}. 
Figure \ref{fig:Wr_dproj} shows that detection mainly occur within $0.6\rvir$ and the $W_{\rm r}$ values of the detected absorbers are consistent among various work. Results of \SiIII\ and \SiIV\ are similar. \citetalias{bordoloi14} found a power law decline in \CIV's equivalent widths out to $\sim0.5\rvir$; we do not observe such a trend in \CIV\ detected near IC1613, likely due to the sparse data points in our sample. Regarding the total metal content, we find that the LG dwarf galaxies ($\mstar\sim10^{7-8}~\msun$) host a reservoir of metals with mass similar to those low-mass galaxies at low redshifts.

In the LG, thus far there have been limited studies of the CGM of dwarf galaxies. \cite{Zheng19b} reported a tentative detection of CGM absorber at 0.5$\rvir$ in WLM (Figure \ref{fig:Wr_dproj}). The uncertainty in their diagnosis of the absorber's origin is also due to the chance alignment with the foreground Magellanic System in the $\ml$--$\vlsr$ diagram, as is shown in Figure \ref{fig:ms_lg_dwarfs}. Our investigation that $\ml$--$\vlsr$ diagram does not yield robust  diagnosis on an absorber's connection to the Magellanic System (see \S\ref{sec:pv_ms}) now has provided stronger argument for the absorber's association with WLM's CGM. The Si mass derived for WLM's CGM is $\sim(0.2-1.0)\times10^5~\msun$, which is similar to what we derive for IC1613.

Furthermore, in a study of Milky Way's ionized high-velocity gas, \cite{richter17} also looked for metal absorption line features along QSO sightlines within impact parameters of $\sim$0.5--2 virial radii of 19 LG dwarf galaxies with or without gas, but did not find significant detection near the systemic velocities of host galaxies. They concluded that there was no compelling evidence of CGM gas near LG dwarf galaxies. However, it is worth noting that the detectability of the CGM absorbers in their data could be compromised because of the low signal-to-noise ratio criterion they adopted to choose the spectra (S/N$\geq$6) and the large impact parameters of the sightlines ($>0.5$ virial radius).

Though current observational effort of low-mass galaxies' CGM is limited, upcoming HST/COS programs, such as \href{http://archive.stsci.edu/proposal_search.php?mission=hst&id=16301}{GO-16301} (PI Putman) and \href{http://archive.stsci.edu/proposal_search.php?mission=hst&id=15227}{GO-15227} (PI Burchett), will provide a promising, large sample of nearby low-mass galaxies for statistically significant comparisons on CGM metal content.

\subsection{The Metal Content in Other LG Dwarf Galaxies}

Our estimate of the Si mass fraction locked in the stars of IC1613 ($\sim3\%$) is consistent with what have been measured for some other LG dwarf galaxies. \cite{kirby11, kirby13} show that $\geq96\%$ of the iron ever produced in LG dwarf galaxies is no longer locked in their stars. In addition to WLM as we have compared with in \S\ref{sec:sim_mass_budget}, another interesting galaxy to discuss is Leo P. Discovered by \citet{giovanelli13}, Leo P is also an isolated dwarf irregular galaxy that is far away from a massive host. Therefore, the galaxy is unlikely to lose its gas through stripping; instead, any gas lost was probably pushed out by stellar feedback. \citet{mcquinn15b} find that the mass of oxygen retained in the stars and ISM of Leo P is 5\%, same as IC1613. Interestingly, Leo P has a stellar mass 180 times less than IC1613 (${\rm M_{*, LeoP}=5.6\times10^5~\msun}$; \citealt{mcquinn15a}). The similar metal retention fractions of Leo P and IC1613 challenge the correlation between the metal mass fraction in the stars and the $\mstar$ of the simulated galaxies as shown in the left panel of Figure \ref{fig:metal_fraction}. More simulations on dwarf galaxies at Leo P's mass \citep[e.g.][]{rey20} are needed to further investigate how the metal fractions in stars scale with $\mstar$ at much lower mass regime.

\section{Conclusion}
\label{sec:summary}

With 4 stellar and 6 QSO sightlines observed with HST/COS, we study the CGM and outflows of IC1613, an isolated, low-mass ($\mstar\sim10^8~\msun$) dwarf irregular galaxy on the outskirts of the LG. IC1613 is among the lowest mass galaxies ever studied in the context of CGM metal content and outflows, and it is one of the rare cases whose CGM is probed by more than one QSO sightline except for the Milky Way and M31.

Our stellar and QSO sightlines probe a wide range of impact parameters, from $<0.1\rvir$ to $0.6\rvir$, and detect a number of \SiII, \SiIII, \SiIV, \CII, and \CIV\ ion absorbers. We consider an absorber to be associated with IC1613's CGM, ISM, outflow, or inflow if its velocity is within the escape velocity of the galaxy (thus gravitationally bound). When comparing the IC1613-associated absorbers with those of dwarf galaxies at low-$z$, we find that the absorbers near IC1613 have similar line strengths. 

We estimate a silicon mass of $M_{\rm Si}^{\rm CGM}\approx(1.0-1.3)\times10^5~\msun$ within $0.6\rvir$ of IC1613's CGM, assuming that the majority of the Si is in the ionization states of \SiII, \SiIII, and \SiIV. We also estimate the Si metal content in the stars and ISM based on IC1613's stellar mass, \HI\ mass, theoretical nucleosynthetic yields, and gas-phase metallicity. We find $M_{\rm Si}^{\rm *}\sim8\times10^3~\msun$ for Si locked in the stars and $M_{\rm Si}^{\rm ISM}\sim7\times10^3~\msun$ for Si in the ISM. Overall, of all the Si ever been produced in IC1613, $\sim$3\%, $\sim$2\%, and $\sim$32--42\% of the mass is in the stars, ISM, and within $0.6\rvir$ of the galaxy's CGM (see Figure \ref{fig:metal_fraction}), which accounts for nearly half of the total Si mass budget. The remaining $\sim$50--60\% of the Si mass is either in the outer CGM of IC1613 ($0.6<r/\rvir<1$), or has escaped beyond the virial radius of the galaxy. Our results are largely consistent with predicted values from existing simulations, although large scatters in the ISM and CGM metal fractions are found in simulated galaxies at different masses (see Figure \ref{fig:metal_fraction}). 

Lastly, based on the \SiII\ and \SiIII\ measurements of the outflow-like absorbers toward S1--S4, we find a metal outflow rate of $\moutz\geq1.1\times10^{-5}~\msunyr$ and an instantaneous metal mass loading factor of $\eta_{\rm z}\geq0.004$, consistent with the predicted values for simulated galaxies at similar masses. We highlight that, assuming a constant metal outflow rate throughout IC1613's star formation history, the total Si mass in the galaxy's CGM as enriched by these metal outflows is consistent with the current CGM mass independently measured from the QSO sightlines Q1--Q6.

To conclude, our work shows that there is a large mass reservoir of silicon in the CGM of IC1613, which has been continuously enriched by metal outflows throughout the galaxy's star formation history. Our results are largely consistent with what have been predicted for simulated galaxies at similar masses. We are looking forward to compiling a larger observational sample consisting of nearby low-mass galaxies to yield statistically meaningful assessment on how the CGM and metal outflow properties vary from galaxy to galaxy and from observations to simulations.\\

\textbf{Acknowledgements}: We thank R. Bordoloi for sharing his python code of escape velocity calculation, E. Patel for discussing the projection effect between the Magellanic System and the LG galaxies using hydrodynamic simulations and HST/Gaia proper motion measurements, and D. Weisz for discussing many aspects of this paper and for his great support as a faculty mentor to Y.Z. at Miller Institute at UC Berkeley. We also thank A. Fox and P. Richter for helpful discussion on the manuscript. Y.Z. acknowledges support from the Miller Institute for Basic Research in Science. Support for Program number HST-GO-15156 was provided by NASA through a grant from the Space Telescope Science Institute, which is operated by the Association of Universities for Research in Astronomy, Incorporated, under NASA contract NAS5-26555.  This material is based upon work supported by the National Science Foundation under Grant No.\ AST-1847909.  E.N.K.\ gratefully acknowledges support from a Cottrell Scholar award administered by the Research Corporation for Science Advancement. This research has made use of the HSLA database, developed and maintained at STScI, Baltimore, USA.

\vspace{5mm}
\facilities{Hubble Space Telescope/Cosmic Origins Spectrograph, Mikulski Archive for Space Telescopes (MAST)}

\software{Astropy \citep{astropy2}, Numpy \citep{numpy}, Matplotlib \citep{matplotlib}, CLOUDY \citep{Ferland17}, IDL, the gala package \citep{gala-code}}


\bibliographystyle{aasjournal}
\bibliography{main}

\begin{thebibliography}{}
\expandafter\ifx\csname natexlab\endcsname\relax\def\natexlab#1{#1}\fi
\providecommand{\url}[1]{\href{#1}{#1}}
\providecommand{\dodoi}[1]{doi:~\href{http://doi.org/#1}{\nolinkurl{#1}}}
\providecommand{\doeprint}[1]{\href{http://ascl.net/#1}{\nolinkurl{http://ascl.net/#1}}}
\providecommand{\doarXiv}[1]{\href{https://arxiv.org/abs/#1}{\nolinkurl{https://arxiv.org/abs/#1}}}

\bibitem[{{Andrews} \& {Martini}(2013)}]{andrews13}
{Andrews}, B.~H., \& {Martini}, P. 2013, \apj, 765, 140,
  \dodoi{10.1088/0004-637X/765/2/140}

\bibitem[{{Asplund} {et~al.}(2009){Asplund}, {Grevesse}, {Sauval}, \&
  {Scott}}]{asplund09}
{Asplund}, M., {Grevesse}, N., {Sauval}, A.~J., \& {Scott}, P. 2009, Annual
  Review of Astronomy and Astrophysics, 47, 481,
  \dodoi{10.1146/annurev.astro.46.060407.145222}

\bibitem[{{Barger} {et~al.}(2017){Barger}, {Madsen}, {Fox}, {Wakker},
  {Bland-Hawthorn}, {Nidever}, {Haffner}, {Antwi-Danso}, {Hernand ez},
  {Lehner}, {Hill}, {Curzons}, \& {Tepper-Garc{\'\i}a}}]{barger17}
{Barger}, K.~A., {Madsen}, G.~J., {Fox}, A.~J., {et~al.} 2017, \apj, 851, 110,
  \dodoi{10.3847/1538-4357/aa992a}

\bibitem[{{Bernard} {et~al.}(2010){Bernard}, {Monelli}, {Gallart}, {Aparicio},
  {Cassisi}, {Drozdovsky}, {Hidalgo}, {Skillman}, \& {Stetson}}]{bernard10}
{Bernard}, E.~J., {Monelli}, M., {Gallart}, C., {et~al.} 2010, \apj, 712, 1259,
  \dodoi{10.1088/0004-637X/712/2/1259}

\bibitem[{{Bordoloi} {et~al.}(2014){Bordoloi}, {Tumlinson}, {Werk},
  {Oppenheimer}, {Peeples}, {Prochaska}, {Tripp}, {Katz}, {Dav{\'e}}, {Fox},
  {Thom}, {Ford}, {Weinberg}, {Burchett}, \& {Kollmeier}}]{bordoloi14}
{Bordoloi}, R., {Tumlinson}, J., {Werk}, J.~K., {et~al.} 2014, \apj, 796, 136,
  \dodoi{10.1088/0004-637X/796/2/136}

\bibitem[{{Bouch{\'e}} {et~al.}(2007){Bouch{\'e}}, {Lehnert}, {Aguirre},
  {P{\'e}roux}, \& {Bergeron}}]{bouche07}
{Bouch{\'e}}, N., {Lehnert}, M.~D., {Aguirre}, A., {P{\'e}roux}, C., \&
  {Bergeron}, J. 2007, \mnras, 378, 525,
  \dodoi{10.1111/j.1365-2966.2007.11740.x}

\bibitem[{{Bowen} {et~al.}(1997){Bowen}, {Tolstoy}, {Ferrara}, {Blades}, \&
  {Brinks}}]{bowen97}
{Bowen}, D.~V., {Tolstoy}, E., {Ferrara}, A., {Blades}, J.~C., \& {Brinks}, E.
  1997, \apj, 478, 530, \dodoi{10.1086/303823}

\bibitem[{{Bresolin} {et~al.}(2007){Bresolin}, {Urbaneja}, {Gieren},
  {Pietrzy{\'n}ski}, \& {Kudritzki}}]{bresolin07}
{Bresolin}, F., {Urbaneja}, M.~A., {Gieren}, W., {Pietrzy{\'n}ski}, G., \&
  {Kudritzki}, R.-P. 2007, \apj, 671, 2028, \dodoi{10.1086/522571}

\bibitem[{{Brooks} {et~al.}(2007){Brooks}, {Governato}, {Booth}, {Willman},
  {Gardner}, {Wadsley}, {Stinson}, \& {Quinn}}]{brooks07}
{Brooks}, A.~M., {Governato}, F., {Booth}, C.~M., {et~al.} 2007, \apjl, 655,
  L17, \dodoi{10.1086/511765}

\bibitem[{{Bryan} \& {Norman}(1998)}]{bryan98}
{Bryan}, G.~L., \& {Norman}, M.~L. 1998, \apj, 495, 80, \dodoi{10.1086/305262}

\bibitem[{{Calura} {et~al.}(2009){Calura}, {Pipino}, {Chiappini}, {Matteucci},
  \& {Maiolino}}]{calura09}
{Calura}, F., {Pipino}, A., {Chiappini}, C., {Matteucci}, F., \& {Maiolino}, R.
  2009, \aap, 504, 373, \dodoi{10.1051/0004-6361/200911756}

\bibitem[{{Chabrier}(2003)}]{chabrier03}
{Chabrier}, G. 2003, \pasp, 115, 763, \dodoi{10.1086/376392}

\bibitem[{{Chisholm} {et~al.}(2016){Chisholm}, {Tremonti}, {Leitherer}, {Chen},
  \& {Wofford}}]{chisholm16}
{Chisholm}, J., {Tremonti}, C.~A., {Leitherer}, C., {Chen}, Y., \& {Wofford},
  A. 2016, \mnras, 457, 3133, \dodoi{10.1093/mnras/stw178}

\bibitem[{{Christensen} {et~al.}(2018){Christensen}, {Dav{\'e}}, {Brooks},
  {Quinn}, \& {Shen}}]{christensen18}
{Christensen}, C.~R., {Dav{\'e}}, R., {Brooks}, A., {Quinn}, T., \& {Shen}, S.
  2018, \apj, 867, 142, \dodoi{10.3847/1538-4357/aae374}

\bibitem[{{Christensen} {et~al.}(2016){Christensen}, {Dav{\'e}}, {Governato},
  {Pontzen}, {Brooks}, {Munshi}, {Quinn}, \& {Wadsley}}]{christensen16}
{Christensen}, C.~R., {Dav{\'e}}, R., {Governato}, F., {et~al.} 2016, \apj,
  824, 57, \dodoi{10.3847/0004-637X/824/1/57}

\bibitem[{{Cole} {et~al.}(1999){Cole}, {Tolstoy}, {Gallagher}, {Hoessel},
  {Mould}, {Holtzman}, {Saha}, {Ballester}, {Burrows}, {Clarke}, {Crisp},
  {Griffiths}, {Grillmair}, {Hester}, {Krist}, {Meadows}, {Scowen},
  {Stapelfeldt}, {Trauger}, {Watson}, \& {Westphal}}]{cole99}
{Cole}, A.~A., {Tolstoy}, E., {Gallagher}, John~S., I., {et~al.} 1999, \aj,
  118, 1657, \dodoi{10.1086/301042}

\bibitem[{{Dalcanton} {et~al.}(2012){Dalcanton}, {Williams}, {Lang}, {Lauer},
  {Kalirai}, {Seth}, {Dolphin}, {Rosenfield}, {Weisz}, {Bell}, {Bianchi},
  {Boyer}, {Caldwell}, {Dong}, {Dorman}, {Gilbert}, {Girardi}, {Gogarten},
  {Gordon}, {Guhathakurta}, {Hodge}, {Holtzman}, {Johnson}, {Larsen}, {Lewis},
  {Melbourne}, {Olsen}, {Rix}, {Rosema}, {Saha}, {Sarajedini}, {Skillman}, \&
  {Stanek}}]{dalcanton12}
{Dalcanton}, J.~J., {Williams}, B.~F., {Lang}, D., {et~al.} 2012, \apjs, 200,
  18, \dodoi{10.1088/0067-0049/200/2/18}

\bibitem[{{Danforth} {et~al.}(2010){Danforth}, {Keeney}, {Stocke}, {Shull}, \&
  {Yao}}]{danforth10}
{Danforth}, C.~W., {Keeney}, B.~A., {Stocke}, J.~T., {Shull}, J.~M., \& {Yao},
  Y. 2010, \apj, 720, 976, \dodoi{10.1088/0004-637X/720/1/976}

\bibitem[{{D'Onghia} \& {Fox}(2016)}]{donghia16}
{D'Onghia}, E., \& {Fox}, A.~J. 2016, \araa, 54, 363,
  \dodoi{10.1146/annurev-astro-081915-023251}

\bibitem[{{El-Badry} {et~al.}(2018){El-Badry}, {Bradford}, {Quataert}, {Geha},
  {Boylan-Kolchin}, {Weisz}, {Wetzel}, {Hopkins}, {Chan}, {Fitts},
  {Kere{\v{s}}}, \& {Faucher-Gigu{\`e}re}}]{el-badry18}
{El-Badry}, K., {Bradford}, J., {Quataert}, E., {et~al.} 2018, \mnras, 477,
  1536, \dodoi{10.1093/mnras/sty730}

\bibitem[{{Emerick} {et~al.}(2018){Emerick}, {Bryan}, {Mac Low},
  {C{\^o}t{\'e}}, {Johnston}, \& {O'Shea}}]{emerick18b}
{Emerick}, A., {Bryan}, G.~L., {Mac Low}, M.-M., {et~al.} 2018, \apj, 869, 94,
  \dodoi{10.3847/1538-4357/aaec7d}

\bibitem[{{Ferland} {et~al.}(2017){Ferland}, {Chatzikos}, {Guzm{\'a}n},
  {Lykins}, {van Hoof}, {Williams}, {Abel}, {Badnell}, {Keenan}, {Porter}, \&
  {Stancil}}]{Ferland17}
{Ferland}, G.~J., {Chatzikos}, M., {Guzm{\'a}n}, F., {et~al.} 2017, \rmxaa, 53,
  385.
\newblock \doarXiv{1705.10877}

\bibitem[{{Fox} {et~al.}(2020){Fox}, {Frazer}, {Bland-Hawthorn}, {Wakker},
  {Barger}, \& {Richter}}]{fox20}
{Fox}, A.~J., {Frazer}, E.~M., {Bland-Hawthorn}, J., {et~al.} 2020, \apj, 897,
  23, \dodoi{10.3847/1538-4357/ab92a3}

\bibitem[{{Fox} {et~al.}(2014){Fox}, {Wakker}, {Barger}, {Hernandez},
  {Richter}, {Lehner}, {Bland-Hawthorn}, {Charlton}, {Westmeier}, {Thom},
  {Tumlinson}, {Misawa}, {Howk}, {Haffner}, {Ely}, {Rodriguez-Hidalgo}, \&
  {Kumari}}]{fox14}
{Fox}, A.~J., {Wakker}, B.~P., {Barger}, K.~A., {et~al.} 2014, \apj, 787, 147,
  \dodoi{10.1088/0004-637X/787/2/147}

\bibitem[{{Gallazzi} {et~al.}(2005){Gallazzi}, {Charlot}, {Brinchmann},
  {White}, \& {Tremonti}}]{gallazzi05}
{Gallazzi}, A., {Charlot}, S., {Brinchmann}, J., {White}, S. D.~M., \&
  {Tremonti}, C.~A. 2005, \mnras, 362, 41,
  \dodoi{10.1111/j.1365-2966.2005.09321.x}

\bibitem[{{Garrison-Kimmel} {et~al.}(2014){Garrison-Kimmel}, {Boylan-Kolchin},
  {Bullock}, \& {Lee}}]{gk14}
{Garrison-Kimmel}, S., {Boylan-Kolchin}, M., {Bullock}, J.~S., \& {Lee}, K.
  2014, \mnras, 438, 2578, \dodoi{10.1093/mnras/stt2377}

\bibitem[{{Garrison-Kimmel} {et~al.}(2017){Garrison-Kimmel}, {Bullock},
  {Boylan-Kolchin}, \& {Bardwell}}]{gk17}
{Garrison-Kimmel}, S., {Bullock}, J.~S., {Boylan-Kolchin}, M., \& {Bardwell},
  E. 2017, \mnras, 464, 3108, \dodoi{10.1093/mnras/stw2564}

\bibitem[{{Gehrels}(1986)}]{gehrels86}
{Gehrels}, N. 1986, \apj, 303, 336, \dodoi{10.1086/164079}

\bibitem[{{Gil de Paz} {et~al.}(2007){Gil de Paz}, {Boissier}, {Madore},
  {Seibert}, {Joe}, {Boselli}, {Wyder}, {Thilker}, {Bianchi}, {Rey}, {Rich},
  {Barlow}, {Conrow}, {Forster}, {Friedman}, {Martin}, {Morrissey}, {Neff},
  {Schiminovich}, {Small}, {Donas}, {Heckman}, {Lee}, {Milliard}, {Szalay}, \&
  {Yi}}]{galex}
{Gil de Paz}, A., {Boissier}, S., {Madore}, B.~F., {et~al.} 2007, \apjs, 173,
  185, \dodoi{10.1086/516636}

\bibitem[{{Giovanelli} {et~al.}(2013){Giovanelli}, {Haynes}, {Adams}, {Cannon},
  {Rhode}, {Salzer}, {Skillman}, {Bernstein-Cooper}, \&
  {McQuinn}}]{giovanelli13}
{Giovanelli}, R., {Haynes}, M.~P., {Adams}, E. A.~K., {et~al.} 2013, \aj, 146,
  15, \dodoi{10.1088/0004-6256/146/1/15}

\bibitem[{{Grcevich} \& {Putman}(2009)}]{grcevich09}
{Grcevich}, J., \& {Putman}, M.~E. 2009, \apj, 696, 385,
  \dodoi{10.1088/0004-637X/696/1/385}

\bibitem[{{Haardt} \& {Madau}(2001)}]{hm01}
{Haardt}, F., \& {Madau}, P. 2001, in Clusters of Galaxies and the High
  Redshift Universe Observed in X-rays, ed. D.~M. {Neumann} \& J.~T.~V. {Tran},
  64.
\newblock \doarXiv{astro-ph/0106018}

\bibitem[{{Hafen} {et~al.}(2019){Hafen}, {Faucher-Gigu{\`e}re},
  {Angl{\'e}s-Alc{\'a}zar}, {Stern}, {Kere{\v{s}}}, {Hummels}, {Esmerian},
  {Garrison-Kimmel}, {El-Badry}, {Wetzel}, {Chan}, {Hopkins}, \&
  {Murray}}]{Hafen19}
{Hafen}, Z., {Faucher-Gigu{\`e}re}, C.-A., {Angl{\'e}s-Alc{\'a}zar}, D.,
  {et~al.} 2019, \mnras, 488, 1248, \dodoi{10.1093/mnras/stz1773}

\bibitem[{{Haffner} {et~al.}(2003){Haffner}, {Reynolds}, {Tufte}, {Madsen},
  {Jaehnig}, \& {Percival}}]{haffner03}
{Haffner}, L.~M., {Reynolds}, R.~J., {Tufte}, S.~L., {et~al.} 2003, \apjs, 149,
  405, \dodoi{10.1086/378850}

\bibitem[{Harris {et~al.}(2020)Harris, Millman, van~der Walt, Gommers,
  Virtanen, Cournapeau, Wieser, Taylor, Berg, Smith, Kern, Picus, Hoyer, van
  Kerkwijk, Brett, Haldane, del R{'{\i}}o, Wiebe, Peterson,
  G{'{e}}rard-Marchant, Sheppard, Reddy, Weckesser, Abbasi, Gohlke, \&
  Oliphant}]{numpy}
Harris, C.~R., Millman, K.~J., van~der Walt, S.~J., {et~al.} 2020, Nature, 585,
  357, \dodoi{10.1038/s41586-020-2649-2}

\bibitem[{{HI4PI Collaboration} {et~al.}(2016){HI4PI Collaboration}, {Ben
  Bekhti}, {Fl{\"o}er}, {Keller}, {Kerp}, {Lenz}, {Winkel}, {Bailin},
  {Calabretta}, {Dedes}, {Ford}, {Gibson}, {Haud}, {Janowiecki}, {Kalberla},
  {Lockman}, {McClure-Griffiths}, {Murphy}, {Nakanishi}, {Pisano}, \&
  {Staveley-Smith}}]{hi4pi16}
{HI4PI Collaboration}, {Ben Bekhti}, N., {Fl{\"o}er}, L., {et~al.} 2016, \aap,
  594, A116, \dodoi{10.1051/0004-6361/201629178}

\bibitem[{{Howk} {et~al.}(2017){Howk}, {Wotta}, {Berg}, {Lehner}, {Lockman},
  {Hafen}, {Pisano}, {Faucher-Gigu{\`e}re}, {Wakker}, {Prochaska}, {Wolfe},
  {Ribaudo}, {Barger}, {Corlies}, {Fox}, {Guhathakurta}, {Jenkins}, {Kalirai},
  {O'Meara}, {Peeples}, {Stewart}, \& {Strader}}]{howk17}
{Howk}, J.~C., {Wotta}, C.~B., {Berg}, M.~A., {et~al.} 2017, \apj, 846, 141,
  \dodoi{10.3847/1538-4357/aa87b4}

\bibitem[{{Hunter} \& {Elmegreen}(2004)}]{hunter04}
{Hunter}, D.~A., \& {Elmegreen}, B.~G. 2004, \aj, 128, 2170,
  \dodoi{10.1086/424615}

\bibitem[{{Hunter} {et~al.}(2012){Hunter}, {Ficut-Vicas}, {Ashley}, {Brinks},
  {Cigan}, {Elmegreen}, {Heesen}, {Herrmann}, {Johnson}, {Oh}, {Rupen},
  {Schruba}, {Simpson}, {Walter}, {Westpfahl}, {Young}, \& {Zhang}}]{hunter12}
{Hunter}, D.~A., {Ficut-Vicas}, D., {Ashley}, T., {et~al.} 2012, \aj, 144, 134,
  \dodoi{10.1088/0004-6256/144/5/134}

\bibitem[{Hunter(2007)}]{matplotlib}
Hunter, J.~D. 2007, Computing in Science \& Engineering, 9, 90,
  \dodoi{10.1109/MCSE.2007.55}

\bibitem[{{Johnson} {et~al.}(2017){Johnson}, {Chen}, {Mulchaey}, {Schaye}, \&
  {Straka}}]{johnson17}
{Johnson}, S.~D., {Chen}, H.-W., {Mulchaey}, J.~S., {Schaye}, J., \& {Straka},
  L.~A. 2017, \apjl, 850, L10, \dodoi{10.3847/2041-8213/aa9370}

\bibitem[{{Keeney} {et~al.}(2012){Keeney}, {Danforth}, {Stocke}, {France}, \&
  {Green}}]{keeney12}
{Keeney}, B.~A., {Danforth}, C.~W., {Stocke}, J.~T., {France}, K., \& {Green},
  J.~C. 2012, Publications of the Astronomical Society of the Pacific, 124,
  830, \dodoi{10.1086/667392}

\bibitem[{{Kirby} {et~al.}(2013){Kirby}, {Cohen}, {Guhathakurta}, {Cheng},
  {Bullock}, \& {Gallazzi}}]{kirby13}
{Kirby}, E.~N., {Cohen}, J.~G., {Guhathakurta}, P., {et~al.} 2013, \apj, 779,
  102, \dodoi{10.1088/0004-637X/779/2/102}

\bibitem[{{Kirby} {et~al.}(2011){Kirby}, {Martin}, \& {Finlator}}]{kirby11}
{Kirby}, E.~N., {Martin}, C.~L., \& {Finlator}, K. 2011, \apjl, 742, L25,
  \dodoi{10.1088/2041-8205/742/2/L25}

\bibitem[{{Kroupa}(2001)}]{kroupa01}
{Kroupa}, P. 2001, \mnras, 322, 231, \dodoi{10.1046/j.1365-8711.2001.04022.x}

\bibitem[{{Lake} \& {Skillman}(1989)}]{lake89}
{Lake}, G., \& {Skillman}, E.~D. 1989, \aj, 98, 1274, \dodoi{10.1086/115215}

\bibitem[{{Lee} {et~al.}(2006){Lee}, {Skillman}, {Cannon}, {Jackson}, {Gehrz},
  {Polomski}, \& {Woodward}}]{lee06}
{Lee}, H., {Skillman}, E.~D., {Cannon}, J.~M., {et~al.} 2006, \apj, 647, 970,
  \dodoi{10.1086/505573}

\bibitem[{{Lehner} {et~al.}(2020){Lehner}, {Berek}, {Howk}, {Wakker},
  {Tumlinson}, {Jenkins}, {Prochaska}, {Augustin}, {Ji}, {Faucher-Gigu{\`e}re},
  {Hafen}, {Peeples}, {Barger}, {Berg}, {Bordoloi}, {Brown}, {Fox}, {Gilbert},
  {Guhathakurta}, {Kalirai}, {Lockman}, {O'Meara}, {Pisano}, {Ribaudo}, \&
  {Werk}}]{lehner20}
{Lehner}, N., {Berek}, S.~C., {Howk}, J.~C., {et~al.} 2020, \apj, 900, 9,
  \dodoi{10.3847/1538-4357/aba49c}

\bibitem[{{Liang} \& {Chen}(2014)}]{liang14}
{Liang}, C.~J., \& {Chen}, H.-W. 2014, \mnras, 445, 2061,
  \dodoi{10.1093/mnras/stu1901}

\bibitem[{{Lucchini} {et~al.}(2020){Lucchini}, {D'Onghia}, {Fox}, {Bustard},
  {Bland-Hawthorn}, \& {Zweibel}}]{lucchini20}
{Lucchini}, S., {D'Onghia}, E., {Fox}, A.~J., {et~al.} 2020, arXiv e-prints,
  arXiv:2009.04368.
\newblock \doarXiv{2009.04368}

\bibitem[{{Ma} {et~al.}(2016){Ma}, {Hopkins}, {Faucher-Gigu{\`e}re}, {Zolman},
  {Muratov}, {Kere{\v{s}}}, \& {Quataert}}]{ma16}
{Ma}, X., {Hopkins}, P.~F., {Faucher-Gigu{\`e}re}, C.-A., {et~al.} 2016,
  \mnras, 456, 2140, \dodoi{10.1093/mnras/stv2659}

\bibitem[{{Mac Low} \& {Ferrara}(1999)}]{maclow99}
{Mac Low}, M.-M., \& {Ferrara}, A. 1999, \apj, 513, 142, \dodoi{10.1086/306832}

\bibitem[{{Madau} \& {Dickinson}(2014)}]{madau14}
{Madau}, P., \& {Dickinson}, M. 2014, \araa, 52, 415,
  \dodoi{10.1146/annurev-astro-081811-125615}

\bibitem[{{Markwardt}(2009)}]{markwardt09}
{Markwardt}, C.~B. 2009, in Astronomical Society of the Pacific Conference
  Series, Vol. 411, Astronomical Data Analysis Software and Systems XVIII, ed.
  D.~A. {Bohlender}, D.~{Durand}, \& P.~{Dowler}, 251.
\newblock \doarXiv{0902.2850}

\bibitem[{{Mathewson} {et~al.}(1974){Mathewson}, {Cleary}, \&
  {Murray}}]{mathewson74}
{Mathewson}, D.~S., {Cleary}, M.~N., \& {Murray}, J.~D. 1974, \apj, 190, 291,
  \dodoi{10.1086/152875}

\bibitem[{{McConnachie}(2012)}]{mcconnachie12}
{McConnachie}, A.~W. 2012, \aj, 144, 4, \dodoi{10.1088/0004-6256/144/1/4}

\bibitem[{{McQuinn} {et~al.}(2019){McQuinn}, {van Zee}, \&
  {Skillman}}]{mcquinn19}
{McQuinn}, K. B.~W., {van Zee}, L., \& {Skillman}, E.~D. 2019, \apj, 886, 74,
  \dodoi{10.3847/1538-4357/ab4c37}

\bibitem[{{McQuinn} {et~al.}(2015{\natexlab{a}}){McQuinn}, {Skillman},
  {Dolphin}, {Cannon}, {Salzer}, {Rhode}, {Adams}, {Berg}, {Giovanelli}, \&
  {Haynes}}]{mcquinn15b}
{McQuinn}, K. B.~W., {Skillman}, E.~D., {Dolphin}, A., {et~al.}
  2015{\natexlab{a}}, \apjl, 815, L17, \dodoi{10.1088/2041-8205/815/2/L17}

\bibitem[{{McQuinn} {et~al.}(2015{\natexlab{b}}){McQuinn}, {Skillman},
  {Dolphin}, {Cannon}, {Salzer}, {Rhode}, {Adams}, {Berg}, {Giovanelli},
  {Girardi}, \& {Haynes}}]{mcquinn15a}
---. 2015{\natexlab{b}}, \apj, 812, 158, \dodoi{10.1088/0004-637X/812/2/158}

\bibitem[{{Moster} {et~al.}(2010){Moster}, {Somerville}, {Maulbetsch}, {van den
  Bosch}, {Macci{\`o}}, {Naab}, \& {Oser}}]{moster10}
{Moster}, B.~P., {Somerville}, R.~S., {Maulbetsch}, C., {et~al.} 2010, \apj,
  710, 903, \dodoi{10.1088/0004-637X/710/2/903}

\bibitem[{{Muratov} {et~al.}(2017){Muratov}, {Kere{\v{s}}},
  {Faucher-Gigu{\`e}re}, {Hopkins}, {Ma}, {Angl{\'e}s-Alc{\'a}zar}, {Chan},
  {Torrey}, {Hafen}, {Quataert}, \& {Murray}}]{muratov17}
{Muratov}, A.~L., {Kere{\v{s}}}, D., {Faucher-Gigu{\`e}re}, C.-A., {et~al.}
  2017, \mnras, 468, 4170, \dodoi{10.1093/mnras/stx667}

\bibitem[{{Nidever} {et~al.}(2008){Nidever}, {Majewski}, \& {Butler
  Burton}}]{nidever08}
{Nidever}, D.~L., {Majewski}, S.~R., \& {Butler Burton}, W. 2008, \apj, 679,
  432, \dodoi{10.1086/587042}

\bibitem[{{Nidever} {et~al.}(2010){Nidever}, {Majewski}, {Butler Burton}, \&
  {Nigra}}]{nidever10}
{Nidever}, D.~L., {Majewski}, S.~R., {Butler Burton}, W., \& {Nigra}, L. 2010,
  \apj, 723, 1618, \dodoi{10.1088/0004-637X/723/2/1618}

\bibitem[{{Patel} {et~al.}(2020){Patel}, {Kallivayalil}, {Garavito-Camargo},
  {Besla}, {Weisz}, {van der Marel}, {Boylan-Kolchin}, {Pawlowski}, \&
  {G{\'o}mez}}]{patel20}
{Patel}, E., {Kallivayalil}, N., {Garavito-Camargo}, N., {et~al.} 2020, \apj,
  893, 121, \dodoi{10.3847/1538-4357/ab7b75}

\bibitem[{{Peek} {et~al.}(2011){Peek}, {Heiles}, {Douglas}, {Lee}, {Grcevich},
  {Stanimirovi{\'c}}, {Putman}, {Korpela}, {Gibson}, {Begum}, {Saul},
  {Robishaw}, \& {Kr{\v{c}}o}}]{peek11}
{Peek}, J.~E.~G., {Heiles}, C., {Douglas}, K.~A., {et~al.} 2011, \apjs, 194,
  20, \dodoi{10.1088/0067-0049/194/2/20}

\bibitem[{{Peek} {et~al.}(2018){Peek}, {Babler}, {Zheng}, {Clark}, {Douglas},
  {Korpela}, {Putman}, {Stanimirovi{\'c}}, {Gibson}, \& {Heiles}}]{peek18}
{Peek}, J.~E.~G., {Babler}, B.~L., {Zheng}, Y., {et~al.} 2018, \apjs, 234, 2,
  \dodoi{10.3847/1538-4365/aa91d3}

\bibitem[{{Peeples} {et~al.}(2014){Peeples}, {Werk}, {Tumlinson},
  {Oppenheimer}, {Prochaska}, {Katz}, \& {Weinberg}}]{peeples14}
{Peeples}, M.~S., {Werk}, J.~K., {Tumlinson}, J., {et~al.} 2014, \apj, 786, 54,
  \dodoi{10.1088/0004-637X/786/1/54}

\bibitem[{Price-Whelan {et~al.}(2017)Price-Whelan, Sipocz, Major, \&
  Oh}]{gala-code}
Price-Whelan, A., Sipocz, B., Major, S., \& Oh, S. 2017, adrn/gala: v0.2.1,
  \dodoi{10.5281/zenodo.833339}

\bibitem[{{Putman} {et~al.}(2012){Putman}, {Peek}, \& {Joung}}]{putman12}
{Putman}, M.~E., {Peek}, J.~E.~G., \& {Joung}, M.~R. 2012, \araa, 50, 491,
  \dodoi{10.1146/annurev-astro-081811-125612}

\bibitem[{{Putman} {et~al.}(2003){Putman}, {Staveley-Smith}, {Freeman},
  {Gibson}, \& {Barnes}}]{putman03}
{Putman}, M.~E., {Staveley-Smith}, L., {Freeman}, K.~C., {Gibson}, B.~K., \&
  {Barnes}, D.~G. 2003, \apj, 586, 170, \dodoi{10.1086/344477}

\bibitem[{{Putman} {et~al.}(1998){Putman}, {Gibson}, {Staveley-Smith}, {Banks},
  {Barnes}, {Bhatal}, {Disney}, {Ekers}, {Freeman}, {Haynes}, {Henning},
  {Jerjen}, {Kilborn}, {Koribalski}, {Knezek}, {Malin}, {Mould}, {Oosterloo},
  {Price}, {Ryder}, {Sadler}, {Stewart}, {Stootman}, {Vaile}, {Webster}, \&
  {Wright}}]{putman98}
{Putman}, M.~E., {Gibson}, B.~K., {Staveley-Smith}, L., {et~al.} 1998, \nat,
  394, 752, \dodoi{10.1038/29466}

\bibitem[{{Rey} {et~al.}(2020){Rey}, {Pontzen}, {Agertz}, {Orkney}, {Read}, \&
  {Rosdahl}}]{rey20}
{Rey}, M.~P., {Pontzen}, A., {Agertz}, O., {et~al.} 2020, \mnras, 497, 1508,
  \dodoi{10.1093/mnras/staa1640}

\bibitem[{{Richter} {et~al.}(2017){Richter}, {Nuza}, {Fox}, {Wakker}, {Lehner},
  {Ben Bekhti}, {Fechner}, {Wendt}, {Howk}, {Muzahid}, {Ganguly}, \&
  {Charlton}}]{richter17}
{Richter}, P., {Nuza}, S.~E., {Fox}, A.~J., {et~al.} 2017, \aap, 607, A48,
  \dodoi{10.1051/0004-6361/201630081}

\bibitem[{{Ritter} {et~al.}(2018{\natexlab{a}}){Ritter}, {C{\^o}t{\'e}},
  {Herwig}, {Navarro}, \& {Fryer}}]{ritter18a}
{Ritter}, C., {C{\^o}t{\'e}}, B., {Herwig}, F., {Navarro}, J.~F., \& {Fryer},
  C.~L. 2018{\natexlab{a}}, \apjs, 237, 42, \dodoi{10.3847/1538-4365/aad691}

\bibitem[{{Ritter} {et~al.}(2018{\natexlab{b}}){Ritter}, {Herwig}, {Jones},
  {Pignatari}, {Fryer}, \& {Hirschi}}]{ritter18b}
{Ritter}, C., {Herwig}, F., {Jones}, S., {et~al.} 2018{\natexlab{b}}, \mnras,
  480, 538, \dodoi{10.1093/mnras/sty1729}

\bibitem[{{Romano} {et~al.}(2019){Romano}, {Calura}, {D'Ercole}, \&
  {Few}}]{romano19}
{Romano}, D., {Calura}, F., {D'Ercole}, A., \& {Few}, C.~G. 2019, \aap, 630,
  A140, \dodoi{10.1051/0004-6361/201935328}

\bibitem[{{Rubin} {et~al.}(2012){Rubin}, {Prochaska}, {Koo}, \&
  {Phillips}}]{rubin12}
{Rubin}, K. H.~R., {Prochaska}, J.~X., {Koo}, D.~C., \& {Phillips}, A.~C. 2012,
  \apjl, 747, L26, \dodoi{10.1088/2041-8205/747/2/L26}

\bibitem[{{Rubin} {et~al.}(2014){Rubin}, {Prochaska}, {Koo}, {Phillips},
  {Martin}, \& {Winstrom}}]{rubin14}
{Rubin}, K. H.~R., {Prochaska}, J.~X., {Koo}, D.~C., {et~al.} 2014, \apj, 794,
  156, \dodoi{10.1088/0004-637X/794/2/156}

\bibitem[{{Salpeter}(1955)}]{salpeter55}
{Salpeter}, E.~E. 1955, \apj, 121, 161, \dodoi{10.1086/145971}

\bibitem[{{Savage} \& {Sembach}(1991)}]{savage91}
{Savage}, B.~D., \& {Sembach}, K.~R. 1991, \apj, 379, 245,
  \dodoi{10.1086/170498}

\bibitem[{{Savage} \& {Sembach}(1996)}]{savage96}
---. 1996, Annual Review of Astronomy and Astrophysics, 34, 279,
  \dodoi{10.1146/annurev.astro.34.1.279}

\bibitem[{{Silich} {et~al.}(2006){Silich}, {Lozinskaya}, {Moiseev},
  {Podorvanuk}, {Rosado}, {Borissova}, \& {Valdez-Gutierrez}}]{silich06}
{Silich}, S., {Lozinskaya}, T., {Moiseev}, A., {et~al.} 2006, \aap, 448, 123,
  \dodoi{10.1051/0004-6361:20053326}

\bibitem[{{Skillman} {et~al.}(2003){Skillman}, {Tolstoy}, {Cole}, {Dolphin},
  {Saha}, {Gallagher}, {Dohm-Palmer}, \& {Mateo}}]{skillman03}
{Skillman}, E.~D., {Tolstoy}, E., {Cole}, A.~A., {et~al.} 2003, \apj, 596, 253,
  \dodoi{10.1086/377635}

\bibitem[{{Skillman} {et~al.}(2014){Skillman}, {Hidalgo}, {Weisz}, {Monelli},
  {Gallart}, {Aparicio}, {Bernard}, {Boylan-Kolchin}, {Cassisi}, {Cole},
  {Dolphin}, {Ferguson}, {Mayer}, {Navarro}, {Stetson}, \&
  {Tolstoy}}]{skillman14}
{Skillman}, E.~D., {Hidalgo}, S.~L., {Weisz}, D.~R., {et~al.} 2014, \apj, 786,
  44, \dodoi{10.1088/0004-637X/786/1/44}

\bibitem[{{Stocke} {et~al.}(2013){Stocke}, {Keeney}, {Danforth}, {Shull},
  {Froning}, {Green}, {Penton}, \& {Savage}}]{stocke13}
{Stocke}, J.~T., {Keeney}, B.~A., {Danforth}, C.~W., {et~al.} 2013, \apj, 763,
  148, \dodoi{10.1088/0004-637X/763/2/148}

\bibitem[{{Telford} {et~al.}(2019){Telford}, {Werk}, {Dalcanton}, \&
  {Williams}}]{Telford19}
{Telford}, O.~G., {Werk}, J.~K., {Dalcanton}, J.~J., \& {Williams}, B.~F. 2019,
  \apj, 877, 120, \dodoi{10.3847/1538-4357/ab1b3f}

\bibitem[{{The Astropy Collaboration} {et~al.}(2018){The Astropy
  Collaboration}, {Price-Whelan}, {Sip{\H o}cz}, {G{\"u}nther}, {Lim},
  {Crawford}, {Conseil}, {Shupe}, {Craig}, {Dencheva}, {Ginsburg},
  {VanderPlas}, {Bradley}, {P{\'e}rez-Su{\'a}rez}, {de Val-Borro}, {Aldcroft},
  {Cruz}, {Robitaille}, {Tollerud}, {Ardelean}, {Babej}, {Bachetti}, {Bakanov},
  {Bamford}, {Barentsen}, {Barmby}, {Baumbach}, {Berry}, {Biscani}, {Boquien},
  {Bostroem}, {Bouma}, {Brammer}, {Bray}, {Breytenbach}, {Buddelmeijer},
  {Burke}, {Calderone}, {Cano Rodr{\'{\i}}guez}, {Cara}, {Cardoso},
  {Cheedella}, {Copin}, {Crichton}, {D{\'A}vella}, {Deil}, {Depagne},
  {Dietrich}, {Donath}, {Droettboom}, {Earl}, {Erben}, {Fabbro}, {Ferreira},
  {Finethy}, {Fox}, {Garrison}, {Gibbons}, {Goldstein}, {Gommers}, {Greco},
  {Greenfield}, {Groener}, {Grollier}, {Hagen}, {Hirst}, {Homeier}, {Horton},
  {Hosseinzadeh}, {Hu}, {Hunkeler}, {Ivezi{\'c}}, {Jain}, {Jenness}, {Kanarek},
  {Kendrew}, {Kern}, {Kerzendorf}, {Khvalko}, {King}, {Kirkby}, {Kulkarni},
  {Kumar}, {Lee}, {Lenz}, {Littlefair}, {Ma}, {Macleod}, {Mastropietro},
  {McCully}, {Montagnac}, {Morris}, {Mueller}, {Mumford}, {Muna}, {Murphy},
  {Nelson}, {Nguyen}, {Ninan}, {N{\"o}the}, {Ogaz}, {Oh}, {Parejko}, {Parley},
  {Pascual}, {Patil}, {Patil}, {Plunkett}, {Prochaska}, {Rastogi}, {Reddy
  Janga}, {Sabater}, {Sakurikar}, {Seifert}, {Sherbert}, {Sherwood-Taylor},
  {Shih}, {Sick}, {Silbiger}, {Singanamalla}, {Singer}, {Sladen}, {Sooley},
  {Sornarajah}, {Streicher}, {Teuben}, {Thomas}, {Tremblay}, {Turner},
  {Terr{\'o}n}, {van Kerkwijk}, {de la Vega}, {Watkins}, {Weaver}, {Whitmore},
  {Woillez}, \& {Zabalza}}]{astropy2}
{The Astropy Collaboration}, {Price-Whelan}, A.~M., {Sip{\H o}cz}, B.~M.,
  {et~al.} 2018, ArXiv e-prints.
\newblock \doarXiv{1801.02634}

\bibitem[{{Tremonti} {et~al.}(2004){Tremonti}, {Heckman}, {Kauffmann},
  {Brinchmann}, {Charlot}, {White}, {Seibert}, {Peng}, {Schlegel}, {Uomoto},
  {Fukugita}, \& {Brinkmann}}]{tremonti04}
{Tremonti}, C.~A., {Heckman}, T.~M., {Kauffmann}, G., {et~al.} 2004, \apj, 613,
  898, \dodoi{10.1086/423264}

\bibitem[{{Tumlinson} {et~al.}(2013){Tumlinson}, {Thom}, {Werk}, {Prochaska},
  {Tripp}, {Katz}, {Dav{\'e}}, {Oppenheimer}, {Meiring}, {Ford}, {O'Meara},
  {Peeples}, {Sembach}, \& {Weinberg}}]{tumlinson13}
{Tumlinson}, J., {Thom}, C., {Werk}, J.~K., {et~al.} 2013, \apj, 777, 59,
  \dodoi{10.1088/0004-637X/777/1/59}

\bibitem[{{Vogelsberger} {et~al.}(2014){Vogelsberger}, {Zavala}, {Simpson}, \&
  {Jenkins}}]{vogelsberger14}
{Vogelsberger}, M., {Zavala}, J., {Simpson}, C., \& {Jenkins}, A. 2014, \mnras,
  444, 3684, \dodoi{10.1093/mnras/stu1713}

\bibitem[{{Wakker} {et~al.}(2015){Wakker}, {Hernandez}, {French}, {Kim},
  {Oppenheimer}, \& {Savage}}]{wakker15}
{Wakker}, B.~P., {Hernandez}, A.~K., {French}, D.~M., {et~al.} 2015, \apj, 814,
  40, \dodoi{10.1088/0004-637X/814/1/40}

\bibitem[{{Weisz} {et~al.}(2014){Weisz}, {Dolphin}, {Skillman}, {Holtzman},
  {Gilbert}, {Dalcanton}, \& {Williams}}]{weisz14}
{Weisz}, D.~R., {Dolphin}, A.~E., {Skillman}, E.~D., {et~al.} 2014, \apj, 789,
  147, \dodoi{10.1088/0004-637X/789/2/147}

\bibitem[{{Werk} {et~al.}(2013){Werk}, {Prochaska}, {Thom}, {Tumlinson},
  {Tripp}, {O'Meara}, \& {Peeples}}]{werk13}
{Werk}, J.~K., {Prochaska}, J.~X., {Thom}, C., {et~al.} 2013, \apjs, 204, 17,
  \dodoi{10.1088/0067-0049/204/2/17}

\bibitem[{{Werk} {et~al.}(2014){Werk}, {Prochaska}, {Tumlinson}, {Peeples},
  {Tripp}, {Fox}, {Lehner}, {Thom}, {O'Meara}, {Ford}, {Bordoloi}, {Katz},
  {Tejos}, {Oppenheimer}, {Dav{\'e}}, \& {Weinberg}}]{werk14}
{Werk}, J.~K., {Prochaska}, J.~X., {Tumlinson}, J., {et~al.} 2014, \apj, 792,
  8, \dodoi{10.1088/0004-637X/792/1/8}

\bibitem[{{Zheng} {et~al.}(2019{\natexlab{a}}){Zheng}, {Peek}, {Putman}, \&
  {Werk}}]{zheng19a}
{Zheng}, Y., {Peek}, J.~E.~G., {Putman}, M.~E., \& {Werk}, J.~K.
  2019{\natexlab{a}}, \apj, 871, 35, \dodoi{10.3847/1538-4357/aaf6eb}

\bibitem[{{Zheng} {et~al.}(2017){Zheng}, {Peek}, {Werk}, \& {Putman}}]{zheng17}
{Zheng}, Y., {Peek}, J.~E.~G., {Werk}, J.~K., \& {Putman}, M.~E. 2017, \apj,
  834, 179, \dodoi{10.3847/1538-4357/834/2/179}

\bibitem[{{Zheng} {et~al.}(2019{\natexlab{b}}){Zheng}, {Putman}, {Emerick},
  {McQuinn}, {Werk}, {Lockman}, {Oppenheimer}, {Fox}, {Kirby}, \&
  {Burchett}}]{Zheng19b}
{Zheng}, Y., {Putman}, M.~E., {Emerick}, A., {et~al.} 2019{\natexlab{b}},
  \mnras, 490, 467, \dodoi{10.1093/mnras/stz2563}

\end{thebibliography}

\appendix
\restartappendixnumbering
\section{Spectral Co-addition}
\label{sec:app_coadd}

\begin{figure*}[t!]
\centering
\includegraphics[width=0.8\textwidth]{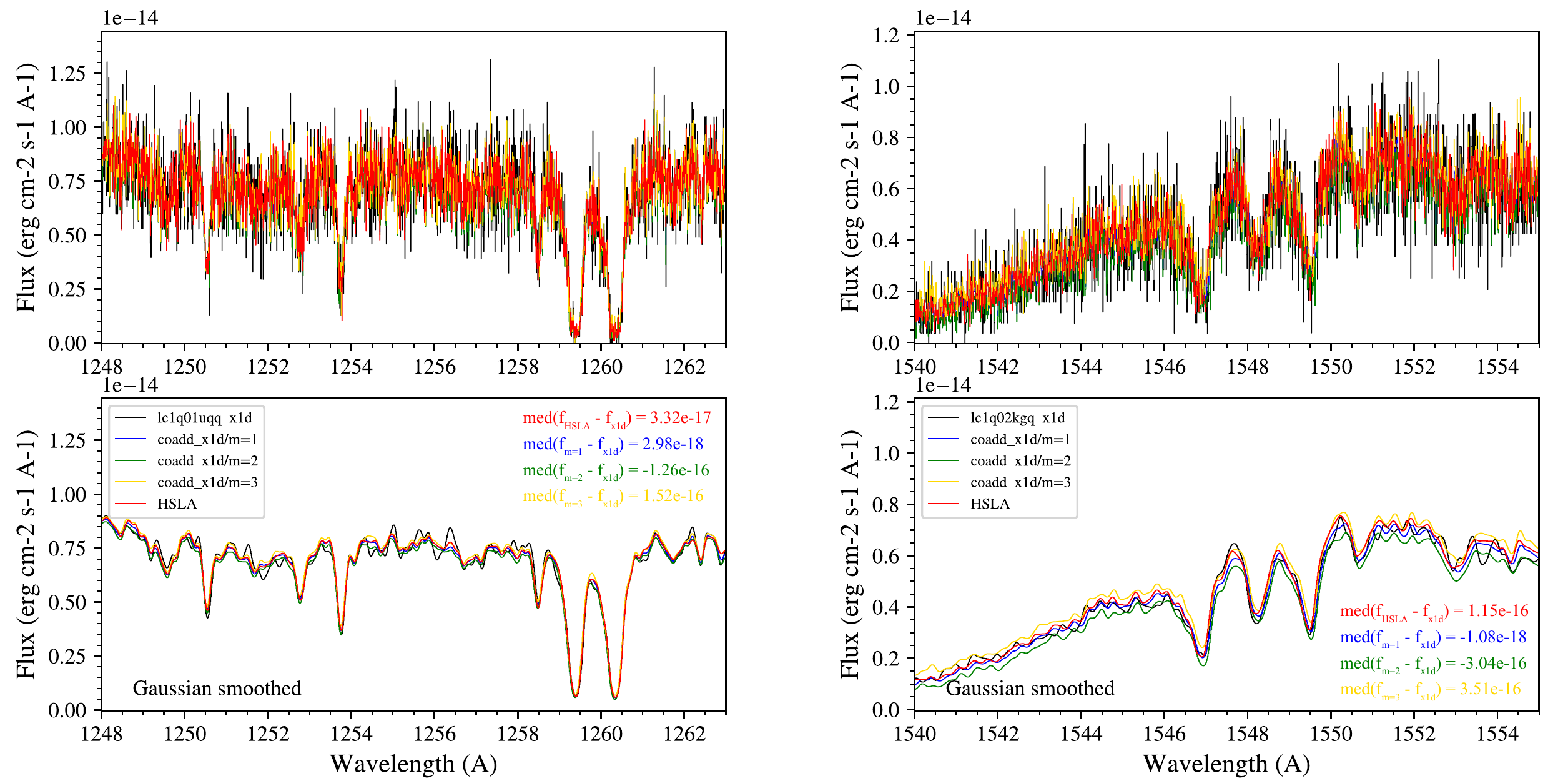}
\caption{The flux level comparison among the \hsla\ (red), \textsl{coadd\_x1d} (green, blue, gold), and the original individual exposure files (black) for S3, which is typical among all the targets we analyzed. The top panels are for data without binning, and the bottom panels for data that are Gaussian smoothed to 6 pixels (i.e., per resolution element). The left and right panels show two different spectral regions. The indicated values show median flux offsets between a given method and one of the original x1d.fits exposure files. We generate similar figures for every exposure to check line profile and flux consistency. Overall, we find that the \hsla\ and \textsl{coadd\_x1d} within method 1 show much better flux consistency with the original exposures, with a typical median flux offset $\lesssim10^{-17}$ \fluxunit. Meanwhile, the \textsl{coadd\_x1d} with method 2 (3) often gives too low (high) flux values, with absolute flux offset often higher than $10^{-16}$--$10^{-15}$ \fluxunit. } 
\label{fig:appA_flux_offset}
\end{figure*}

Spectra observed with HST/COS are processed by the standard {\it CalCOS} pipeline up to visit level; however, those taken with different grating setups remain separate. \cite{wakker15} point out that the {\it CalCOS} pipeline often overestimates the errors of co-added spectra for faint targets with fluxes $\lesssim10^{-14}$ erg cm$^{-2}$ s$^{-1}$ \AA$^{-1}$. A number of authors have written their own co-adding codes \citep[e.g.][]{danforth10, keeney12, tumlinson13, wakker15}. To produce science-ready co-added spectra in our work, here we focus on two publicly available resources, \href{https://archive.stsci.edu/missions-and-data/hst-spectroscopic-legacy-archive-hsla}{the HSLA} and \textsl{coadd\_x1d} \citep{danforth10, keeney12}. We describe how they work and compare the spectral co-addition products.

The second data release of the \hsla\ publishes co-added spectra for targets observed with HST/COS that went public as of April 2017. In their algorithm, multi-exposure spectra were co-added using photon counts from each file \citep{gehrels86}, and then the total counts were converted to flux density based on the flux calibration ratio from the keyword FLUXFACTOR recorded in the original fits file header. Flux errors were handled using Poisson statistics. Because of the large data volume and the diverse target types of the HST/COS database, \href{https://archive.stsci.edu/missions-and-data/hst-spectroscopic-legacy-archive-hsla}{HSLA} did not perform wavelength calibration and instead adopted the original wavelengths provided by the {\it CalCOS} pipeline for each file. This may result in artificial line profiles if spectra from different exposures had systemic velocity shifts.

Meanwhile, the \textsl{coadd\_x1d} code \citep{danforth10, keeney12} chooses to co-add multi-exposure spectra based on fluxes instead of photon counts. 
Users running the code can decide among three different weighting options to co-add spectra: (1) exposure time, (2) inverse variance, or (3) the square of signal-to-noise ratio (S/N) per exposure. As pointed out by \cite{wakker15}, the inverse-variance weighting in option (2) may give rise to potential line-shape distortion if different data files are observed with different exposure times. Similar to the \href{https://archive.stsci.edu/missions-and-data/hst-spectroscopic-legacy-archive-hsla}{HSLA}, the \textsl{coadd\_x1d} also handles error arrays based on Poisson statistics. For wavelength calibration, the \textsl{coadd\_x1d} derives constant velocity shifts using a number of strong interstellar lines over 10 \AA\ windows among all input exposures. Then it manually applies the velocity shifts to all exposures to align their wavelengths with a randomly selected reference exposure. As noted by \cite{zheng17}, such an alignment procedure may introduce a velocity offset of $\sim10~\kms$, which is smaller than the COS wavelength accuracy of 15--20 $\kms$ (see the COS Instrument Handbook).

\begin{figure}
\centering
\includegraphics[width=\columnwidth]{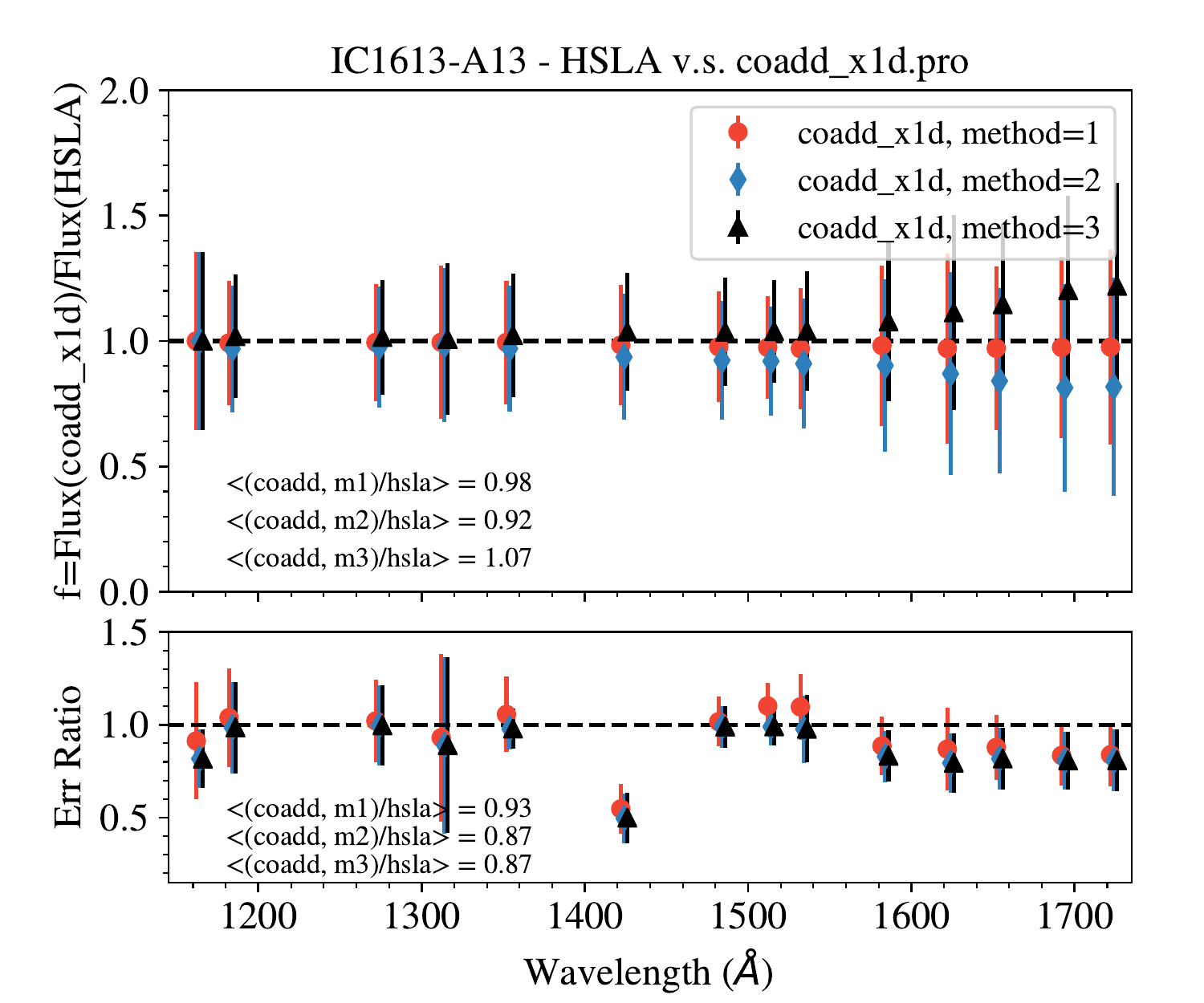} 
\caption{Flux ratios of absorption-free spectra from \textsl{coadd\_x1d} method 1 (red circle), 2 (blue diamond), and 3 (black triangle) to \hsla's at different wavelengths. At each wavelength, the flux ratio is sampled over a 10\AA\ spectral window. Ratio of 1.0 indicates consistent co-added spectral fluxes between a given \textsl{coadd\_x1d} method and \hsla. Overall, \textsl{coadd\_x1d} with method 1 yields much more consistent co-added spectral fluxes with \hsla\ than method 2 or 3.}
\label{fig:appA_flux_ratio}
\end{figure}

We run the \textsl{coadd\_x1d} code to process the spectra for all the targets using the three weighting options mentioned above and compare the difference in terms of the flux levels. Moreover, for five of the ten targets, S3, S4, Q4, Q5, Q6 that have co-added spectra from \hsla, we also compare the results between \textsl{coadd\_x1d} and \hsla. We design two steps to evaluate the performance of the two co-addition routines:  
\begin{enumerate}

\item We compare the co-added spectra with each of the original exposure files (i.e., **x1d.fits) to check if line profiles and fluxes are consistent after co-addition. For G130M grating, we check two 15\AA\ wide spectral regions: [1248, 1263]\AA\ for \SII\ 1250/1259/1260\AA\ and [1390, 1405]\AA\ for \SiIV\ 1393/1402\AA. For G160M, we check another two 15\AA\ wide spectral regions: [1540, 1555]\AA\ for \CIV\ 1548/1550\AA\ and [1600, 1615]\AA\ for \FeII\ 1608\AA. We only compare the flux levels because the errors of the co-added spectra will be reduced, thus lower than those of each individual exposure file. We show a typical flux level comparison in Figure \ref{fig:appA_flux_offset}. Among all the targets we analyze, we find that all the co-added spectra show visibly similar line profiles as each individual exposure, however, the flux levels differ depending on the method in use. Generally speaking, \textsl{coadd\_x1d} with method 3 (2) often yield higher (lower) fluxes than those of individual exposures, with absolute flux offset larger than 10$^{-16}$--10$^{-15}$ \fluxunit. \textsl{coadd\_x1d} with method 1 and HSLA-co-added spectra (when available) show consistent flux levels with individual exposures in most cases, with absolute flux offset less than $\lesssim10^{-17}$ \fluxunit. 

\item We further quantify the differences between \hsla\ and \textsl{coadd\_x1d} co-added spectra by calculating flux ratios of \textsl{coadd\_x1d} spectra to \hsla's at a number of absorption-line free regions. This step is only applied to S3, S4, Q4, Q5, and Q6 because they were included in the recent \hsla\ co-added spectra release. In Figure \ref{fig:appA_flux_ratio}, we show an example of the flux ratio comparison using the same target (S3) as in Figure \ref{fig:appA_flux_offset}. Overall, for S3, S4, and Q4, we find consistent fluxes between {\it coadd\_x1d} with method 1 and \hsla, with flux ratios nearly 1.0. For Q5 and Q6, we find the flux ratios deviate from 1.0 by less than 15\%. The co-added spectra with method 2 and 3 show less consistent results with \hsla's, especially at longer wavelengths. 
\end{enumerate}

In all, we find that the line profiles are not significantly altered during the co-added procedures of \hsla\ or \textsl{coadd\_x1d}. When comparing co-added flux levels, we find that \hsla\ and \textsl{coadd\_x1d} with method 1 provide the most consistent co-added fluxes in comparison with the original individual exposure files. \textsl{coadd\_x1d} with method 2 (3) often produce spectral with too low (high) flux values. Therefore, we decide to use the \hsla\ co-added spectra for our analyses when available (i.e., S3, S4, Q4, Q5, Q6). For targets without \hsla\ co-added spectra (i.e., S1, S2, Q1, Q2, Q3), we process the data using \textsl{coadd\_x1d} with method 1. 

\section{Dwarf Galaxies' CGM Absorbers Near the Magellanic System} 
\label{sec:app_ms_lg}

In \S\ref{sec:pv_ms}, we have shown that the LG galaxies at $d_\odot>300$ kpc are coincidentally aligned with the \HI\ emission from the Magellanic System on the position ($\ml$) -- velocity ($\vlsr$) diagram, with which we argue that the ionized cross section of the Magellanic System should be revisited using more robust methods other than the position-velocity diagram. Here we further show that potential CGM absorbers originated from the \HI-rich members of these LG galaxies would appear on a similar $\ml$--$\vlsr$ parameter space, further complicating the diagnosis of an absorber's origin.

Among the 81 LG dwarf galaxies at $d_\odot>300$ kpc as shown in Figure \ref{fig:ms_lg_dwarfs}, we find 40 \HI-rich galaxies (36 dwarfs and M31, M33, NGC55, NGC300) that potentially have extended CGM that could be confused with the Magellanic ionized gas in projection. We show the angular extents of the dark matter halos (as approximated by $\rvir$) of these \HI-rich galaxies as circles in the top panel of Figure \ref{fig:ms_lg_dwarfs} and highlight them as red dots in the middle panel. Given that CGM absorbers are commonly found within $\pm100~\kms$ of the host galaxies' systemic velocities \citep[e.g., ][]{werk13}, if these \HI-rich LG galaxies contain CGM gas in their dark matter halos, the CGM absorbers would be located at similar locations as the host galaxies on the position-velocity diagram. Indeed, as we show in the bottom panel of Figure \ref{fig:ms_lg_dwarfs}, absorbers detected near IC1613 (this work) are found to be mostly aligned with the \HI\ from the Magellanic System, so do a large fraction of ion absorbers detected near M31 \citep{lehner20}.

We estimate the surface area of the CGM of each gas-rich galaxy with $A=2\pi(1-{\rm cos}\theta)(180/\pi)^2$ deg$^2$, where $\theta$ is the projected CGM radius in radians. The total surface area of the CGM of these galaxies is $\sim3500$ deg$^2$ if assuming 100\% detection rate within $\rvir$. In particular, the CGM of M31 accounts for nearly half of the total surface area ($\sim1500$ deg$^2$). Here we have taken into account the overlap of the CGM cross sections of adjacent galaxies.
Given that the detection rate of CGM absorbers in low-mass galaxies is found to be significantly reduced beyond $0.5\rvir$ \citep{bordoloi14}, if we only consider the CGM detection within $0.5\rvir$ for the gas-rich galaxies in our sample but include the full CGM size of M31, the total surface area is $\sim2000$ deg$^2$. Our estimate shows that the cross sections of the extended CGM of LG gas-rich galaxies occupy a non-negligible fraction of the sky near the Magellanic System in projection. Therefore, when considering the ionized cross section of the Magellanic system, one should take into account the contamination of potential CGM absorbers from distant gas-rich galaxies in the LG.  


Though it is beyond the scope of this work to further investigate the true ionized extent of the Magellanic System or the origins of the ion absorbers, the overall ionized gas and dwarf galaxy environment in the Milky Way as well as in the LG should be examined closely in the future. We attempted to differentiate the Magellanic ionized gas from other sources using measurements such as detection rates, ion line ratios, and velocities in other rest frames (e.g., $v_{\rm GSR}$, $v_{\rm LGSR}$). 
None of the attempts led to conclusive answers on the actual extension of the Magellanic ionized gas. The similar kinematics of the Magellanic \HI, \citetalias{fox14}'s ion absorbers, and the LG dwarf galaxies indicates that the coincidence may be partially subject to the co-rotation of the Solar System with the Milky Way. In fact, \cite{richter17} have also noted this coincidental alignment between the LG galaxies and ionized HVCs' absorption velocities. More investigation is needed to further understand the underlying physics of the coincidental alignments among different component in the LG. 


\begin{figure*}[t]
\centering
\includegraphics[width=\textwidth]{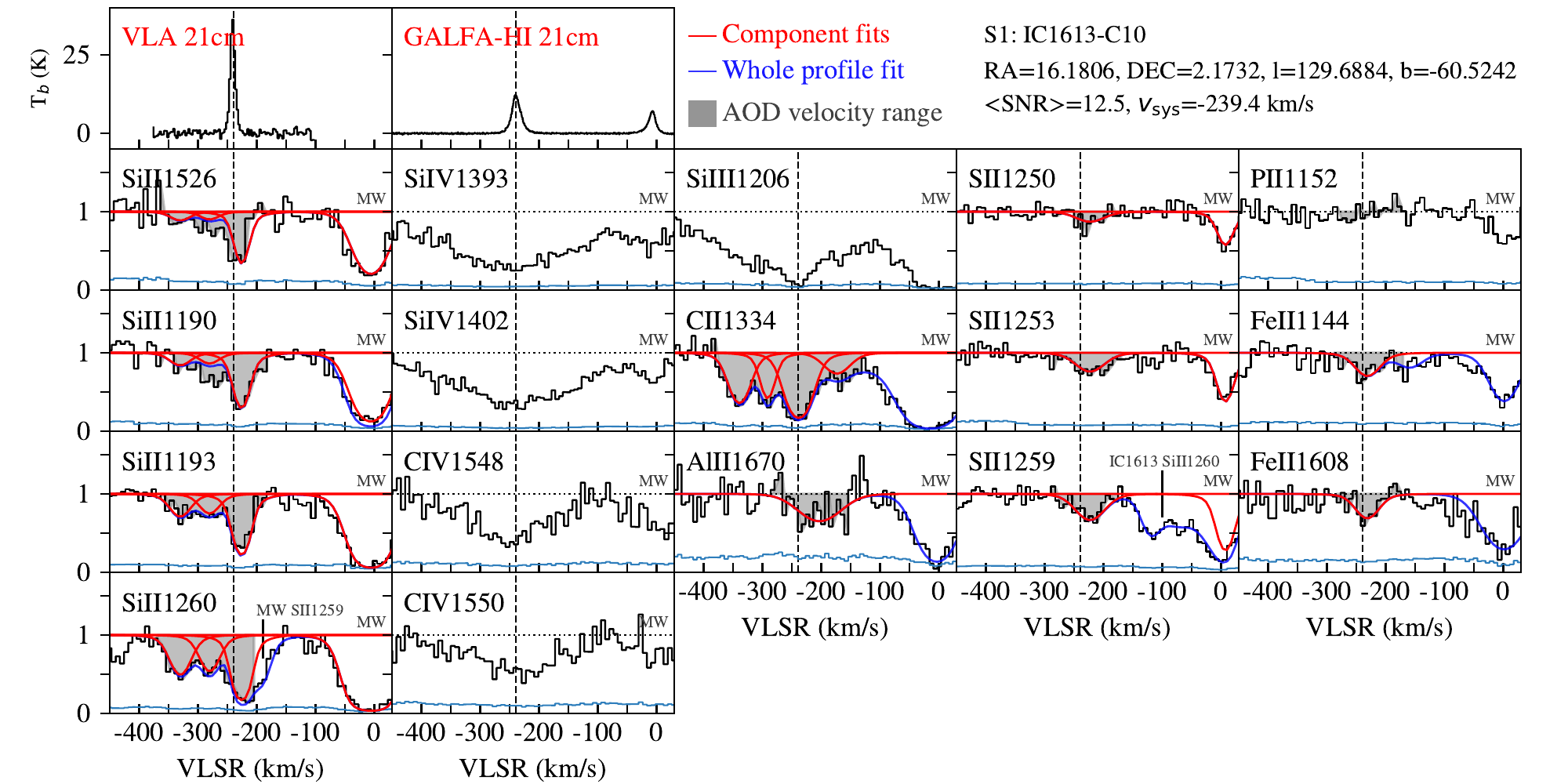}
\caption{S1: IC1613-C10. The top two panels show \HI\ data from VLA \citep{hunter12} and from the GALFA-\HI\ DR2 \citep{peek18}. The rest panels show continuum-normalized ion lines and their Voigt-profile fits when available. The red curves are for individual component fits; the blue curves indicate the whole profile fits, which are the sum of the red curves and nuisance components (e.g., the MW's ISM). The dashed line in each panel indicates the systemic velocity of IC1613, $v_{\rm sys}$. 
The \SiIII, \SiIV, and \CIV\ lines are broad without distinguishable individual components; we attempted Voigt-profile fitting for these lines, but could not find converging results with realistic component widths of $b\leq50$ \mkms. We decide not to use \SiIII, \SiIV, and \CIV\ lines in this case even though extended wind features can be seen in the line profiles. }
\label{fig:s1_spec}
\end{figure*}

\begin{figure*}[t]
\centering
\includegraphics[width=\textwidth]{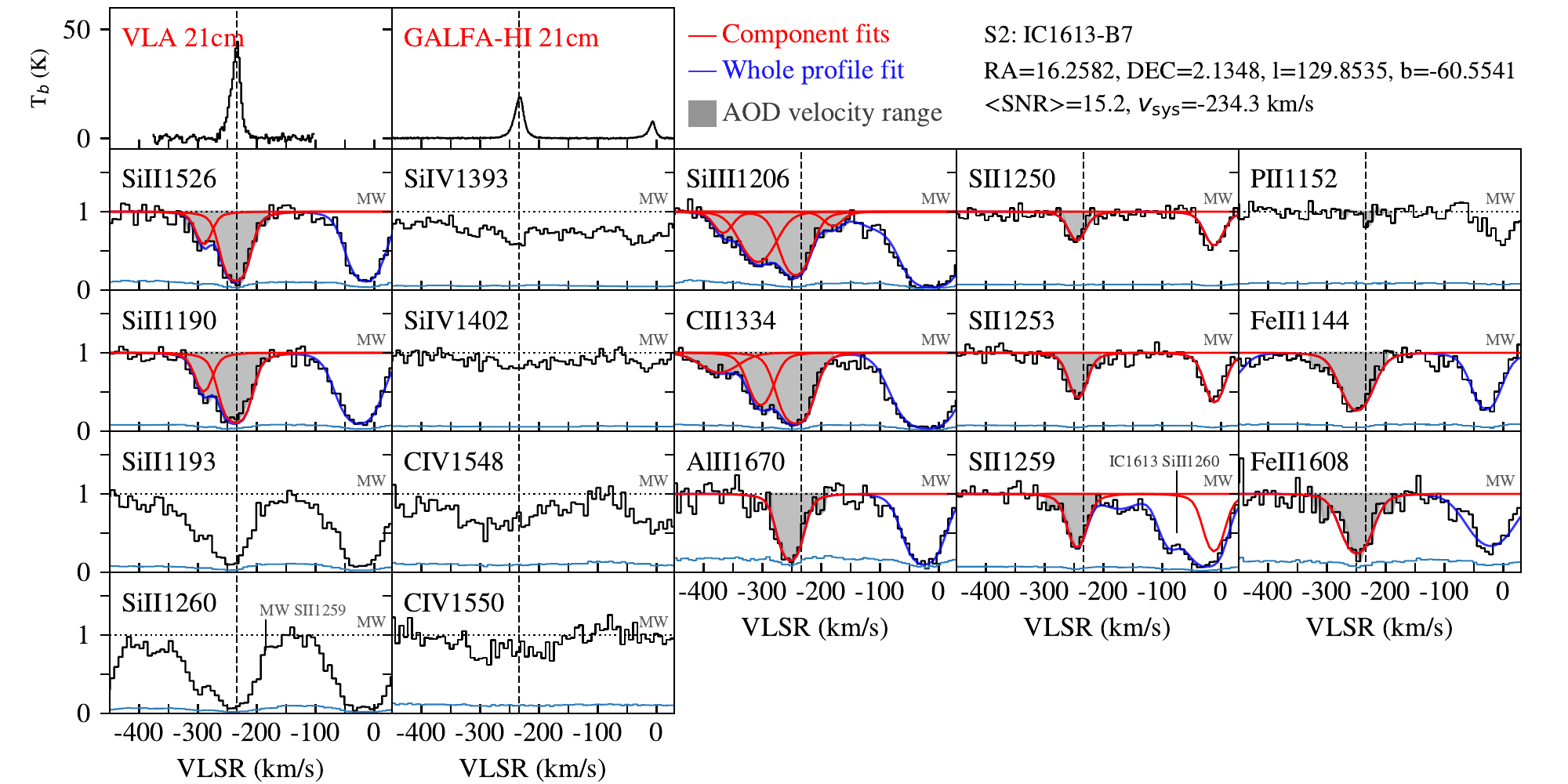}
\caption{S2: IC1613-B7. See Figure \ref{fig:s1_spec} for figure legend description. Even though there are detectable absorption features in \SiIV\ 1393/1402 and \CIV\ 1548/1550 lines, we cannot find robust Voigt-profile fitting results for these lines. And because they are blended with ISM absorption from IC1613, and there is not efficient way to separate the ISM and wind signals, we decide to not to use these lines. In addition, we do not use \SiII\ 1193 in our fitting because the line is heavily saturated. We do not use \SiII\ 1260 either because of its high saturated and contamination from the MW's \SII\ 1259 lines. } 
\label{fig:s2_spec} 
\end{figure*}

\begin{figure*}[t]
\centering
\includegraphics[width=\textwidth]{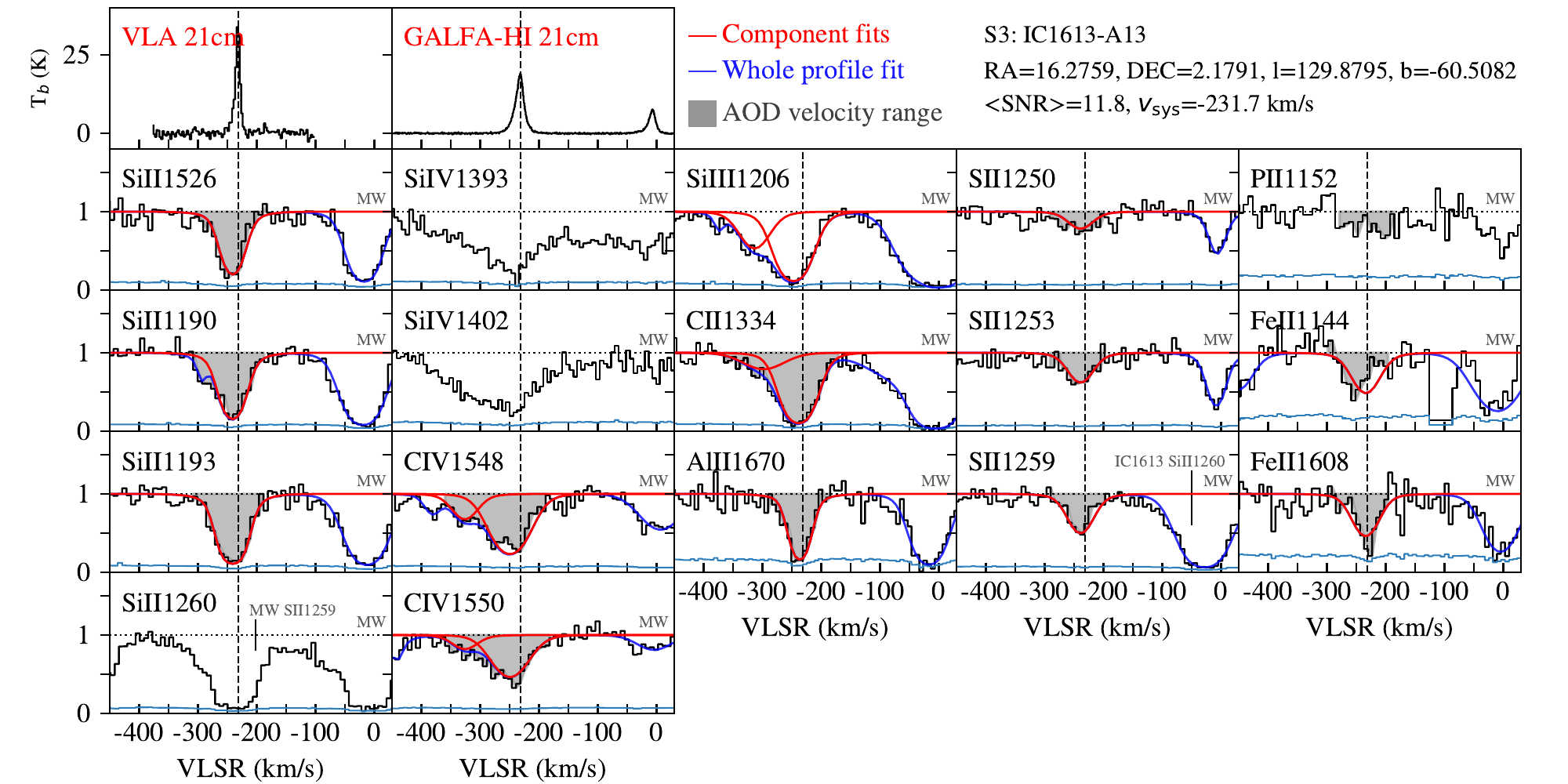}
\caption{S3: IC1613-A13. See Figure \ref{fig:s1_spec} for figure legend description. Most of the lines can be successfully fitted with Voigt profiles except the \SiIV\ doublets, which appear to be broad without apparent individual line components. Meanwhile, \SiIV\ 1402\AA\ is blended with an unknown feature which seems stronger than the corresponding part in \SiIV\ 1393\AA.}
\label{fig:s3_spec}
\end{figure*}

\begin{figure*}[t]
\centering
\includegraphics[width=\textwidth]{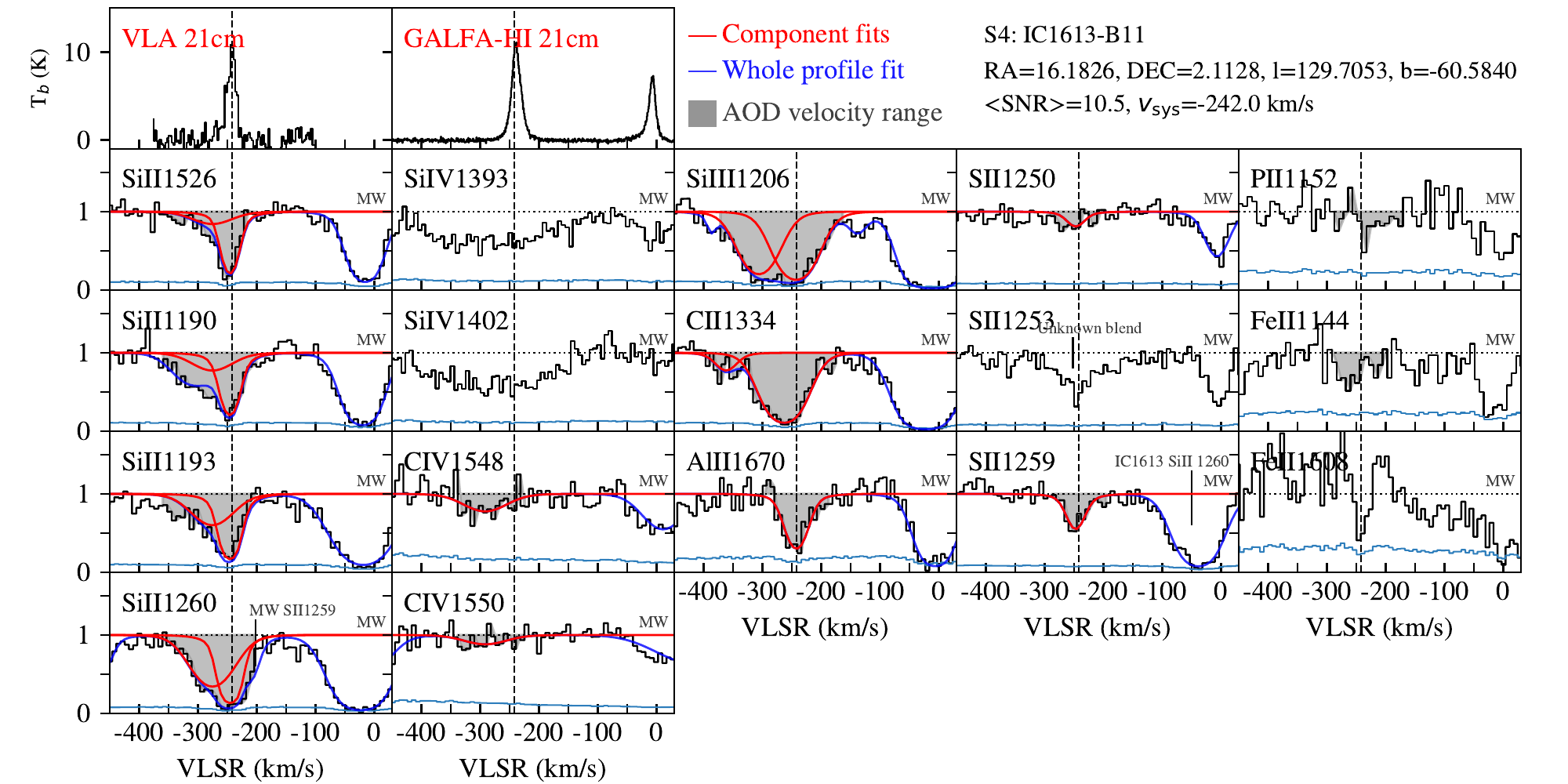}
\caption{S4: IC1613-B11. See Figure \ref{fig:s1_spec} for figure legend description. \SII\ 1253\AA\ appears to be abnormally broader and stronger than \SII\ 1250\AA\ and \SII\ 1259\AA. \FeII\ 1608\AA\ is highly noisy and the MW component of the line is likely to be contaminated by stellar lines. We do not find robust Voigt-profile fits for \SiIV\ 1393/1402 lines.}
\label{fig:s4_spec}
\end{figure*}

\begin{figure*}[t]
\centering
\includegraphics[width=\textwidth]{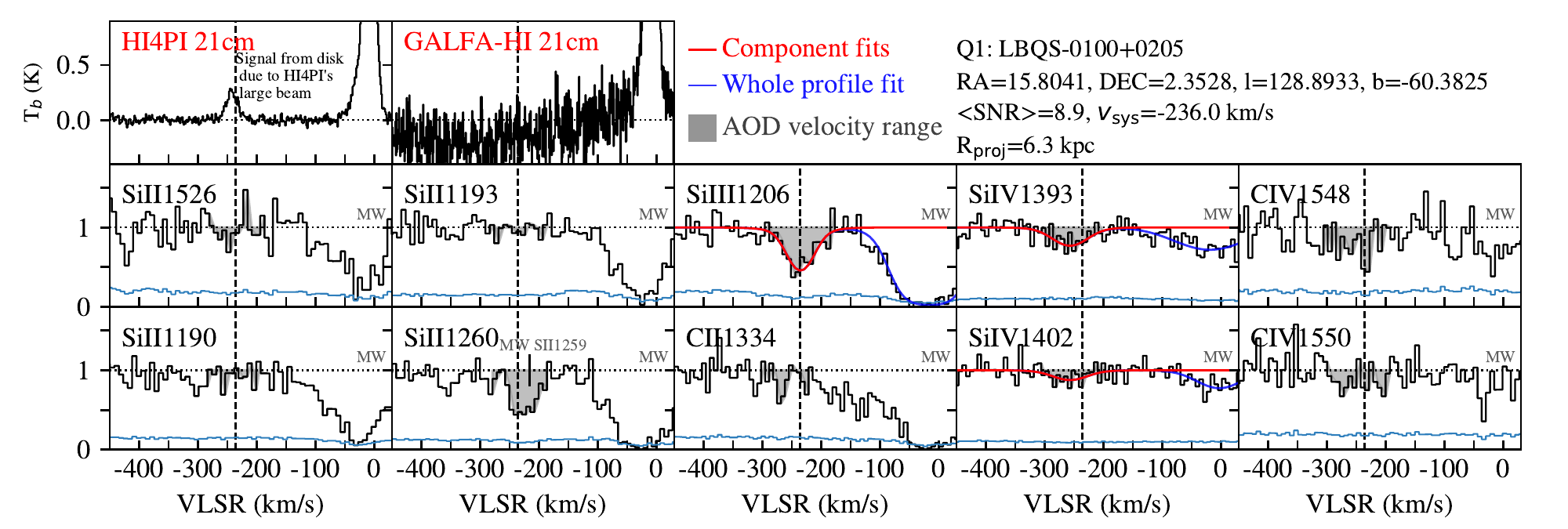}
\caption{Q1: LBQS-0100+0205. See Figure \ref{fig:s1_spec} for figure legend description. The \HI\ 21cm signal in HI4PI is from the disk due to the large beam size (16.2 arcmin) of the data. For all the QSO sightlines, $v_{\rm sys}$ is from the galaxy's systemic velocity determined from \HI\ 21cm observation by \cite{lake89} and \citetalias{mcconnachie12}. \CII\ 1334\AA\ is blended with the wing from the same line of the MW's ISM, as well as \CII* 1335. No detection in \SiII\ 1526/1190/1193 \AA. The absorption feature in \SiII\ 1260\AA\ line is in fact due to \SII\ 1259\AA\ from the MW's ISM. Unlike our stellar sightlines S1-S4, we do not study \PII, \FeII, \SII, and \AlII\ along QSO sightlines Q1-Q6 because these ions are uncommon in a galaxy's CGM due to their low ionization states. }
\label{fig:q1_spec}
\end{figure*}

\begin{figure*}[t]
\centering
\includegraphics[width=\textwidth]{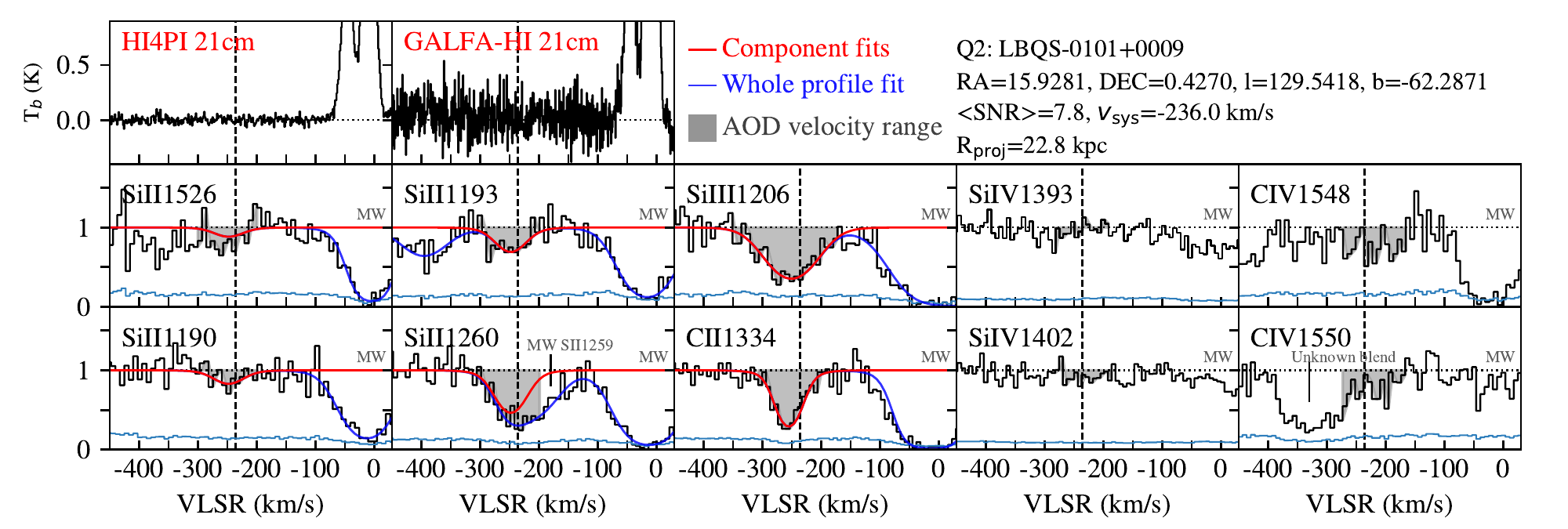}
\caption{Q2: LBQS-0101+0009. See Figure \ref{fig:s1_spec} for figure legend description. the left wing of \CIV\ 1550\AA\ is blended with a broad feature that cannot be identified. }
\label{fig:q2_spec}
\end{figure*}

\begin{figure*}[t]
\centering
\includegraphics[width=\textwidth]{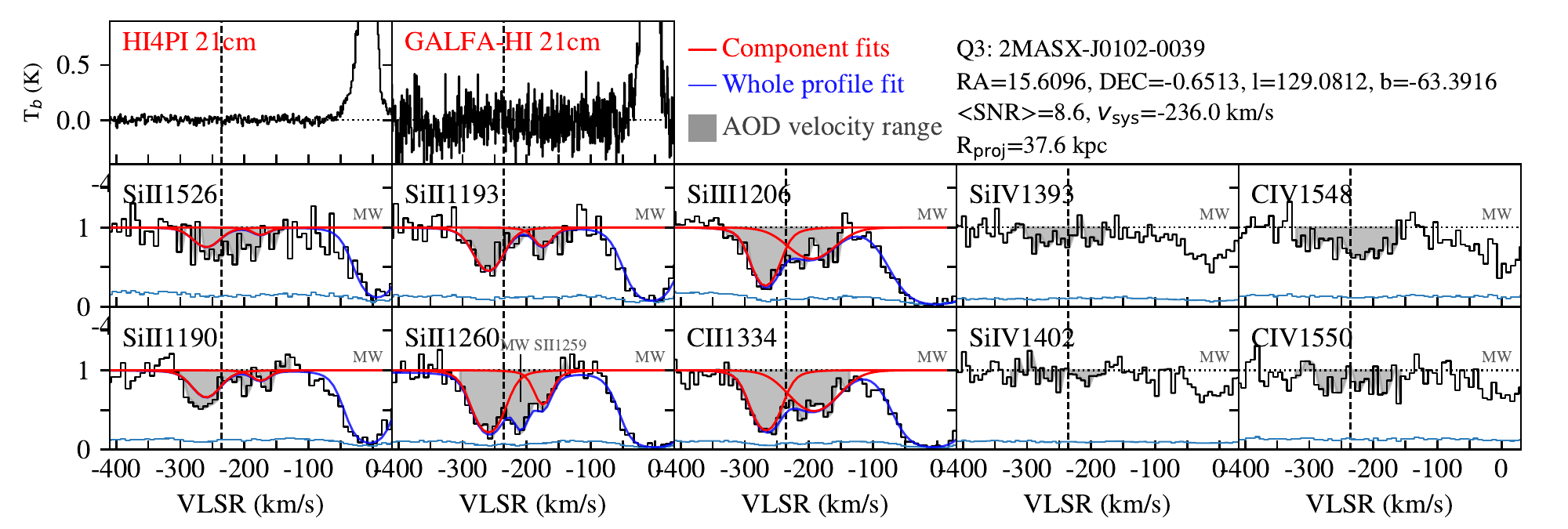}
\caption{Q3: 2MASX-J0102-0039. See Figure \ref{fig:s1_spec} for figure legend description. No converging Voigt-profile solutions can be found in \CIV\ doublets without invoking $b$ values larger than 50 \mkms.  }
\label{fig:q3_spec}
\end{figure*}

\begin{figure*}[t]
\centering
\includegraphics[width=\textwidth]{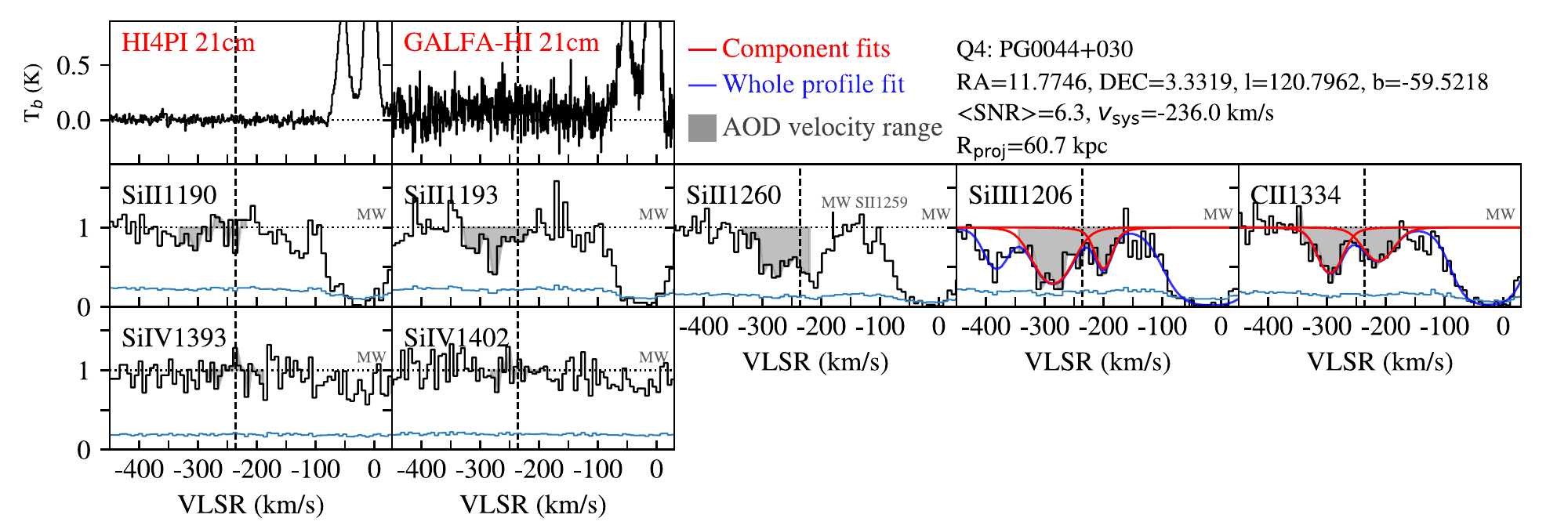}
\caption{Q4: PG0044+030. See Figure \ref{fig:s1_spec} for figure legend description. This target only has G130M grating, so there are no data for \SiII\ 1526\AA\ and \CIV\ doublet. }
\label{fig:q4_spec}
\end{figure*}

\begin{figure*}[t]
\centering
\includegraphics[width=\textwidth]{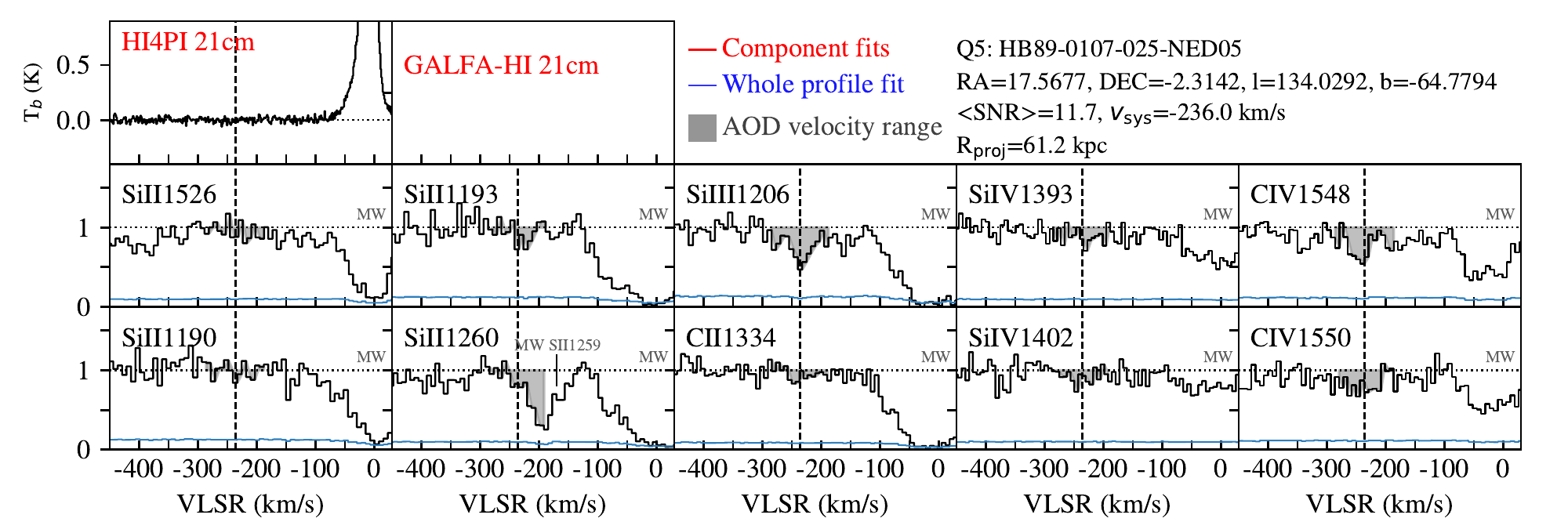}
\caption{Q5: HB89-0107-025-NED05. See Figure \ref{fig:s1_spec} for figure legend description. Most of the lines do not have detection or $1-2\sigma$ absorption signals. }
\label{fig:q5_spec}
\end{figure*}

\begin{figure*}[t]
\centering
\includegraphics[width=\textwidth]{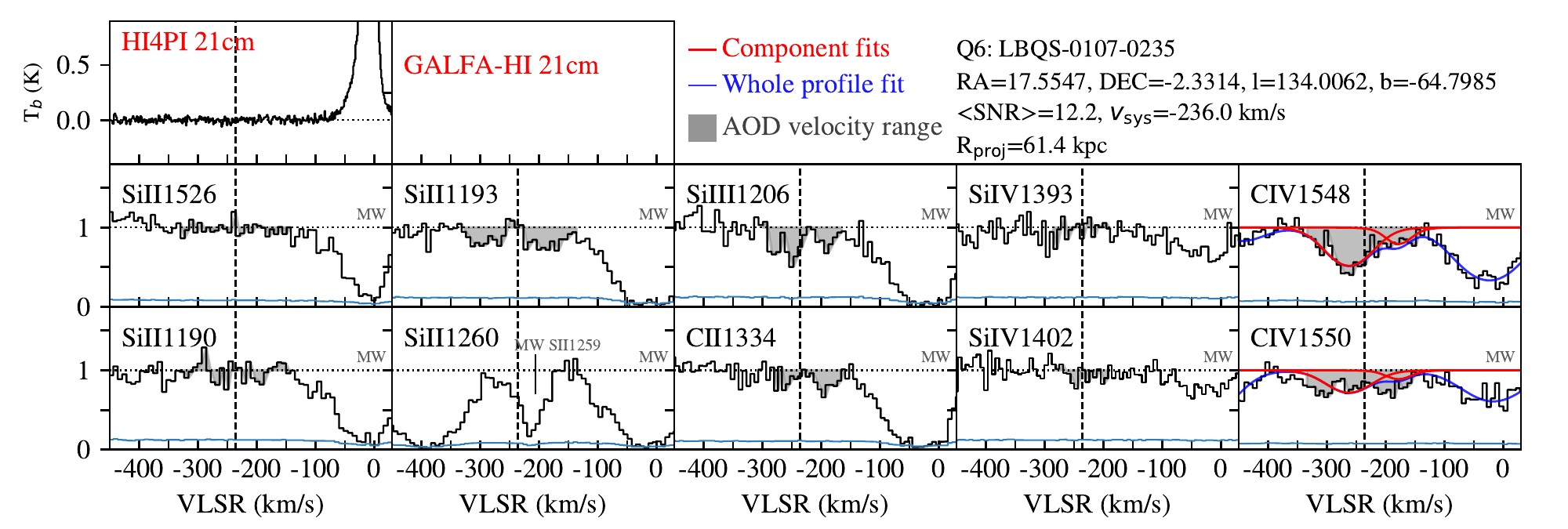}
\caption{Q6: LBQS-0107-0235. See Figure \ref{fig:s1_spec} for figure legend description. \CIV\ 1550\AA\ has different line profile from its 1548\AA\ counterpart. }
\label{fig:q6_spec}
\end{figure*}




\end{document}